\documentclass[fleqn,usenatbib,twocolumn]{mnras}

\usepackage{newtxtext,newtxmath}

\usepackage[T1]{fontenc}
\usepackage{ae,aecompl}

\usepackage{graphicx}	
\usepackage{amsmath}	
\usepackage{nccmath}
\usepackage{bm}
\usepackage{xcolor}
\usepackage{upgreek}
\usepackage{cleveref}

\newcommand{\tianqing}[1]{{\color{black}{#1}}}

\newcommand{\jsrv}[1]{{\color{black}{#1}}}
\newcommand{\hmrv}[1]{{\color{black}{#1}}}
\newcommand{\hmcm}[1]{{\color{black}{(HM: #1)}}}
\newcommand{\dnnz}{\texttt{dNNz}}

\newcommand{\beq}{\begin{equation}}
\newcommand{\beqa}{\begin{eqnarray}}
\newcommand{\eeq}{\end{equation}}
\newcommand{\eeqa}{\end{eqnarray}}

\newcommand{\xihm}{\xi_\mathrm{hm}}
\newcommand{\xihh}{\xi_\mathrm{hh}}
\newcommand{\mr}[1]{\mathrm{#1}}

\defcitealias{projeff}{SP20}

\newcommand{\simgt}{\lower.5ex\hbox{$\; \buildrel > \over \sim \;$}}
\newcommand{\simlt}{\lower.5ex\hbox{$\; \buildrel < \over \sim \;$}}

\defcitealias{Park_2022}{PS22}
\defcitealias{projeff}{SP20}

\title[HSCxSDSS Cluster Cosmology]{Optical Cluster Cosmology with SDSS redMaPPer clusters and HSC-Y3 lensing measurements}

\author[T. Sunayama et al.]{Tomomi Sunayama,$^{1,2}$\thanks{E-mail: tsunayama@arizona.edu}
Hironao Miyatake,$^{2,3,4}$
Sunao Sugiyama,$^{4}$
Surhud More,$^{4,5}$
Xiangchong Li,$^{4,6}$
\newauthor
Roohi Dalal,$^{7}$
Markus Michael Rau,$^{8}$
Jingjing Shi,$^{4,9}$
I-Non~Chiu,$^{10}$
Masato Shirasaki,$^{11,12}$
\newauthor
Tianqing Zhang,$^{8,13}$
Atsushi J. Nishizawa,$^{2,14}$
\\
$^{1}$Steward Observatory, University of Arizona, Tucson, AZ, 85719, U.S.A\\
$^{2}$Kobayashi-Maskawa Institute for the Origin of Particles and the Universe (KMI), Nagoya University, Nagoya, 464-8602, Japan\\
$^{3}$Institute for Advanced Research, Nagoya University, Nagoya 464-8601, Japan \\
$^{4}$Kavli Institute for the Physics and Mathematics of the Universe (WPI), The University of Tokyo Institutes for Advanced Study (UTIAS), The University of Tokyo, \\Chiba 277-8583, Japan \\
$^{5}$Inter-University Centre for Astronomy and Astrophysics, Ganeshkhind, Pune, 411007, India \\
$^{6}$McWilliams Center for Cosmology, Department of Physics, Carnegie Mellon University, 5000 Forbes Ave, Pittsburgh, PA 15213, USA\\
$^{7}$Department of Astrophysical Sciences, Princeton University, Princeton, NJ 08544, USA\\
$^{8}$High Energy Physics Division, Argonne National Laboratory, Lemont, IL 60439, USA\\
$^{9}$Center for Data-Driven Discovery (CD3),  Kavli IPMU (WPI), UTIAS, The University of Tokyo, Kashiwa, Chiba 277-8583, Japan \\
$^{10}$Department of Physics, National Cheng Kung University, 70101 Tainan, Taiwan\\
$^{11}$National Astronomical Observatory of Japan (NAOJ), National Institutes of Natural Science, Mitaka, Tokyo 181-8588, Japan \\
$^{12}$The Institute of Statistical Mathematics,Tachikawa, Tokyo 190-8562, Japan \\
$^{13}$Department of Physics and Astronomy and PITT PACC, University of Pittsburgh, Pittsburgh, PA 15260, USA\\
$^{14}$Gifu Shotoku Gakuen University, Yanaizucho, Gifu, 501-6194, Japan\\
}

\date{\today}
\pubyear{2023}

\begin{document}
\label{firstpage}
\pagerange{\pageref{firstpage}--\pageref{lastpage}}
\maketitle

\begin{abstract}

We present cosmology results obtained from a blind joint analysis of the abundance, projected clustering, and weak lensing of galaxy clusters measured from the Sloan Digital Sky Survey (SDSS) redMaPPer cluster catalog and the Hyper-Suprime Cam (HSC) Year3 shape catalog. We present a full-forward model for the cluster observables, which includes empirical modeling for the anisotropic boosts on the lensing and clustering signals of optical clusters. We validate our analysis via mock cluster catalogs which include observational systematics, \hmrv{such as the projection effect and the effect of baryonic feedback}, and find that our analysis can robustly constrain cosmological parameters in an unbiased manner without any informative priors on our model parameters.
The joint analysis of our observables in the context of the flat $\Lambda$CDM model results in cosmological constraints for $S_8\equiv \sigma_8 \sqrt{\Omega_{\rm m} / 0.3}=0.816^{+0.041}_{-0.039}$.
Our result is consistent with the $S_8$ inference from other cosmic microwave background- and large scale structure-based cosmology analyses, including the result from the \emph{Planck} 2018 primary CMB analysis. 
\end{abstract}




\section{Introduction}

Galaxy clusters are the most massive and gravitationally self-bound objects in the Universe, forming at rare high peaks in the initial density field \citep{bbks, kravtsovborgani}. Their abundance and time evolution are highly sensitive to the growth of structure in the Universe \citep{2001ApJ...553..545H, limahu}, making them an important tool for constraining cosmological parameters (\citealt{PlanckSZ:16, SPT2019, DES2020, KiDS2022}, see also \citealt{weinberg13} for a review). Many current and future galaxy surveys, including the Hyper Suprime-Cam survey \citep[][HSC]{HSCOverview:17}, the Dark Energy Survey\footnote{\url{ https://www.darkenergysurvey.org}} \citep[][DES]{DES2005}, the Kilo Degree Survey\footnote{\url{http://kids.strw.leidenuniv.nl/}} \citep[][KiDS]{KiDs2015}, the Rubin Observatory Legacy Survey of Space and Time\footnote{\url{https://www.lsst.org}} \citep[][LSST]{LSST2009}, \textit{Euclid}\footnote{\url{ https://sci.esa.int/web/euclid}} \citep{euclid2018}, and the Nancy Grace Roman Telescope\footnote{\url{https://wfirst.gsfc.nasa.gov}} \citep{WFIRST2019}, will provide unprecedented numbers of clusters and enable us to carry out optical cluster cosmology analyses with great precision if all the systematic effects are under control. 

In particular, photometric surveys allow for uniform and complete observations of clusters \citep{rm1, rm2, rm3, rm4, camira}, which makes optically identified clusters from photometric surveys an interesting cosmological probe. These surveys also detect the weak lensing signal around clusters by observing the shapes of galaxies in their background. By combining the observed cluster abundances with the halo mass information measured from the cluster lensing signal \citep{2007arXiv0709.1159J, wtg1, wtg4, simet17, muratasdss, mcclintock18, 2019PASJ...71..107M,Inon2022}, it is possible to carry out a self-contained analysis that both calibrates cluster masses and constrains cosmology \citep{limahu, takadabridle, rozo10, oguritakada11,Inon2023}. Recent studies, such as \cite{costanzisdss} and \cite{desy1cl} have calibrated cluster masses using cluster lensing signals, and used these calibrations to simultaneously constrain cosmology and the mass-observable relation (MOR) using cluster abundances. Both studies (especially the latter) found that the resulting cosmological constraints favored lower values of $\Omega_\mr{m}$ and higher values of $\sigma_8$ compared to other constraints derived from cosmic microwave background (CMB) or large-scale structure (LSS) data. These findings suggest the presence of yet unknown systematic effects for optical clusters.

One of the main systematic effects for optically identified clusters is the so-called {\it projection effect} in which interloper galaxies along the line-of-sight (LOS) to a cluster are mistakenly identified as members of the cluster. 
Projection effects alter the mass-observable relation such that the observable for the optical clusters, which is the probability weighted sum of member galaxies (also called {\it richness}), is boosted with respect to its halo mass \citep{costanzisdss}. 
In addition to the alteration of the mass-richness relation, \citet[][\citetalias{projeff} henceforth]{projeff} found that projection effects boost the amplitude of cluster lensing and clustering signals on large scales due to the preferential identification of clusters lying at the nodes of filaments aligned with the LOS direction. 
This results in an anisotropic distribution of matter around optical clusters, which breaks the isotropic halo model generally assumed to carry out lensing mass calibrations for clusters. Therefore, these anisotropic boosts inevitably lead to errors in the cluster mass calibrations, if not properly modeled.

These anisotropic boosts have been parameterized in a few cluster cosmology analyses. \cite{tokrause2} modeled the boost in their combined cosmology analysis with DES Y1 galaxies and redMaPPer clusters and obtained a similar size of the boost as the study by \citetalias{projeff}. 
In previous work, \citet[][\citetalias{Park_2022} henceforth]{Park_2022} also employed this boost model in a cluster cosmology analysis of the abundance, clustering and lensing signal of the red-sequence Matched-filter Probabilistic Percolation (redMaPPer) cluster catalog \citep{rm1}, constructed from the Sloan Digital Sky Survey (SDSS) DR8 data \citep{Aihara_etal2011}. 
While the value for the boost parameter inferred in \citetalias{Park_2022} was consistent with the inference in \cite{tokrause2}, the resulting cosmological constraints favored low $\Omega_{\rm m}$ and high $\sigma_8$, similar to \cite{costanzisdss} and \cite{desy1cl}. This raises the question of how exactly the boost manifests in the observables of the real data, i.e., in the measured abundance, lensing and clustering signals. 
In particular, \citetalias{Park_2022} found hints of internal tensions among the different sectors of the data vector from a series of post-unblinding analyses.
One of the findings from these tests is significant underprediction of the measured cluster lensing signals under a \emph{Planck} cosmology (see Fig.~10 in \citetalias{Park_2022}).
This finding motivates our study using the cluster lensing signals measured from the HSC Year 3 (HSC-Y3) data, which has significantly deeper photometry and better image quality than the SDSS shape catalog and enables us to select source galaxies more securely and robustly due to reduced systematics related to intrinsic alignment and source-cluster member confusion.

This paper is structured as follows. 
In Section~\ref{sec:data}, we describe the SDSS redMaPPer cluster catalog and the HSC-Y3 shape catalog as well as the HSC mock catalogs.
In Section~\ref{sec:meas}, we describe measurements of the cluster abundance, clustering, and lensing observables used in our analysis.
In Section~\ref{sec:model}, we describe our analysis method and the theoretical model including emulator-based halo model predictions and models for cluster systematics.
In Section~\ref{sec:mock_test}, we discuss the validation tests against possible cluster systematics using mocks  and validate our analysis through these tests.
In Section~\ref{sec:result}, we present the result of our analysis pipeline on real data and a series of internal consistency tests.
We finally summarize our study and discuss its implications in Section~\ref{sec:summary}.

\section{Data}
\label{sec:data}
In this section, we describe the cluster catalog used to measure the cluster abundance and clustering signals in Sec.~\ref{sec:data:redmapper} and the galaxy shape catalog used to measure the cluster lensing signals in Sec.~\ref{sec:data:hsc_source}. In Sec.~\ref{sec:data:hsc_mock}, we describe the mock catalogs used to compute the covariance matrix for the lensing signals.

\subsection{SDSS redMaPPer clusters}
\label{sec:data:redmapper}
The redMaPPer cluster finding algorithm was developed by \citet{rm1} to identify galaxy clusters using red-sequence galaxies in optical imaging surveys \citep{rm1,rm2,rm3,rm4}.
We use the redMaPPer galaxy cluster catalog v6.3 derived from the SDSS DR8 photometric galaxy catalog \citep{Aihara_etal2011}.
The cluster finder uses the $ugriz$ magnitudes and their
errors to identify overdensities of red-sequence galaxies with
similar colors as galaxy clusters. 
For each cluster, the catalog contains an optical richness 
estimate $\lambda$, a centering probability $p_{\rm cen}$, RA, Dec 
as well as a photometric redshift 
$z_\lambda$ and a spectroscopic redshift $z_{\rm spec}$, if available.

This catalog covers 10,401 $\mathrm{deg}^2$, and we use a total of 8,379 clusters within the richness range $20 \leq \lambda \leq 200$ and the redshift range $0.1\leq z_{\lambda} \leq 0.33$. 
For model predictions discussed in Sec.~\ref{sec:mod:emu}, we choose a single representative redshift for this range, $z=0.251$, which corresponds to the mean redshift of the redMaPPer clusters and comment on the impact of our choice.
This redshift range ensures that the catalog is very nearly volume-limited. At higher redshifts, the clusters near our lower richness threshold will require uncertain incompleteness corrections due to the flux limit of SDSS.
Throughout this paper, we use the position of the most probable central galaxy in each cluster region as a proxy of the cluster center.
We also use the random catalogs provided along with the redMaPPer cluster catalog. 
These catalogs contain corresponding position information, redshift, richness, and weights for each random cluster. When using the randoms for any further subsamples based on richness that we adopt in our analysis, we carry out appropriate cuts in richness, and redshifts while constructing the corresponding random subsamples.

\subsection{HSC-Y3 galaxy shape catalog}
\label{sec:data:hsc_source}
The HSC-Y3 galaxy shape catalog is based on the HSC-Y3 internal data release processed with \texttt{hscPipe} v7 \citep{Li_2022}, which consists of multi-band $grizy$ imaging.
Briefly, the galaxies for the catalog were selected from the full-depth-full-color region with basic quality cuts regarding pixel level information and consist of a magnitude-limited sample in the $i$-band ($i<$24.5 mag using \texttt{cmodel} magnitude \citep{Bosch_2017}).
The galaxy shapes are measured in the $i$-band, with a $5\sigma$ depth of $i\sim26$, using the re-Gaussianization shape measurement method \citep{hirataseljak}. We refer the reader to \cite{Li_2022} for further details on the selection of galaxies that make up the shape catalog.
The shape catalog consists of about 35.7 million galaxies in an area of 433.48~deg$^2$ with an effective number density of 19.9 arcmin$^{-2}$. The area is divided into six fields: XMM, GAMA09H, GAMA15H, HECTOMAP, VVDS, and WIDE12H.
The weight $w_s$ for each source galaxy is given by the inverse variance of the shape noise:
\begin{equation}
    w_s = (\sigma_{e,s}^2+e_{{\rm rms},s}^2)^{-1},
\end{equation}
where $\sigma_{e,s}$ is the shape measurement error and $e_{{\rm rms},s}$ is the intrinsic root mean square ellipticity per component for a source galaxy.

The shape measurements were calibrated based on detailed image simulations, which adapt Hubble Space Telescope images from the COSMOS region to the quality expected from ground based imaging in the HSC survey \citep{Mandelbaum_2018,Li_2022}, and the multiplicative and additive biases in the shape measurement were inferred for each object depending upon the similarity of their properties to galaxies in these simulations. The uncertainty in the calibration of the multiplicative bias is assessed to be less than $\sim 10^{-2}$.

The HSC-Y3 
data provide photometric redshift estimates based on three different methods: \dnnz \citep{nishizawa2020photometric},
DEMPz \citep{DEMP}, and MIZUKI \citep{Tanaka_2017}.
The first method is neural network-based conditional density estimation algorithms and the second is an empirical method to find the local relation between color and redshift, while the third method is a template fitting-based method.
All three methods provide an estimation of the redshift posterior distribution $P(z_s)$ for each galaxy.
We select a sample of source galaxies with
\begin{equation}
    \int_{z_{s,\hmrv{\rm min}}}^{7} d z_s P(z_s) \geq 0.99,
\end{equation}
where $z_{s,\hmrv{\rm{min}}}$ is the minimum redshift of the source galaxies. We chose $z_{s,\hmrv{\rm{min}}}=0.6$ to ensure that the redshift distribution of the selected source galaxies does not overlap with our cluster sample. We use the \dnnz photoz catalog as our fiducial choice. The effective number density of our source galaxies is about 16~arcmin$^{-2}$ and the mean redshift of the source sample is \hmrv{$\langle z_s \rangle\sim 1.21$}.

\subsection{HSC Mock Catalogs}
\label{sec:data:hsc_mock}
We use a suite of 800 mock shape catalogs as well as mock cluster catalogs to measure the covariance matrix of our cluster lensing signals. 
The mock shape catalogs are generated following the method described in \cite{Shirasaki_2019}, which accounts for the survey footprint, galaxy shape noise, shape measurement error, and photometric redshift error of the HSC-Y3 shear catalog. 
Briefly, galaxies are populated in the mock catalog using their actual angular positions in the HSC-Y3 shape catalog and redshifts estimated from the \texttt{dNNz}, based on the method described in \cite{Shirasaki_2014} and \cite{Shirasaki_2017}. Then, the shapes of these galaxies are randomly rotated to erase the real lensing signals in their shapes. 
Lastly, the lensing distortion at each position of the galaxy is interpolated from the full-sky ray-tracing lensing simulations \citep{Takahashi_2017}. These simulations calculate the light-ray deflection using the projected matter density field in 38 spherical shells with a radial thickness of 150$h^{-1}$Mpc up to redshift $z=2.4$. The angular resolution of the shear map is 0.43~arcmin.
After the lensing distortion is added to each galaxy's intrinsic shape, a shape measurement noise, which is generated from a zero-mean Gaussian distribution with the standard deviation measured in the HSC-Y3 shear catalog, is added. For more details about the HSC-Y3 mock shape catalog generation, please refer to \citep{More_etal2023}.

To construct the mock cluster catalogs, we use the resultant halo catalogs from the N-body light-cone simulations \citep{Takahashi_2017}, which are also used for the full-sky lensing simulations, where halos are identified from the N-body simulation outputs at each redshift by the \texttt{Rockstar} \hmrv{halo finder} \citep{rockstar}.
To generate a hypothetical redMaPPer cluster catalog from the halo catalogs of the light-cone simulation, we follow the method to infer the intrinsic mass-richness relation of the SDSS redMaPPer clusters by \cite{muratasdss} and obtain the parameter values of $A=3.636$, $B=0.983$, $\sigma_0=0.212$, and $q=-0.140$.
These parameters are explained in Eqs.~\ref{eq:mor1} and ~\ref{eq:mor2}.
We follow the same notation used in \cite{muratasdss} to describe the mass-richness relation (see Eqs.~\ref{eq:mor1} and \ref{eq:mor2}), and the best-fit parameters for $P(\ln \lambda|M)$ are estimated from MCMC analysis for the sample of $20 \leq \lambda \leq 200$: A =3.207, B= 0.993, $\sigma_0$ = 0.456 and q = -0.169. 
Using the estimated probability distribution $P(\ln \lambda|M)$,
we randomly assign a mock richness $\lambda$ to each halo that
resides inside the HSC-Y3 footprint in the redshift range of $0.1 \leq z \leq 0.33$ and the richness range of $20 \leq \lambda \leq \lambda$, in each light-cone realization.

For an estimation of the covariance matrix for the stacked lensing signals, we construct the mock random catalogs in a way that each catalog reproduces the redshift and richness distributions of the mock clusters with random angular distribution within the HSC-Y3 survey footprint.

\section{Measurements}
\label{sec:meas}
In this section, we describe how we measure the abundance, clustering, and lensing signal around our sample of galaxy clusters. Note that we follow the same methodology employed in \citetalias{Park_2022} for the measurements of abundance and clustering signals. For the lensing measurements, we follow the methodology of \citet{More_etal2023} using the shape catalog from \citet{Li_2022}. For completeness, in this section, we describe the procedure we adopt for these measurements.

\subsection{Abundance}
\label{sec:meas:nc}
Our data vector includes the abundance of the redMaPPer clusters, i.e. the total number of clusters detected across the SDSS footprint in a given richness bin.
To measure cluster abundances, we first count the number of clusters in our choice of richness bins, i.e, for $\lambda \in [20,30), ~[30,40), ~[40,55), ~[55,200)$ in redshift range $0.1\leq z \leq 0.33$. 
Since the measured cluster abundance is affected by survey geometries such as masks, we correct the raw cluster abundance $N_\mr{raw}(\lambda_{\alpha})$ for such missed clusters using an effective area $\Omega_{\rm eff}(\lambda,z_{\lambda})$ in the following manner.
\begin{equation}
    N_\mr{corr}(\lambda_{\alpha}) = \sum_{l;\lambda_l \in \lambda_{\alpha}}\frac{\Omega_{\rm tot}}{\Omega_{\rm eff}(\lambda_l,z_l)}\,,
\end{equation}
with the survey area without any masks $\Omega_{\rm tot}$.
The effective area $\Omega_{\rm eff}$ is a function of the richness and redshift of a cluster because the projected size of a cluster depends on richness and redshift and is defined through a detection efficiency $w_{\rm rand}(z,\lambda)$
\begin{equation}
    \Omega_{\rm eff}(\lambda,z_{\lambda})=\frac{\Omega_{\rm tot}}{w_{\rm rand}(\lambda,z_{\lambda})}\,,
\end{equation}
with $w_{\rm rand}$ given by \cite{muratasdss}, calculated on the same catalog as ours, to correct for the missed detections. The resultant ratios of $N_\mr{raw}/N_\mr{corr}$ are 92.7\%, 98.4\%, 98.5\%, and 98.6\% for the cluster abundance of $\lambda \in [20,30), ~[30,40), ~[40,55), ~[55,200)$ in redshift range $0.1\leq z \leq 0.33$.

\subsection{Projected Correlation Functions}
\label{sec:meas:wp}
We compute the projected auto-correlation functions for each cluster sample as
\begin{equation}
w_{\rm p}(R)=2 \int_{0}^{\pi_{\rm max}} d\pi \,\xi_{\rm cc}(R,\pi),
\end{equation}
where $R$ is the projected comoving separation, $\pi$ is the line-of-sight (LOS) separation. 
We choose $\pi_{\rm max}=100h^{-1}{\rm Mpc}$ to avoid a large suppression in amplitude with a small $\pi_{\rm max}$ due to the redshift-space distortion \citep{vdBosch2013} and use the Landy-Szalay estimator \citep{landyszalay} to compute $\xi_{\rm cc}$, such that
\begin{equation}
\xi_{\rm cc}=\frac{\left<D D  \right>-2 \left<D R \right>+\left<R R \right>}{\left<R R \right>}\,,\label{eq:landy-szalay}
\end{equation}
where $DD$, $DR$, and $RR$ are the normalized number of cluster-cluster, cluster-random, and random-random pairs, respectively. We use the random catalog provided together with the SDSS redMaPPer cluster catalog and we selected randoms using the same richness and redshift selection as the redMaPPer clusters.

\subsection{Lensing Signals}
\label{sec:meas:dsigma}

\begin{figure*}
	\includegraphics[width=0.6\columnwidth]{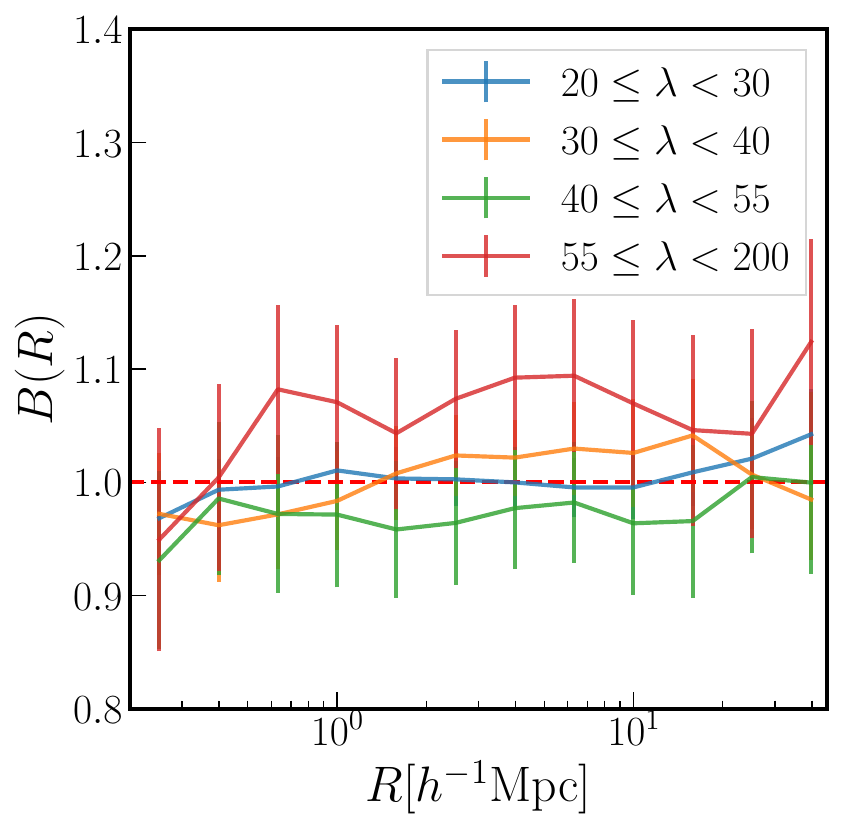}
 \includegraphics[width=0.6\columnwidth]{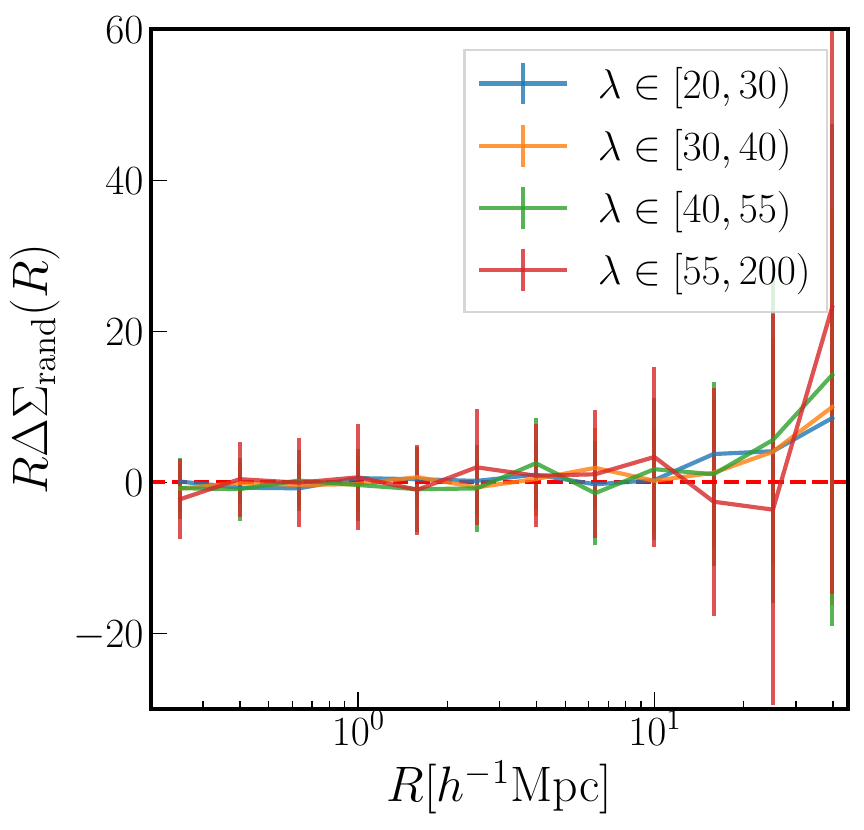}
 \includegraphics[width=0.6\columnwidth]{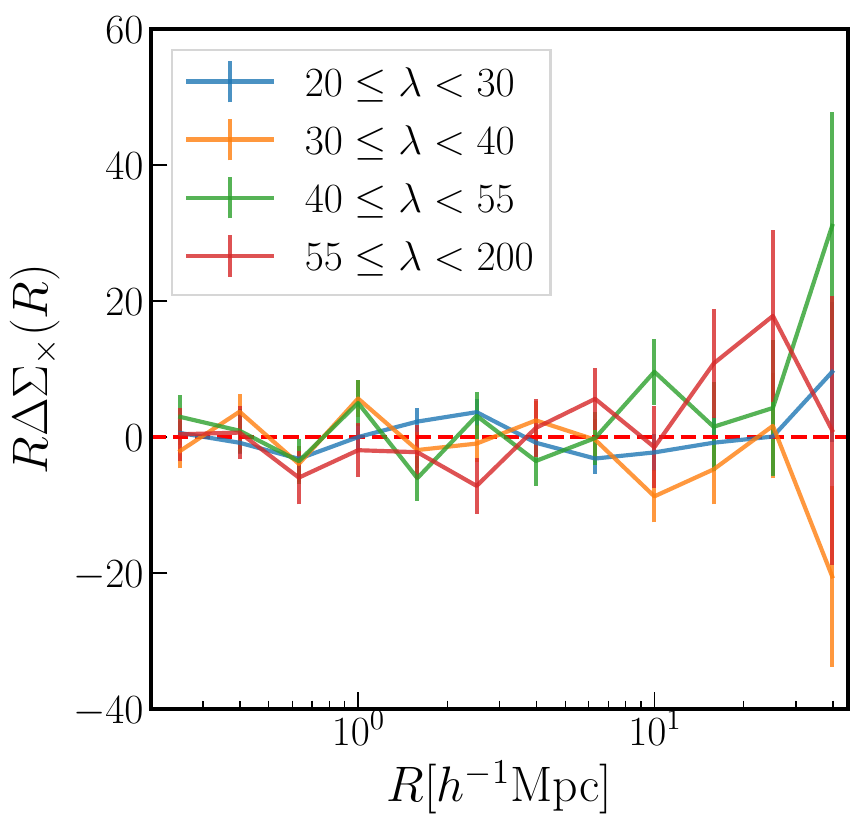}
    \caption{\textit{Left:} The  ``boost factor'' $B(R)$, which measures an excess in the number of the SDSS redMaPPer clusters (lens) and HSC-Y3 source galaxy pairs using random points and the source pairs. The error bars are estimated from the boost factors measured from 800 HSC mock clusters and source catalogs. \textit{Middle:} The  random signals $\Delta \Sigma_{\rm rand}(R)$. \textit{Right:} The 45-degree rotated component of the lensing measurements $R \Delta \Sigma_{\times}(R)$.}
    \label{fig:boost_cross}
\end{figure*}


The weak gravitational lensing signal at the projected separation from the cluster center $R$ is given by
\begin{equation}
\Delta \Sigma(R) = \bar{\Sigma}(<R)-\Sigma(R)=\Sigma_{\rm crit} \gamma_{\rm t}(R),
\end{equation}
where $\Delta \Sigma(R)$ is the excess surface density, $\Sigma(R)$ is the projected mass density profile, $\bar{\Sigma}(<R)$ is the average mass density within radius $R$, $\Sigma_{\rm crit}$ is the critical surface mass density, and $\gamma_{\rm t}$ is the tangential shear at radius $R$. 

The critical density $\Sigma_{\rm crit}$ for lens and source galaxies depends on the redshifts $z_l$ and $z_s$, respectively, and is given by
\begin{equation}
\Sigma_{\rm crit} = \frac{c^2}{4\pi G}\frac{d_A(z_s)}{(1+z_l)^2 d_A(z_l)d_A(z_l,z_s)},
\end{equation}
where $d_A(z_l)$, $d_A(z_s)$, and $d_A(z_l,z_s)$ are the comoving angular diameter distances for the lens at $z_l$, the source at $z_s$,and the lens-source pair from $z_l$ to $z_s$, respectively. 
The factor $(1+z_l)^2$ in the \hmrv{denominator} 
arises from the choice of comoving coordinates.

To compute the weak lensing signal for our cluster sample, we follow the methodology described in \cite{More_etal2023} and use the HSC-Y3 shape catalog \cite{Li_2022}. We measure the average projected mass density profile $\Delta \Sigma(R)$ as 
\begin{equation}
\Delta \Sigma(R)=\frac{\Sigma_{ls} w_{ls} e_{t,ls}\left<\Sigma_{\rm crit}^{-1}\right>_{ls}^{-1}}{2{\cal R}(1+\hat{m}(R))\Sigma_{ls} w_{ls}},
\label{eq:DeltaSigma}
\end{equation}
where ${\cal R}$ is the shear responsivity \citep[defined in Eq.~17 in ][]{More_etal2023}, and the summation runs over all the lens-source pairs separated by R and the subscript ``l'' and ``s'' represent lens and source galaxies, respectively. The tangential components of ellipticities are denoted by $e_{t,ls}$ and the weight $w_{ls}$ is given by
\begin{equation}
w_{ls} = w_l w_s \left \langle \Sigma_{\rm crit}^{-1}\right\rangle_{ls}^{2},
\end{equation}
where $w_l$ is one for all the redMaPPer clusters and $w_s$ is the weight listed in the source catalog, respectively.
The overall factor $(1+\hat{m}(R))$ in the denominator is used to correct for a multiplicative shear bias \citep{miller13} and is computed as an average of the lens-source pairs at a distance $R$. \hmrv{In addition, after computing the lensing signal following Eq.~\eqref{eq:DeltaSigma}, we apply a correction for the additive and multiplicative selection bias as described in \cite{More_etal2023}.}

To measure the lensing signal accurately, we apply two corrections: boost and random corrections, to remove possible contamination from residual systematics. 
For these corrections, we measure the lensing signals around random points and we use 20 times more number of random points than the number of the redMaPPer clusters.

The boost correction is for a dilution of the lensing signal due to source galaxies that are physically associated with the lens \hmrv{cluster.} 
This effect can be corrected by a boost factor $B(R)$ given by
\begin{equation}
B(R)=\frac{\sum_r w_r}{\sum_l w_l} \frac{\Sigma_{ls} w_l w_s}{\Sigma_{rs} w_r w_s},
\end{equation}
where the superscript ``r'' stands for random catalogs and $w_r$ is the weight of the random provided by the catalog. 
$N_r$ and $N_l$ are the numbers of randoms and lenses, respectively. If there is no physical correlation between lens and source galaxies, $B(R)=1$. The boost factor $B(R)$ generally deviates from 1 on small scales but converges to 1 on large scales. By multiplying the measured signal by the boost factor, it corrects the dilution of the signal. 
The left panel of Fig.~\ref{fig:boost_cross} shows that boost factors for all the richness bins are almost one on all scales.
This implies that the distributions of the redMaPPer clusters and the selected source galaxies do not have any spatial correlations, and therefore no contamination from the foreground galaxies on our lensing measurements.
This is because of our secure selection of source galaxies with $z_{s}\geq 0.6$, well above our cluster high-$z$ limit of 0.33. 
Note that the $\chi^2$ value of the boost factor is at most 8 with the degree of freedom of 12 for the case of the cluster sample with $\lambda \in ~[55,200)$\hmrv{, because of the strong correlation of the boost factor between radial bins}.
Even though applying the boost correction would not change the measured lensing signals, we still apply the correction to the raw estimated $\Delta\Sigma(R)$.

We correct for possible residual systematics in the shape measurement due to imperfect corrections of optical distortion across the field of view of the camera by using the lensing signal around random points in the survey area. The measurement of the lensing signal around such points should not result in any coherent tangential distortions. 
On the contrary presence of such signals around the randoms indicates possible systematics.
The middle panel of Fig.~\ref{fig:boost_cross} shows that random signals $\Delta \Sigma_{\rm rand}(R)$ for all the richness bins are consistently zero on all scales.
This implies that there are no residual systematics in the shape measurement.
Even though we do not need to apply random corrections for our case, we still apply the correction to the boost corrected $\Delta \Sigma(R)$ by subtracting the lensing signal around the randoms from the one around the lenses.
The right panel of Fig.~\ref{fig:boost_cross} shows the 45-degree rotated component of the lensing signals after the boost and random corrections.
For all the cluster samples, the values of $R \Delta \Sigma_{\times}(R)$ are consistently zero indicating that the lensing signals are not contaminated by residual systematic errors.
The cluster sample with $\lambda \in [30,40)$ has the largest $\chi^2$ for $R\Delta\Sigma_{\times}(R)$ and it is 10.2 with the degree of freedom of 12, statistically consistent with a null detection.

Note that we need to assume a background cosmology for the clustering and lensing measurement calculations. We have assumed a flat $\Lambda$CDM model with $\Omega_\mr{m}=0.27$.  In our likelihood analyses, we correct our theoretical predictions following the framework presented in \citet[Sec.~III.B.2 in][]{hscy3_3x2_minimal} which is based on the dependence of the corresponding observables on cosmological parameters from \citet{2013ApJ...777L..26M}.

\subsection{Covariance Matrix}
\label{sec:meas:cov}

For the abundance and clustering components of a covariance matrix, we estimate the covariance matrix directly from the SDSS redMaPPer catalog using jackknife resampling. 
We split into 83 independent jackknife regions and assume a block-diagonal covariance matrix where each of the five components of our data vector ($\mathbf{N}_\mathrm{c}, \mathbf{w}_\mr{p,1}, \mathbf{w}_\mr{p,2}, \mathbf{w}_\mr{p,3}, \mathbf{w}_\mr{p,4}$) are considered uncorrelated.
This assumption is validated in previous studies (e.g., \cite{simet17}). We additionally test this in Appendix~\ref{app:off-diag} and reach the same conclusion that the off-diagonal components of the full covariance matrix have negligible impact on cosmology analysis.
Note that we apply the Hartlap correction \citep{hartlap} to perform a debiasing of our inverse covariance matrix used in the likelihood analyses.

\begin{figure}
	\includegraphics[width=0.8\columnwidth]{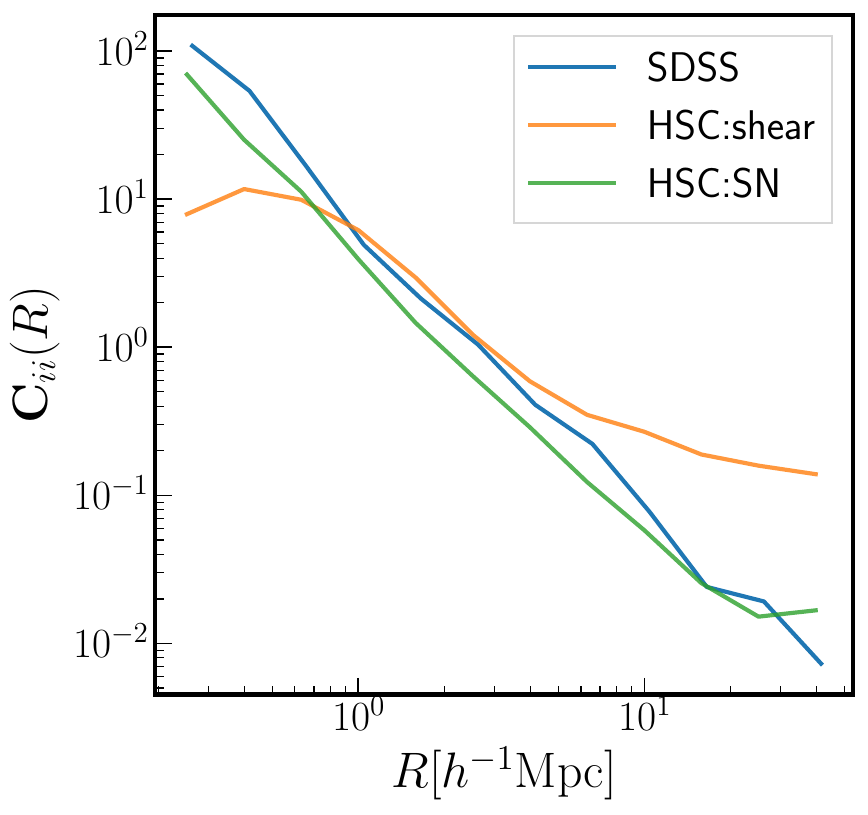}
    \caption{The diagonal components of the covariance matrix for the stacked cluster lensing profiles measured from the SDSS data and the HSC-Y3 data. For the covariance matrix of the HSC-Y3, we show the shear and shape-noise \jsrv{(SN)} components of the covariance separately to show the relative contribution of each component. The results shown here are from the redMaPPer clusters with $20 \leq \lambda < 30$.}
    \label{fig:cov_diag}
\end{figure}

For the lensing component of the covariance matrix, we use 800 HSC mock catalogs described in Sec.~\ref{sec:data:hsc_mock} to measure mock lensing signals and to compute the covariance matrix. 
This is because we cannot use jackknife resampling of the small area covered by the HSC-Y3 data.
The HSC-Y3 lensing covariance consists of two components \citep[see][ for a similar methodology]{More_etal2023}:
\begin{equation}
\mathbf{C}_{\Delta \Sigma}=\mathbf{C}_{\rm SN}+\mathbf{C}_{\rm shear},
\end{equation}
where the shape noise covariance $\mathbf{C}_{\rm SN}$ is due to intrinsic shapes of source galaxies, and the shear-component of covariance $\mathbf{C}_{\rm shear}$ is due to a finite survey volume (i.e., sample variance).
We compute the shape noise covariance by repeatedly measuring the lensing signals for the source galaxy positions by randomly rotating their orientations 500 times. 
In this way, the shape noise covariance can correctly account for the inhomogeneity in the source galaxy distribution and the survey geometry.
For the shear-component covariance $\mathbf{C}_{\rm shear}$, we use the 800 mock cluster catalogs and mock source galaxies described in Sec.~\ref{sec:data:hsc_mock}. 
These mock catalogs share the same footprints of the shear and cluster catalogs as the real data catalogs, and we measure the mock lensing signals from the shear values without shape noise. 
We also measure the lensing signals of random catalogs to correct for the boost factor and the random signals described in Sec.~\ref{sec:meas:dsigma}.
Fig.~\ref{fig:cov_diag} shows the diagonal components of each covariance $\mathbf{C}_{\rm SN}$ and $\mathbf{C}_{\rm shear}$ for the HSC-Y3 data in comparison with the SDSS lensing covariance.
A typical number density of the SDSS source galaxies is about 1~arcmin$^{-2}$, while the number density of the HSC-Y3 source galaxies is about 16~arcmin$^{-2}$ for our selection of the source galaxies. 
Due to the superb depth of the HSC data, the shape noise is smaller for HSC than the one from the SDSS data.
On the other hand, the sample variance term is much larger for the HSC-Y3 data due to the small area ($\sim 430~{\rm deg}^2$).
As is clear from Fig.~\ref{fig:cov_diag}, the lensing covariance for the HSC-Y3 lensing measurements is not shape-noise dominated on large scales, and the diagonal components of $\mathbf{C}_{\rm shear}$ is much larger than the SDSS covariance.
Note that the turn-over of $\mathbf{C}_{\rm shear}$ at $R \sim 1h^{-1}{\rm Mpc}$ in Fig.~\ref{fig:cov_diag} is due to the finite resolution of the lightcone mocks, and for this reason, we do not use the sample variance covariance at $R<1^{-1}{\rm Mpc}$ with this reason.

\section{Analysis Methods}
\label{sec:model}
In this section, we outline how we model the cluster abundance, clustering, and lensing signals including systematics such as projection effects, mis-centering, and photometric redshift uncertainties of the clusters.
In Sec.~\ref{sec:mod:emu}, we introduce the Dark Emulator \citep{darkemu} software and explain how we model cluster observables using the Dark Emulator. 
The Dark Emulator is built by using an ensemble of N-body simulations for different cosmologies and provides halo observables, which resolves the difficulties in modeling small-scale clustering and lensing signals.
Then, we explain how we model projection effects on top of the cluster observables modeled by the Dark Emulator in Sec.~\ref{sec:mod:boost}.
This explicitly accounts for the anisotropic boost on the cluster clustering and lensing signals caused by the preferential selection of clusters on aligned filaments along the LOS.
In Sec.~\ref{sec:mod:photoz}-\ref{sec:mod:res}, we discuss the rest of systematics related to clusters such as photometric redshift uncertainties of the clusters, mis-centering, and the corrections to the measured clustering and lensing signals.
Lastly, we outline the parameter inference in Sec.~\ref{sec:mod:param}.

\subsection{Modeling with Dark Emulator}
\label{sec:mod:emu}
In our analysis, we use the Dark Emulator \citep{darkemu} software 
to generate predictions for the following cluster observables for a given set of cosmological parameters $\bm\uptheta$ and redshift $z$:
\begin{itemize}
    \item the halo mass function $\mathrm{d}n/\mathrm{d}\ln M$
    \item the 3D halo-halo correlation function $\xihh(r|M,M')$
    \item the 3D halo-matter correlation function $\xi_\mathrm{hm}(r|M)$\hmrv{,}
\end{itemize}
\hmrv{where $M$ and $M'$ are the halo masses, which are defined within a boundary, $r_{\rm 200m}$, that encloses an average density which is $200$ times the mean matter density at present, and $r$ is the 3D separation.}
To calculate model predictions, we choose a single representative redshift $z=0.25$, which corresponds to the mean redshift of the cluster sample.
To make sure that this choice does not cause any biased results in our cosmology analysis, we compared the clustering and lensing predictions at $z=0.25$ and at multiple reshift bins considering the redshift distribution of the cluster sample and we did not find any differences.
We have also confirmed that using a halo mass function at a single representative redshift does not bias our cosmology analysis.

\hmrv{The Dark Emulator} was trained based on the measurement of the above quantities in a suite of 101 numerical simulations for distinct $w$CDM cosmological models.
The emulator outputs are typically accurate to better than a few percent within the sampled $w$CDM cosmologies over the redshift range and mass scales we are interested in ($0<z<1.48$ and $M\gtrsim10^{12}h^{-1}M_\odot$), including the quasi-linear and the fully nonlinear regimes of matter clustering, but assume statistical isotropy for the computed halo statistics.
In our model predictions, we assume a flat $\Lambda$CDM model of cosmology with the \hmrv{present-day} neutrino density fixed to $\Omega_\nu h^2 = 0.00064$\hmrv{.}

We then map these quantities to cluster observables as follows. First, for cluster abundances in a given richness bin (labelled by $i$), we have
\begin{eqnarray}
    N_{\mathrm{c},i} & = & \Omega \int\! \mathrm{d}z~ \frac{\mathrm{d}^2V}{\mathrm{d}z\mathrm{d}\Omega} \int_{\ln \lambda_\mathrm{i,min}}^{\ln \lambda_\mathrm{i,max}}\! \mathrm{d}\ln \lambda \int\!\mathrm{d} \ln M \frac{\mathrm{d}n}{\mathrm{d}\ln M} P(\ln \lambda|M) \nonumber \\
    & = & \Omega \int\! \mathrm{d}z \frac{\chi^2(z)}{H(z)} \int\! \mathrm{d}\ln M \frac{\mathrm{d}n}{\mathrm{d}\ln M}S_i(M),
\end{eqnarray}
\jsrv{where $\lambda$ is the richness, $\Omega$ stands the effective survey area, $\chi$ is
comoving distance, $H$ is the Hubble constant, and $S_i$ is the mass selection function.} We have assumed that the mass-richness relation does not evolve with redshift, and that we use the halo mass function at the effective redshift which makes the above equation separable in terms of an effective volume and the integral of the mass selection function. 
We follow \cite{muratasdss} to define our mass-richness relation $P(\ln \lambda|M)$ as a log-normal distribution, i.e. $\ln \lambda (M) \sim \mathcal{N}(\mu_\mathrm{M}, \sigma_\mathrm{M}^2)$, with mean $\mu_\mathrm{M}$ and standard deviation $\sigma_\mathrm{M}$ given by
\begin{eqnarray}
    \mu_\mathrm{M} & = & A + B \ln \left( \frac{M}{M_\mathrm{piv}} \right)
\label{eq:mor1}, \\
    \sigma_\mathrm{M} & = & \sigma_0 + q \ln \left( \frac{M}{M_\mathrm{piv}} \right) 
\label{eq:mor2},
\end{eqnarray}
where these four parameters $A$, $B$, $\sigma_0$, and $q$ are marginalized over.
We set our pivot mass $M_\mr{piv} = 3\times10^{14}M_\odot/h$. With this, the mass selection function $S_i(M)$ is given by,
\begin{eqnarray}
    S_i(M) & = & \int_{\ln \lambda_\mathrm{i,min}}^{\ln \lambda_\mathrm{i,max}} d\ln \lambda ~ P(\ln \lambda|M) \nonumber \\
    & = & \Phi(\ln \lambda_\mathrm{i,max}|\mu_\mathrm{M},\sigma_\mathrm{M}) - \Phi(\ln \lambda_\mathrm{i,min}|\mu_\mathrm{M},\sigma_\mathrm{M}),
\end{eqnarray}
where $\Phi(x|\mu,\sigma)$ is the cumulative distribution function of a Gaussian distribution with mean $\mu$ and standard deviation $\sigma$:
\begin{equation}
    \Phi(x|\mu,\sigma) = \frac{1}{2}\left[1+\mathrm{erf}\left(\frac{x-\mu}{\sqrt{2}\sigma}\right)\right],
\end{equation}
with $\mathrm{erf}(x)$ being the error function.

Next, we model the cluster clustering signal. First, we derive the cluster auto-correlation function for the $i$-th bin as
\begin{equation}
\begin{split}
    \xi_{\mathrm{cc},i}(r) =   \frac{1}{(n_{\mathrm{c},i})^2}
    \int\!\mathrm{d}M \int\!\mathrm{d}M'~   & \frac{\mathrm{d}n}{\mathrm{d}\ln M} 
    \frac{\mathrm{d}n}{\mathrm{d}\ln M' }  ~ \times \\
    & S_i(M) S_i(M') \xihh(r|M,M'),
\end{split}
\end{equation}
with $n_{\mathrm{c},i}$ is the number density of clusters per $(h^{-1}{\rm Mpc})^3$. 
We then model the cluster clustering observable, i.e., the projected cluster auto-correlation function $w_\mr{p}(R)$, as
\begin{equation}
    w_\mr{p,i}(R) = \int_{-\pi_\mathrm{max}}^{\pi_\mathrm{max}}\!\mathrm{d}\pi ~  \xi_{\mathrm{cc},i}\left(\sqrt{R^2+\pi^2}\right),
\label{eq:xicctowp}
\end{equation}
where $\pi_\mr{max}$ is the maximum line-of-sight projection width, we 
use $\pi_{\rm max}=100\,h^{-1}{\rm Mpc}$ as our fiducial choice. Note that in principle there could be residual redshift space distortions that may manifest themselves on large scales due to the use of a finite $\pi_\mr{max}$ \citep{vdBosch2013}. We have explicitly checked that accounting for such differences does not have any impact on our cosmological results. 

Lastly, we model the cluster lensing signal as follows. Similar to the clustering case, we first obtain the cluster-matter cross-correlation function for the $i$-th bin as
\begin{equation}
    \xi_{\mathrm{cm}, i}(r) = \frac{1}{n_{\mathrm{c},i}}
    \int\! \mathrm{d}\ln M~ \frac{\mathrm{d}n}{\mathrm{d}\ln M} S_i(M) \xihm(r|M),
\end{equation}
where the cluster number density for the $i$-th cluster richness bin, $n_{\mr{c},i}$, is given by
\begin{equation}
    n_{\mr{c},i} =  \int\! \mathrm{d}\ln M~ \frac{\mathrm{d}n}{\mathrm{d}\ln M} S_i(M).
\end{equation}
We then obtain the surface density of matter around clusters $\Sigma(R)$  as
\begin{equation}
    \Sigma_i(R) = \bar{\rho}_{\mathrm{m},0} \int_{-\infty}^{\infty}\!\mathrm{d}\pi ~ \xi_{\mathrm{cm}, i}\left(\sqrt{R^2+\pi^2}\right),
\end{equation}
where we only consider the matter density in the Universe that contributes to lensing and $\bar{\rho}_{\mathrm{m},0}$ is the mean comoving matter density of the Universe. Here, $R$ is the separation perpendicular to the LOS, and $\pi$ is the separation parallel. 
Finally, we obtain the cluster lensing observable, i.e., the excess surface density $\Delta\Sigma(R)$ as
\begin{eqnarray}
    \Delta\Sigma_i(R) & \equiv & \left<\Sigma_i\right>(<R)-\Sigma_{i}(R) \nonumber\\
    & = &  \frac{2}{R^2} \int_0^R \mathrm{d}R' R' \Sigma_i(R') - \Sigma_i(R).   
\end{eqnarray}
Equivalently, $\Delta\Sigma(R)$ is given by
\begin{eqnarray}
\Delta\Sigma(R) & = & 
\bar{\rho}_{\mathrm{m},0} \int\! \frac{KdK}{2\pi} J_2(KR)C_\mr{cm}(K) \\
& \simeq & \bar{\rho}_{\mathrm{m},0} \int\! \frac{kdk}{2\pi} J_2(kR) P_\mr{cm}(k),
\end{eqnarray}
where $C_\mr{cm}(K)$ and $P_\mr{cm}(k)$ are the 2D and 3D cluster-matter power spectra with the 2D wavenumber $K$ and the 3D wavenumber $k$, and $J_2$ is the second-order Bessel function. The last line uses the the Limber approximation \citep{limber,limberkaiser}, which becomes exact under our assumption of a single redshift value
\citep[also see][]{2013MNRAS.435.2345H,2021arXiv211102419M}.

\subsection{Anisotropic Boosts}
\label{sec:mod:boost}
Since the predictions from \hmrv{Dark Emulator} 
assume the isotropic distribution of clusters, we need to model the anisotropic boosts on the cluster clustering and lensing signals on top of the \hmrv{Dark Emulator} 
predictions.
We model this boost as a scale-dependent multiplicative factor $\Pi(R)$ applied to the cluster clustering and lensing signals. This then transforms the isotropic predictions from \Cref{sec:mod:emu} as
\begin{eqnarray}
    w_\mr{p}(R) & = & \Pi^2(R) w_\mr{p}^\mr{iso}(R),\\
    \Sigma(R) & = & \Pi(R) \Sigma^\mr{iso}(R).
\end{eqnarray}
The anisotropic boost factor $\Pi(R)$ modifies $\Sigma(R)$ rather than $\Delta \Sigma(R)$ because $\Sigma(R)$ is the quantity that represents the raw cluster-matter correlation function.

This anisotropic boost factor $\Pi(R)$ is given by three model parameters, $\Pi_0, R_0$ and $c$:
\begin{equation}
    \Pi(R) = 
    \begin{cases}
        \Pi_0 (R/R_0) & \text{for } R \leq R_0, \\
        \Pi_0 + c \ln (R/R_0) & \text{for } R > R_0,
    \end{cases}
\label{eq:anisoboost}
\end{equation}
where $R_0$ roughly corresponds to the transition scale from one-halo to two-halo regime (i.e., a few Mpc$/h$). 
\citetalias{projeff} and \citetalias{Park_2022} show that the anisotropic boost $\Pi(R)$ behaves differently on small scales ($R<R_0$) and large scales ($R>R_0$).
While the boost is roughly linear in $R$ \jsrv{on small scales}, the boost slowly falls off logarithmically \jsrv{with increasing $R$}, and our modeling of $\Pi(R)$ captures these behaviors well (see Fig.~1 in \citetalias{Park_2022}).
Note that the parameters $\Pi_0$ and $R_0$ are defined for each richness bin in \citetalias{Park_2022}, while we decide to use the same parameters $\Pi_0$ and $R_0$ for all the richness bins. This choice is explained in Sec.~\ref{sec:mock_test}.

\subsection{Cluster photo-$z$ scatter}
\label{sec:mod:photoz}
The photometric redshift uncertainties of clusters produce dilutions in cluster clustering measurements.
To model this effect, we assume a Gaussian scatter in the LOS positions of clusters with the standard deviation $\sigma_\mr{ph}$. This scatter is then taken into account in the calculation of the cluster auto-correlation function $w_\mr p(R)$ as a modification to \Cref{eq:xicctowp},
\begin{equation}
    w_\mr{p}(R) = 2\int_0^\infty\! \mathrm{d}\pi ~ S_\mr{ph}(\pi|\pi_\mr{max}, \sigma_\mr{ph}) ~ \xi_{\mathrm{cc},i}\left(\sqrt{R^2+\pi^2}\right),
\end{equation}
where the LOS separation selection function $S_\mr{ph}(\pi, \sigma_\mr{ph})$ is given by 
\begin{equation}
    S_\mr{ph}(\pi | \pi_\mr{max}, \sigma_\mr{ph}) = 
    \Phi(\pi | -\pi_\mr{max}, \sqrt{2}\sigma_\mr{ph}) - 
    \Phi(\pi | \pi_\mr{max}, \sqrt{2}\sigma_\mr{ph})
\end{equation}
with $\Phi(x|\mu,\sigma)$ being the cumulative distribution function for a Gaussian distribution with mean $\mu$ and standard deviation $\sigma$.
We only model the impact of photo-$z$ for clustering signals, but not for abundance and lensing signals. This is because the effects of photo-$z$ scatter of the SDSS redMaPPer clusters assumed to be roughly 1\% on abundance and lensing signals are estimated in \citetalias{Park_2022} to be less than 0.2\% and 0.4\% respectively.

\subsection{Miscentering}
\label{sec:mod:mis}
For optical clusters, cluster centers are defined by the location of the brightest cluster galaxy (BCG) within each cluster.
However, the location of the BCG is not always the true center of the corresponding dark matter halo for the cluster \citep{oguritakada11,2013MNRAS.435.2345H,miyatakecl}.
When this happens, the measured cluster lensing signal around the BCG will have a characteristic dilution on small scales. 
To model this effect (called ``mis-centering'' ), we assume that a fraction ($f_\mr{mis}$) of clusters which are mis-centered by a random variable $R'$ that follows the Rayleigh distribution such as
\begin{equation}
    P_\mr{mis}(R'|R_\mr{mis}) = \frac{R'}{R_\mr{mis}^2} \exp \left( - \frac{R'^2}{2R_\mr{mis}^2}\right).
\label{eq:miscen}
\end{equation}
This assumption simplifies the calculation of the averaged lensing profile for the mis-centered clusters in Fourier space \citep{oguritakada11}:
\begin{eqnarray}
    C_\mr{cm}^\mr{mis} (K) & = & C_\mr{cm}(K) \int\!\mathrm{d}R' J_0 (KR') P_\mr{mis}(R'|R_\mr{mis}) \nonumber \\
    & = & C_\mr{cm}(K) \exp \left(-\frac{1}{2}K^2R_\mr{mis}^2\right).
\end{eqnarray}
Then, the global average is given by
\begin{equation}
    C_\mr{cm}^\mr{tot} (K) = (1-f_\mr{mis}) C_\mr{cm}(K) + f_\mr{mis} C_\mr{cm}^\mr{mis}(K).
\end{equation}
This can be Fourier transformed back to real space for the cluster lensing signal with the mis-centering effect included.

\subsection{Modeling Residual Systematic Errors}
\label{sec:mod:res}
In this section, we present a method to account for the effects of residual systematic errors on the measured lensing signals: residual multiplicative shear error and residual systematic photo-$z$ uncertainty.
Note that these corrections are only relevant for the internal consistency tests \hmrv{for the real data described in Sec~\ref{sec:res:consistency}}.

The residual multiplicative shear error is the possible residual error in the shape calibration factor. To account for this error, we use a nuisance parameter of the residual multiplicative bias $\Delta m$ and shift the theoretical templates of the lensing signals as
\begin{equation}
\Delta\Sigma^{\rm corr}(R,\Delta m) = (1+\Delta m)\Delta\Sigma(R),
\end{equation}
with $\Delta m=0.0$ for our fiducial analysis. 
As one of our internal consistency tests, we vary $\Delta m$ using a flat prior from -0.05 to 0.05 and investigate its effects on our cosmological parameter constraints as described in Sec.~\ref{sec:res:consistency}.

There is a possibility of a residual systematic bias in the photometric redshifts inferred for the population \hmrv{ of source galaxies} we use. Cosmic shear cosmology analyses using the HSC-Y3 source galaxies \citep{hscy3_shear_fourier,hscy3_shear_real} report a residual photo-$z$ error for the HSC source galaxies at high redshifts, consistent with the HSC 3x2pt analyses \citep{hscy3_3x2_minimal,hscy3_3x2_small}.
Since we use the same source galaxies at high redshifts, we need to consider this residual photo-$z$ uncertainty.
To self-calibrate the residual photo-$z$ uncertainty, the analysis has to be done in a tomographic manner like the analyses mentioned above \citep{hscy3_3x2_minimal,hscy3_3x2_small,hscy3_shear_fourier,hscy3_shear_real}.
However, our analysis uses the cluster sample at a single redshift bin.
This implies that we cannot use a wide flat prior on the residual photo-$z$ uncertainty in our analysis and carry out a self-calibration of the photo-$z$ distribution.
For this reason, we measure the residual photo-$z$ uncertainty of our source galaxies using the photo-$z$ uncertainties measured from \cite{hscy3_shear_real}.
We describe the details of this measurement in Appendix~\ref{app:photo-z}.
For our fiducial analysis, we set the residual photo-$z$ uncertainty $\Delta z_{\rm ph}$ of our source galaxies to be $-0.093$ and model the true photo-$z$ distribution as
\begin{equation}
p^{\rm true}(z)=p^{\rm est}(z+\Delta z_{\rm ph}),
\end{equation}
where $p^{\rm est}(z)$ is the estimated distribution from the measured photo-$z$ of the source galaxies. Note that we varied the residual photo-$z$ uncertainty of our source galaxies as one of the internal consistency tests.
Using the above corrected photo-$z$ distribution, we can correct the model template of the lensing signal by recomputing the averaged lensing efficiency $\left<\Sigma_{\rm cr}^{-1} \right>$ and the weight $w_{ls}$,
\begin{equation}
\Delta\Sigma_{\rm corr}(R,\Delta z_{\rm ph})=f_{\Delta\Sigma}(\Delta z_{\rm ph})\Delta\Sigma(R),
\end{equation}
where the correction factor $f_{\Delta \Sigma}(\Delta z_{\rm ph})$ is defined as
\begin{equation}
f_{\Delta\Sigma}(\Delta z_{\rm ph}) \equiv \frac{\sum_{ls} w_{ls}\left< \Sigma_{cr}^{-1} \right>_{ls}^{\rm true}/\left< \Sigma_{cr}^{-1} \right>_{ls}^{\rm est}}{\sum_{ls}w_{ls}}.
\end{equation}
Note that we test random sampling of $\Delta z_{\rm ph}$ from the chains provided by \cite{hscy3_shear_real} as one of internal consistency tests as described in Sec.~\ref{sec:res:consistency}.
For more details of the correction method, we refer the readers to Sec.~III.B.1 of \cite{hscy3_3x2_minimal}.


\subsection{Parameter Inference}
\label{sec:mod:param}

\begin{table}
	\centering
	\caption{The model parameters varied in our fiducial analysis and the respective priors assumed. Note that all the priors are flat.}
	\label{tab:priors}
	\begin{tabular}{cc} 
		\hline\hline
		Parameter & Prior \\
		\hline
		\multicolumn{2}{c}{\bf Cosmological parameters}\\
		$\Omega_\mr{b} h^2$ & $\mathcal{U}(0.0211375, 0.0233625)$\\
		$\Omega_\mr{c} h^2$ & $\mathcal{U}(0.10782, 0.13178)$\\
		$\ln 10^{10} A_\mr{s}$ & $\mathcal{U}(2.5,3.7)$\\
		$n_\mr{s}$ & $\mathcal{U}(0.92, 1.01)$\\
		$\Omega_\Lambda$ & $\mathcal{U}(0.54752, 0.82128)$\\
		\hline
		\multicolumn{2}{c}{\bf Mass-richness relation}\\
        $A$ & $\mathcal{U}(0.5, 5.0)$\\
        $B$ & $\mathcal{U}(0.0, 2.0)$\\
        $\sigma_0$ & $\mathcal{U}(0.0, 1.5)$\\
        $q$ & $\mathcal{U}(-1.5, 1.5)$\\
        \hline
        \multicolumn{2}{c}{\bf Anisotropic boost parameters}\\
        $\Pi_\mathrm{0}$ & $\mathcal{U}(0.0, 1.0)$ \\
        $R_\mathrm{0}$ & $\mathcal{U}(0.1, 5.0)$ \\
        $c$ & $\mathcal{U}(-2.0, 2.0)$\\
        \hline
		\multicolumn{2}{c}{\bf Mis-centering}\\
		$f_\mr{mis,1-4}$ & $\mathcal{U}(0.0, 1.0)$\\
		$R_\mr{mis,1-4}$ & $\mathcal{U}(0.0, 1.0)$\\
        \hline
		\multicolumn{2}{c}{\bf Cluster photo-$z$ scatter}\\
		$\sigma_\mr{ph}$ & $\mathcal{U}(1.0, 100.0)$\\
		\hline \hline
	\end{tabular}
\end{table}

We assume a Gaussian likelihood model and compute the likelihoods with the measurements $\mathbf{d}$ and the model predictions $\bm\upmu(\bm\uptheta)$, where $\bm\uptheta$ are the parameters, with covariance matrix $\mathbf{C}$
\begin{equation}
    \ln P(\mathbf{d}|\bm\uptheta) = 
    -\frac{1}{2} \left[\mathbf{d}-\bm{\upmu}(\bm\uptheta)\right]^\intercal \mathbf{C}^{-1} \left[\mathbf{d}-\bm{\upmu}(\bm\uptheta)\right].
\end{equation}
With this likelihood, we perform Bayesian parameter inference for the posterior $P(\bm\uptheta|\mathbf{d})$ as 
\begin{equation}
    P(\bm\uptheta|\mathbf{d}) \propto P(\mathbf{d}|\bm\uptheta) P(\bm\uptheta),
\end{equation}
with our likelihood $P(\mathbf{d}|\bm\uptheta)$ and prior $P(\bm\uptheta)$. 
We use the \texttt{cosmosis} \citep{cosmosis} software\footnote{\url{https://github.com/joezuntz/cosmosis}} to perform our likelihood analysis, using the \texttt{multinest} \citep{multinest1,multinest2,multinest3} samplers integrated therein to sample our posteriors. 
We turn on ``constant efficiency'' mode to shorten the runtime. 
However, we later find out that the \texttt{multinest} with this mode gives inconsistent estimates of the Bayesian evidence and narrower parameter credible intervals by choosing different seeds.
Therefore, we use the \texttt{multinest} with this mode for the mock tests discussed in Sec.~\ref{sec:mock_test} and the internal consistency tests in Sec.~\ref{sec:res:consistency} with the same seed, since the crucial point of these analyses is to make sure that all the results are consistent.
However, we use the \texttt{multinest} without the ``constant efficiency'' mode for our fiducial analysis to avoid biased constraints on cosmological parameters.
We also use the \texttt{chainconsumer} \citep{chainconsumer} software to plot confidence regions for our posterior samples. 
Under the fiducial analysis setup in \citetalias{Park_2022}, there were a total of 27 independently varied model parameters: 5 for cosmology, 4 for mass-observable relation, 9 for anisotropic boosts, 8 for mis-centering, and 1 for the cluster photo-$z$ scatter. However, we decided to use 3 parameters for anisotropic boosts to model the boosts on clustering and lensing for all richness bins as explained in the next section. The fiducial analysis setup is summarized in Table~\ref{tab:priors}.
Our choice of prior ranges for cosmological parameters is set by the parameter space covered by the Dark Emulator.
We use the same prior ranges as \cite{muratasdss} for mass-richness relation parameters and set wide enough prior ranges for mis-centering and photo-$z$ scatter parameters.
The reason we do not use the Gaussian priors for the mis-centering parameters based on the X-ray studies \citep[e.g.][]{zhang18} is because these studies did not investigate the SDSS redMaPPer clusters which do not have X-ray counterparts and we do not know the mis-centering effect of such clusters.
For the parameters of anisotropic boost, we set the prior ranges based on the previous study done by \citetalias{projeff}.

\section{Mock Tests}
\label{sec:mock_test}

\begin{figure*}
	\includegraphics[width=\columnwidth]{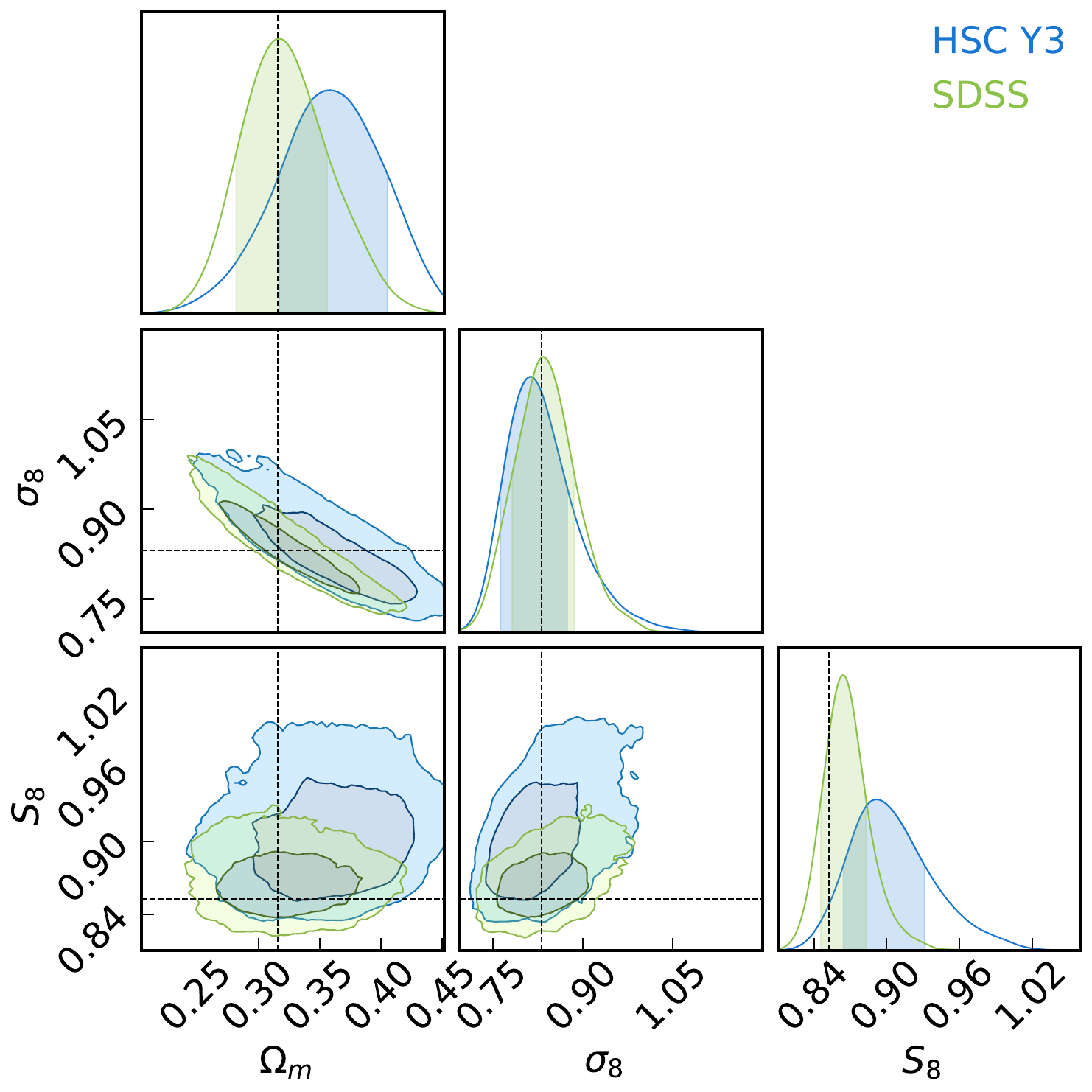}
	\includegraphics[width=\columnwidth]{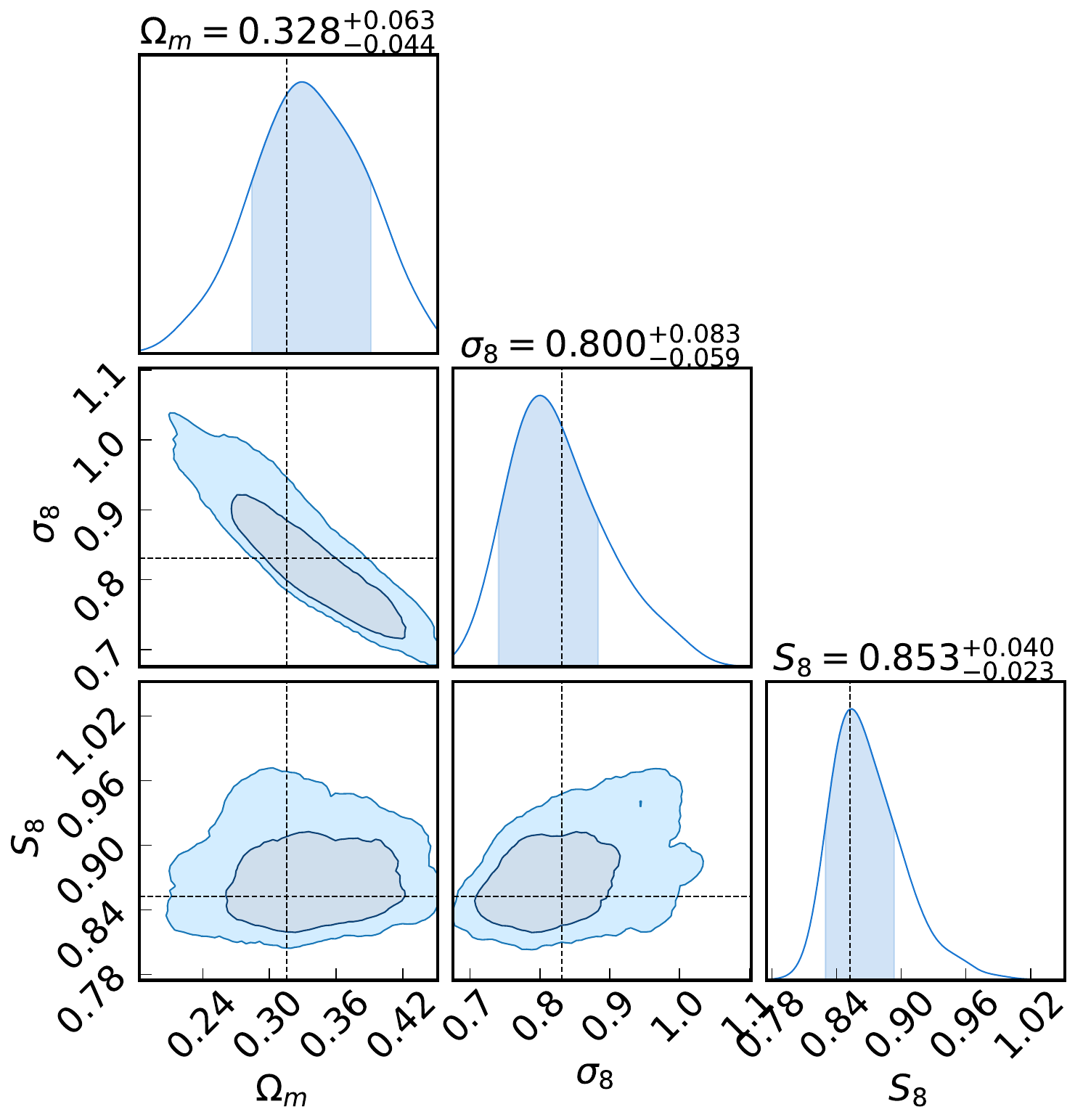}
    \caption{\textit{Left:} The 68\% confidence regions for cosmological parameters of interest, obtained by running our fiducial pipeline on the test mocks assuming no systematics. Different contours indicate the results using the covariance matrix measured from the SDSS data (green) and the HSC-Y3 data (blue). The dashed lines indicate the values of the input cosmological parameters of $\Omega_{\rm m}=0.3156$ and $\sigma_8=0.831$. \textit{Right:} The same as the left figure \jsrv{using the covariance matrix measured from HSC-Y3 data} but with the model using the same anisotropic boost parameters for the cluster samples across all richness bins. Dashed lines represent the input cosmology for the mocks.} 
    \label{fig:mock_sdss2hsc}
\end{figure*}

In the previous study by \citetalias{Park_2022}, we tested several analysis choices including the range and binning of radial scales for clustering and lensing measurements as well as a functional form of the effective bias $\Pi(R)$ modeling the anisotropic boosts on clustering and lensing observables.
To do this, we used the mock cluster catalogs and measurements from \citetalias{projeff} to determine the optimal form of the analysis.
We subsequently validate our model with the mock cluster catalogs (Sec.~3.2 in \citetalias{Park_2022}) to ensure that we could robustly recover the input cosmology within the $0.5~\sigma$ confidence region in the $\Omega_{\rm m}-S_8$ plane for all three versions of the validation measurements including projection effects, mis-centering, and photo-$z$ scatter. 
In principle, we follow this analysis setup in this study as well.
However, we now use the lensing signals measured 
\hmrv{with the HSC-Y3}
data, so the lensing component of the covariance matrix is different from \citetalias{Park_2022}. 
To test how these changes affect the analysis, we repeat the same validation tests done by \citetalias{Park_2022} in Sec.~\ref{sec:mock:proj}.
In Sec.~\ref{sec:mock:baryon}, we additionally consider the effects of baryonic processes on lensing signals. Baryonic effects suppress the amplitude of the lensing signals on small scales as shown in Fig.~\ref{fig:dsigma_baryon}. In this section, we identify the optimal form of the analysis with these changes and effects.


In the analysis of this section, we use the same mock cluster catalogs and noiseless data vector used in \citetalias{Park_2022}. 
In short, the mock catalogs start with N-body simulations and resulting halo catalogs from \cite{darkemu}. 
Halos are identified with masses greater than $10^{12}M_\odot/h$, and they are populated with galaxies using a Halo Occupation Distribution (HOD) model for the number of red-sequence galaxies and a Navarro-Frenk-White \citep[NFW;][]{nfw} profile for the spatial distribution of the galaxies. Lastly, a cluster finder algorithm based on \cite{sunayamamore} is run on the galaxy catalog to generate the mock cluster catalog, assuming that all galaxies within the projection length $d_{\rm proj}$ along the LOS from the central galaxies are member galaxies of the clusters.
Projection effects are built in by this process using $d_{\rm proj}$, and the parameter $d_{\rm proj}$ corresponds to the photo-$z$ uncertainties of the cluster member galaxies.
We use a total of 19 realizations of this mock catalog, each with $1~ (\mathrm{Gpc}/h)^3$ volume and all at redshift $z=0.251$ to measure the abundance and the lensing signals, and additionally 14 realizations of $2 ~(\mathrm{Gpc}/h)^3$ volume to measure the cluster clustering signals.
By averaging all the measurements from the mocks, we obtain a noiseless data vector to test our model. 
To include other systematics in the mock cluster catalogs, we randomly displace the cluster centers in the 2D plane perpendicular to the LOS using the same model as in Sec.\ref{sec:mod:mis} for mis-centering and along the LOS direction following a Gaussian distribution as in Sec.~\ref{sec:mod:photoz} for photometric redshift scatter.
For more details on the test simulations and the measurement procedures for the mock catalogs, we refer the reader to Sections~2.2 -- 2.5 of \citetalias{projeff}.

Our data vector consists of the abundance, lensing signals, and cluster clustering of four cluster samples with $\lambda \in [20,30), [30,40), [40,55),$ and $[55,200)$:
\begin{equation}
    \mathbf{d} = \left\{ \mathbf{N}_\mathrm{c}, \bm{\upDelta\upSigma}_1, \bm{\upDelta\upSigma}_2, \bm{\upDelta\upSigma}_3, \bm{\upDelta\upSigma}_4, \mathbf{w}_\mr{p,1}, \mathbf{w}_\mr{p,2}, \mathbf{w}_\mr{p,3}, \mathbf{w}_\mr{p,4} \right\},
\end{equation}
where $\bm{\upDelta\upSigma}_i$ and $\mathbf{w}_\mr{p,i}$ respectively represent the cluster lensing and cluster clustering signals for the $i$-th richness bin, and $\mathbf{N}_\mathrm{c}=\{N_\mathrm{c,i}\}$ is the vector of cluster counts for the four richness bins.
We start from the lensing data vector with 12 logarithmically spaced bins from $R=0.2h^{-1}{\rm Mpc}$ to $R=50h^{-1}{\rm Mpc}$.
For the clustering measurements, we compute the projected correlation functions by integrating out to $\pi_{\rm max}=100h^{-1}{\rm Mpc}$ along the LOS.


\subsection{Projection Effects}
\label{sec:mock:proj}

We first present the results for the mock cluster samples with projection effects to test whether our pipeline can robustly recover cosmological parameters. 
The purpose is to test our pipeline with the lensing covariance measured from HSC-Y3 mock catalogs.
In the previous analysis done in \citetalias{Park_2022}, we used the lensing covariance measured by the jackknife resampling method using the SDSS data. 
The typical number density of the SDSS source galaxies is about $1 \hmrv{~\rm{arcmin}}^{-2}$, while the number density of the HSC-Y3 source galaxies is about $16\hmrv{~\rm{arcmin}}^{-2}$ for our selection of the
source galaxies. 
Due to the superb depth of the HSC data, the shape
noise is smaller than the one from the SDSS data, but
the sample variance term is much larger for the HSC-Y3 data due
to the small area ($\sim 430~{\rm deg}^2$). 
As is clear from Fig.~\ref{fig:cov_diag}, the lensing
covariance for the HSC-Y3 lensing measurement is not shapenoise
dominated on large scales, and we want to make sure our analysis pipeline can robustly recover the input cosmology with this covariance matrix.


Using the covariance matrix measured for the 
\hmrv{HSC-Y3}
lensing signals, we first test our pipeline used in \citetalias{Park_2022} on a noiseless data vector.
The left panel of Fig.~\ref{fig:mock_sdss2hsc} shows our results for the case using the lensing covariance from the SDSS data and the HSC-Y3 data with the same data vector.
As you can see, our pipeline cannot recover the input cosmology with the 
\hmrv{HSC-Y3} lensing
covariance.
We suspect this is due to a large contribution of the shear component of the HSC-Y3 lensing covariance on large scales as shown in Fig.~\ref{fig:cov_diag}. 
In our analysis setup, we model the anisotropic boost on top of the lensing and clustering predictions from the Dark Emulator.
The amplitudes of predicted clustering and lensing signals are degenerate with the anisotropic boost, and we need both clustering and lensing signals to solve this degeneracy.
Due to the larger error of the HSC-Y3 lensing signal compared to the error of the SDSS lensing signal, the HSC-Y3 lensing covariance has a weaker constraining power on large scales and possibly leads to biased cosmology constraints as shown in the left panel of Fig.~\ref{fig:mock_sdss2hsc}.

To mitigate this issue, we decide to use the same boost parameters for the cluster samples across all the richness bins. \cite{sunayama22} studied and measured the anisotropic boost of the SDSS redMaPPer clusters using the cluster-galaxy cross-correlation functions, and found almost the same strength of the anisotropic boost for all the richness bins.
Based on this finding, we assume that our choice in the modeling of projection effects is safe.
With this parametrization, we robustly recover the input cosmology using the lensing covariance from the HSC-Y3 data as shown in the right panel of Fig.~\ref{fig:mock_sdss2hsc}.

\begin{figure}
	\includegraphics[width=\columnwidth]{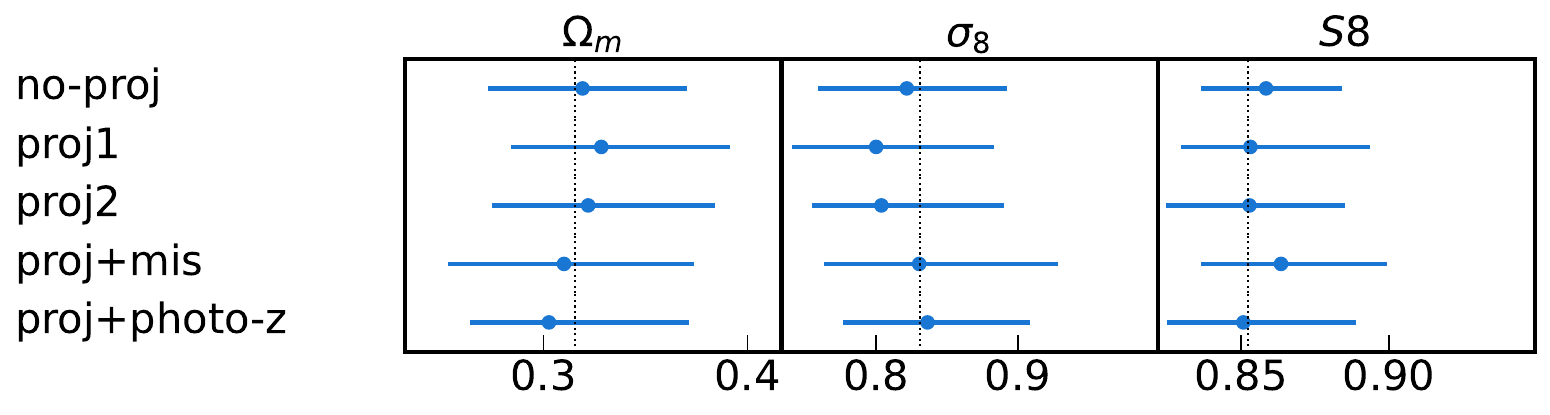}
    \caption{Summary of model validation for the analysis with the HSC-Y3 lensing covariance. Constraints on cosmological parameters $\Omega_{\rm m}$, $\sigma_8$, and $S_8$ are shown in each panel from left to right. The input cosmological parameters used for making synthetic data vectors are indicated by the vertical line in the panels.
    "no-proj" uses the data vector without projection effects and "proj1" and "proj2" use the data vector including projection effects with the projection length of $d_{\rm proj}=60h^{-1}{\rm Mpc}$ and $120h^{-1}{\rm Mpc}$ respectively. "mis" and "photoz" mean including the mis-centering effect and photo-z uncertainty.}
    \label{fig:summary_proj}
\end{figure}

To test the robustness of our pipeline to projection effects, we use measurements that include different strengths of projection effects by choosing the projection length $d_{\rm proj}=60h^{-1}{\rm Mpc}$ (\texttt{proj1}) and $120h^{-1}{\rm Mpc}$  (\texttt{proj2}).
The projection length models the distance uncertainty of cluster member galaxies along the LOS.
We also use measurements including mis-centering (\texttt{proj+mis}) and photometric redshift scatters of clusters (\texttt{proj+photoz}) in addition to projection effects.
We model the mis-centering to the mocks by first randomly selecting 40\% of clusters and re-assigning the position of the clusters perpendicular to the LOS using Eq.~\ref{eq:miscen} with $R_{\rm mis}=0.4$.
Similary, we model the cluster photo-$z$ scatter to the mocks by re-assigning the position of the clusters along the LOS using $\sigma_{\rm ph}=30h^{-1}{\rm Mpc}$.
The results are summarized in Fig.~\ref{fig:summary_proj}.
We do not find any significant deviation from the input cosmological parameters, indicating that our analysis is robust to these effects.

\subsection{Baryonic Effects}
\label{sec:mock:baryon}

\begin{figure}
	\includegraphics[width=0.9\columnwidth]{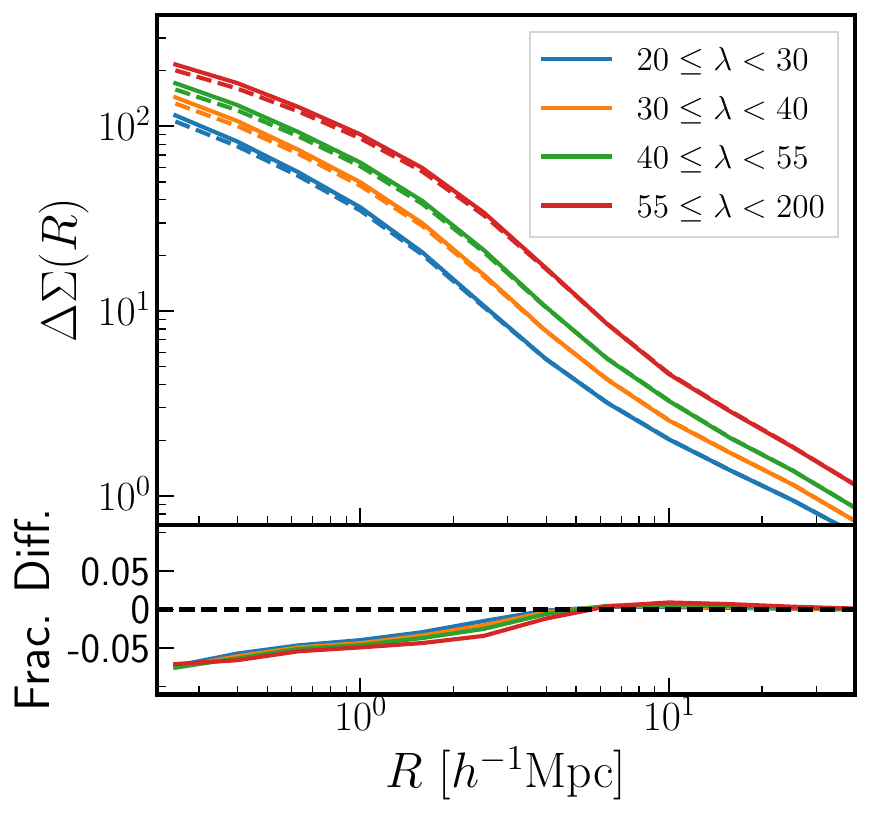}
    \caption{Comparison of the measured cluster lensing signals with (dashed lines) and without (solid lines) the modeling of baryonic effects \jsrv{in the mock}. Different colors correspond to cluster samples in different richness bins. Baryonic effects suppress the lensing signals on small scales, roughly $\geq 5\%$ of $R=0.5h^{-1}{\rm Mpc}$.}
    \label{fig:dsigma_baryon}
\end{figure}

\begin{figure}
	\includegraphics[width=\columnwidth]{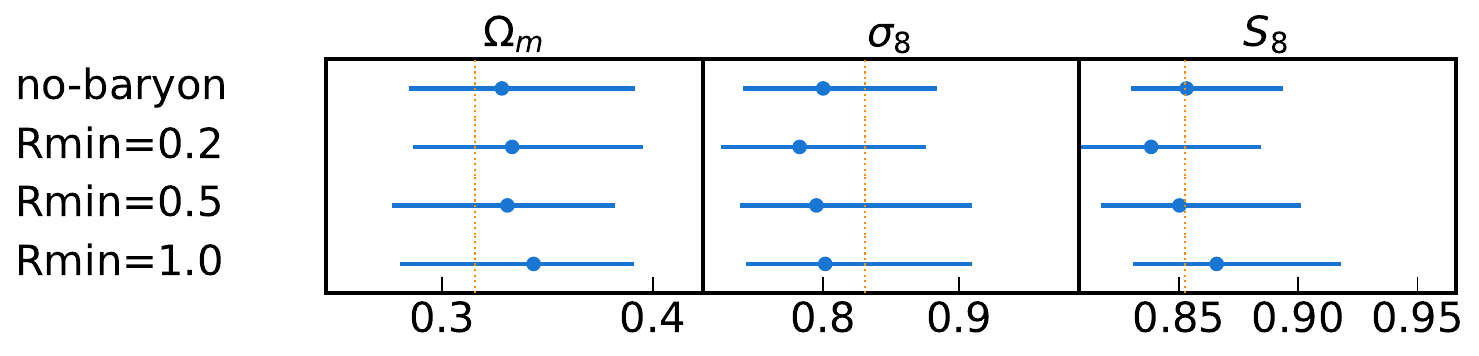}
    \caption{Summary of model validation with baryonic effects. Constraints on cosmological parameters $\Omega_{\rm m}$, $\sigma_8$, and $S_8$ are shown in each panel from left to right. The input cosmological parameters used for making synthetic data vectors are indicated by the vertical line in the panels. The case with "no-baryon" uses the lensing data vector without baryonic effects, while the other cases use the lensing data vector with baryonic effects. "Rmin" implies the minimum radial scale used for the lensing signal. The result shows that using the lensing signal at $R>0.5h^{-1}{\rm Mpc}$ gives the most unbiased result on $S_8$.}
    \label{fig:test_baryon}
\end{figure}

Baryonic feedback affects the mass distribution in and around halos.
On cluster-sized halos, the feedback from active galactic nuclei (AGN) blows intracluster medium away from the cluster center and suppresses the small-scale lensing profile compared to the case without baryonic feedback processes.
To model baryonic effects on the cluster mass profiles and investigate the effect on our cluster cosmology analysis, we employ a semi-analytic model developed by \cite{baryonification1,baryonification2,baryonification3}, known as the \emph{baryonification} method.
This method enables us to include baryonic effects in gravity-only N-body simulations.
We use this baryonification method for the outputs of the Dark Emulator.

Briefly, the baryonification method transforms dark-matter only profiles $\rho_{\rm dm}(r)$ into dark-matter and baryon profiles $\rho_{\rm dm+b}(r)$ including the effects of dark matter, gas, and stars:
\begin{equation}
    \rho_{\rm dm+b}(r,M)=\rho_{\rm DM}(r,M)+\rho_{\rm gas}(r,M)+\rho_{\rm star}(r,M),
\end{equation}
where $\rho_{\rm DM}$, $\rho_{\rm gas}$, and $\rho_{\rm star}$ are the density profiles of dark matter, gas, and stars respectively.
The dark matter component of the profile is simply defined as
\begin{equation}
    \rho_{\rm DM}(r)=\left(1-\frac{\Omega_{\rm b}}{\Omega_{\rm m}}\right)\rho_{\rm dm}(r),
\end{equation}
where the factor of 1-$\Omega_{\rm b}/\Omega_{\rm m}$ is the global mass fraction of cosmic dark matter and $\rho_{\rm dm}(r)$ is the dark-matter profile from the gravity-only N-body simulations.
For the gas density profile, we use the functional form as proposed in \cite{baryonification3}:
\begin{equation}
    \rho_{\rm gas}(r)\propto \left[1+\left(\frac{r}{0.1R_{\rm 200c}}\right)\right]^{-\beta_g}\left[1+\left(\frac{r}{\theta_{\rm ej}R_{\rm 200c}}\right)^{\gamma_g}\right]^{-(\delta_g-\beta_g)/\gamma_g},
\end{equation}
where $\beta_g$, $\gamma_g$, $\delta_g$, and $\theta_{\rm ej}$ are the free parameters in the model and $R_{\rm 200c}$ is the spherical halo boundary radius within the mean mass density is 200 times the critical density. We use $\gamma_g=2$, $\delta_g=7$, $\beta_g=2.5$ and $\theta_{\rm ej}=4$, which are also used in \cite{Shirasaki_etal_prep}. 
The normalization of $\rho_{\rm gas}(r)$ is set by
\begin{equation}
    \int_0^\infty\! \mathrm{d}r~ 4\pi r^2 \rho_{\rm gas}(r)=\left[\frac{\Omega_{\rm b}}{\Omega_{\rm m}}-f_{\rm star}\right]M,
\end{equation}
where $f_{\rm star}$ is the stellar-to-halo mass relation set to 0.01.
For the stellar density profile, there are two components: One is the contribution from the central galaxy $\rho_{\rm cga}(r)$ and another component is from the satellite stars $\rho_{\rm sga}(r)$.
The central galaxy component is described by an exponentially truncated power-law,
\begin{equation}
\rho_{\rm cga}(r)=\frac{f_{\rm cga}M}{4\pi^{3/2}R_{*}r^2}\exp{\left[-\left(\frac{r}{2R_{*}}\right)^2\right]},
\end{equation}
where $f_{\rm cga}$ describes the fraction of stars in the central galaxy, set to 0.001.
The satellite component is defined as
\begin{equation}
\rho_{\rm sga}(r)=(f_{\rm star}-f_{\rm cga})\rho_{\rm dm}(r).
\end{equation}
Note that \cite{baryonification3} concluded that varying these parameters does not significantly alter matter clustering at Mpc scales.
However, the impact of the stellar density profile on the lensing signals can be significant at $R<0.1h^{-1}{\rm Mpc}$.
Since we only consider using the lensing signals at $R>0.2h^{-1}{\rm Mpc}$, the impact of the stellar density profile would not be significant for our analysis.

Fig.~\ref{fig:dsigma_baryon} shows the mock cluster lensing signals with and without baryonic effects using the above-mentioned baryonification method. 
We first compute the cluster lensing signals using the Dark Emulator 
from the halo mass distribution of the mock cluster sample. Similarly, we compute the lensing signals with baryonic effects using the baryonification method described above.
Then, we take the ratio of these lensing signals to obtain the change due to baryonic effects on the lensing signals.
We multiply the mock lensing signals by this ratio to model the suppression.
The suppression due to baryonic effects is roughly 7~\% at $R\simeq 0.2h^{-1}{\rm Mpc}$, and the strength of the suppression does not show a dependence on richness. The lensing signal with baryonic effects slowly converges to the lensing signals without baryonic effects as $R$ gets larger.
This convergence scale correlates with the halo radius of the cluster sample and therefore depends on richness.
The goal of this section is to make sure that the suppression of the lensing signals due to baryonic effects does not bias our cosmology results for the chosen scale of the lensing signals.

Fig.~\ref{fig:test_baryon} shows the the summary of $\Omega_{\rm m}$, $\sigma_8$, and $S_8$ for the measurements including baryonic effects. We vary the minimum radial scale of the lensing signals from $0.2 h^{-1}{\rm Mpc}$ to $1h^{-1}{\rm Mpc}$. As a comparison, we also show the result from the measurement without baryonic effects (\texttt{no-baryon}).
This figure tells us that including baryonic effects does not significantly bias the cosmology result, but broadens the posterior distribution on $S_8$.  
Since the case using the lensing signal at $R>0.5h^{-1}{\rm Mpc}$ gives the most unbiased result compared to other choices of the minimum radial scales, we choose the minimum radial scale of $R_{\rm min}=0.5h^{-1}{\rm Mpc}$ as our fiducial analysis setup. 
From now on, all the results shown use the minimum radial scale of 0.5$h^{-1}{\rm Mpc}$ if not mentioned.


\begin{figure}
	\includegraphics[width=\columnwidth]{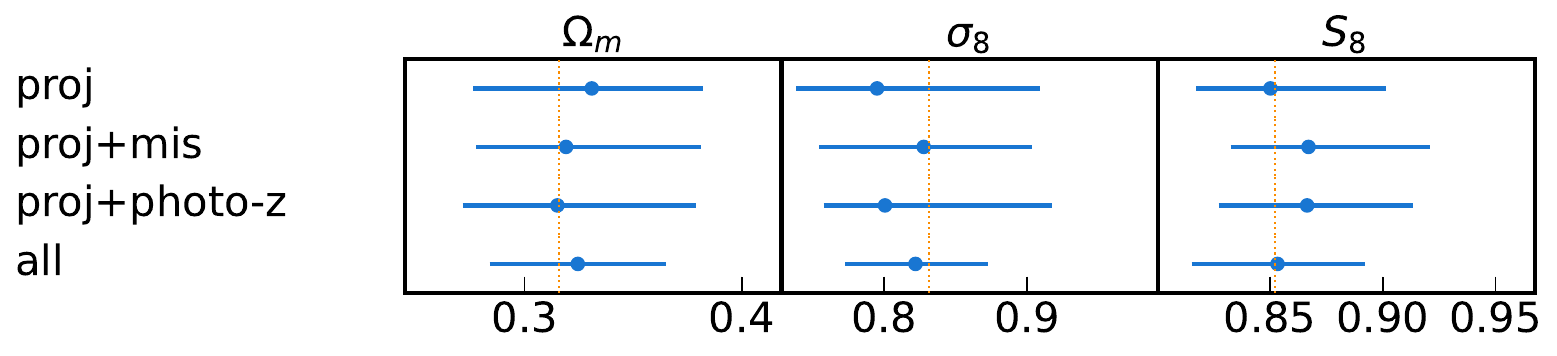}
    \caption{Summary of model validation with baryonic effects. Constraints on cosmological parameters $\Omega_{\rm m}$, $\sigma_8$, and $S_8$ are shown in each panel from left to right. The input cosmological parameters used for making synthetic data vectors are indicated by the vertical line in the panels.
    The case with "proj" uses the data vector including projection effects with a projection length of $d_{\rm proj}=60h^{-1}{\rm Mpc}$. The terms "mis" and "photo-z" imply including the mis-centering effect and photo-z scatter are included in the data vector. We include all the systematics in the data vector for the case of "all". For this analysis, we use the lensing signals at $R>0.5h^{-1}{\rm Mpc}$.}
    \label{fig:summary_baryon}
\end{figure}

Fig.~\ref{fig:summary_baryon} shows the summary of our tests including projection effects (\texttt{proj}), mis-centering (\texttt{proj+mis}), and photometric redshift scatters (\texttt{proj+photo-z}) in addition to baryonic effects. 
The result of \texttt{all} includes all the possible systematics (projection effects, mis-centering, photo-z scatter, and baryonic effects).
The constraints on $S_8$ are not biased in all these cases as also seen in Sec.~\ref{sec:mock:proj}.
Since baryonic effects did not affect the constraint on $S_8$, we decided to move on without modeling baryonic effects.
However, modeling baryonic effects is crucial for future work which includes smaller scales of the lensing profile or to improve the accuracy of the constraints.

\begin{figure}
	\includegraphics[width=\columnwidth]{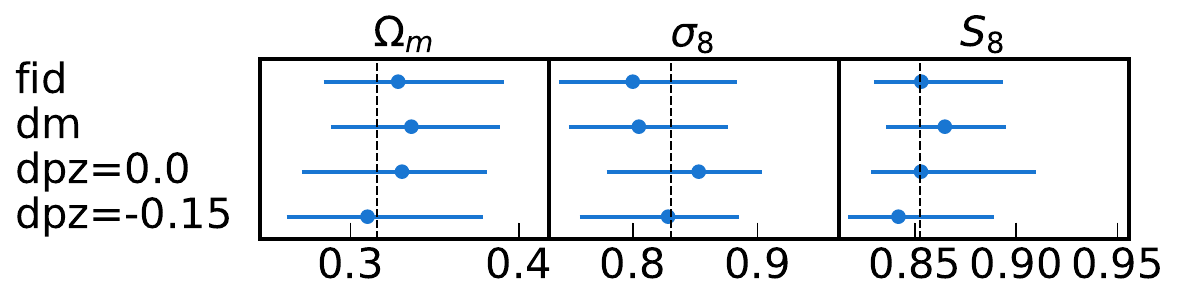}
    \caption{Summary of model validation using different priors for the residual 
    \hmrv{multiplicative} bias and the residual photo-$z$ bias of the source galaxies. Constraints on cosmological parameters $\Omega_{\rm m}$, $\sigma_8$, and $S_8$ are shown in each panel from left to right. The input cosmological parameters used for making synthetic data vectors are indicated by the vertical line in the panels. The case with "dm" uses a flat prior from -0.05 to 0.05 for the residual  
    \hmrv{multiplicative} bias and "dpz=0.0~(-0.15)" uses the values of 0.0~(-0.15) for the residual photo-$z$ bias of the source galaxies. These results demonstrate that even a wrong choice of the resudial photo-$z$ bias does not bias the cosmology constraint.}
    \label{fig:test_baryon2}
\end{figure}

Finally, we test the parameter choices for the residual multiplicative bias and the residual photo-$z$ bias of the source galaxies.
Both parameter values are set to zero in all the analyses presented in this section since the mock lensing signals do not suffer from these measurement-related residual systematics.
However, we want to make sure our choice for these parameter values do not bias our analysis result.
Fig.~\ref{fig:test_baryon2} shows the summary of our tests varying these parameters.
For the case of the residual multiplicative bias (labeled as ``dm''), we use a flat prior from -0.05 to 0.05 and apply the correction to the lensing data vector as described in Sec.~\ref{sec:mod:res}.
For the case of the residual photo-$z$ bias of the source galaxies (denoted as ``dpz''), we cannot use a flat prior and therefore test two different values for the residual photo-$z$ bias, dpz=0.0 and -0.15.
This is because our cluster analysis is done at a single redshift bin and we cannot self-calibrate this systematics.
As is clear from Fig.~\ref{fig:test_baryon2}, both using a flat prior on the residual multiplicative bias and using a wrong estimation of the residual photo-$z$ bias do not bias our analysis results.

We set our fiducial analysis following \citetalias{Park_2022} as discussed at the beginning of this section. The major changes from the fiducial analysis in \citetalias{Park_2022} are the number of parameters used to model the anisotropic boost on clustering and lensing observables and the minimum radial scale of lensing signals set to be $0.5h^{-1}{\rm Mpc}$.

\section{Results}
\label{sec:result}
In this section, we present the results of internal consistency tests in Sec.~\ref{sec:res:consistency} and the main result of this paper, the cosmological parameters constrained from the joint measurements of cluster abundance, clustering, and lensing signals measured from the SDSS redMaPPer cluster and the HSC-Y3 catalogs, in Sec.~\ref{sec:res:result}.
To avoid confirmation bias, we perform our cosmological analysis in a blinded manner using an analysis-level blinding.
The analysis-level blinding means that the values of cosmological parameters used in the theoretical models are not allowed to be seen when comparing to data and the central values of measured parameters are shifted to zero when plotting. Additionally, we do not compare the posterior for cosmological parameters with external results such as the \emph{Planck} CMB result.
During the internal consistency tests, we are allowed to see the ranges of the credible intervals and the relative shift of the cosmological parameters with respect to our fiducial analysis.

\subsection{Internal Consistency Tests}
\label{sec:res:consistency}

\begin{table*}
\caption{A summary of the analysis setups. The first column denotes each analysis setup, and $\mathcal{D}(\bm{\theta})$ and $\mathcal{D}({\bf d})$ denote the dimension of parameters and data vector in each analysis. Our data vector consist of the cluster abundance $\mathbf{N}_\mathrm{c}$, $\bm{\upDelta\upSigma}_i$ in 10 logarithmically spaced bins spanning $R \in [0.5,50] h^{-1}{\rm Mpc}$, and $\mathbf{w}_\mr{p,i}$ in 6 logarithmically spaced bins spanning $R \in [2,50] h^{-1}{\rm Mpc}$ each for four richness bins $\lambda \in [20,30), [30,40), [40,55),$ and $[55,200)$.
 \label{tab:analysis_setups}}
\renewcommand{\arraystretch}{1.2}
\setlength{\tabcolsep}{2pt}
\begin{center}
\begin{tabular}{llc}\hline\hline
setup label & description & $\mathcal{D}(\bm{\theta})$, $\mathcal{D}({\bf d})$ \\
\hline 
fiducial                                               & {\it baseline analysis} $\mathbf{N}_\mathrm{c} + \mathbf{w}_\mr{p,1-4} + \bm{\upDelta\upSigma}_{1-4}$                                  & 21, 68\\
\hline
demp                                               & 
demp  is used to infer the source redshift distribution 
and  for $\bm{\upDelta\upSigma}_{1-4}$ measurement.                                              
& 21, 68\\
mizuki                                                  & mizuki is used for source sample selection and $\bm{\upDelta\upSigma}_{1-4}$ measurement                                                                                   & 21, 68\\
\dnnz                                                   & \dnnz is used for source sample selection and $\bm{\upDelta\upSigma}_{1-4}$ measurement (same as fiducial)                                                                                    & 21, 68\\
\hline
Rmin=0.2                                  & Using the minimum scale cut of $0.2h^{-1}{\rm Mpc}$ for $\bm{\upDelta\upSigma}_{1-4}$                             & 21, 76\\
Rmin=1.0                                 & Using the minimum scale cut of $1.0h^{-1}{\rm Mpc}$ for $\bm{\upDelta\upSigma}_{1-4}$                             & 21, 64\\
\hline
nonc                                  & $\mathbf{w}_\mr{p,1-4} + \bm{\upDelta\upSigma}_{1-4}$ without $\mathbf{N}_\mathrm{c}$                           & 21, 64\\
nods                                 &  $\mathbf{N}_\mathrm{c} + \mathbf{w}_\mr{p,1-4}$ without $\bm{\upDelta\upSigma}_{1-4}$                            & 21, 28\\
nowp                                  & $\mathbf{N}_\mathrm{c} + \bm{\upDelta\upSigma}_{1-4}$ without $\mathbf{w}_\mr{p,1-4}$                             & 21, 44\\
\hline
no20to30                                     & $\mathbf{N}_\mathrm{c,1-3} + \mathbf{w}_\mr{p,1-3} + \bm{\upDelta\upSigma}_{1-3}$ for $\lambda \in [30,40), [40,55),$ and $[55,200)$ but not $\lambda \in [20,30)$                                                                                                                          & 21, 51\\
no30to40                                  & the same as above, without the observables for $\lambda \in [30,40)$                                                                                                                                  & 21, 51\\
no40to55                                  & the same as above, without the observables for $\lambda \in [40,55)$                                                                                                                                 & 21,51\\
no55to200                                 & the same as above, without the observables for $\lambda \in [55,200)$                                                                                                                                 & 21,51\\

\hline
XMM  ($\sim33$~deg$^2$)${}^{\ast}$                                 & similar to "fiducial", but using the lensing signals of the XMM field alone                                                                                              & 21, 68\\
GAMA15H ($\sim41$~deg$^2$)${}^{\ast}$                                & similar to "fiducial", but using the lensing signals of the GAMA15H field alone                                                                                          & 21, 68\\
HECTOMAP ($\sim43$~deg$^2$)${}^{\ast}$                               & similar to "fiducial", but using the lensing signals of the HECTOMAP field alone                                                                                         & 21, 68\\
GAMA09H ($\sim78$~deg$^2$)${}^{\ast}$                                & similar to "fiducial", but using the lensing signals of the GAMA09H field alone                                                                                          & 21, 68\\
VVDS ($\sim96$~deg$^2$)${}^{\ast}$                                 & similar to "fiducial", but using the lensing signals of the VVDS field alone                                                                                             & 21, 68\\
WIDE12H ($\sim121$~deg$^2$)${}^{\ast}$                              & similar to "fiducial", but using the lensing signals of the WIDE12H field alone                                                                                          & 21, 68\\

\hline
dpzs                                        & randomly sampling $\Delta z_{\rm ph}$ from the chain provided by \cite{hscy3_shear_real}                                                                                                        & 22, 69\\

dm                                              & using a uniform prior of $\Delta m=\mathcal{U}(-0.05,0.05)$                                                                                                         & 22, 69\\

\hline\hline
\end{tabular}
\end{center}
\renewcommand{\arraystretch}{1}
\end{table*}

\begin{figure*}
	\includegraphics[width=\textwidth]{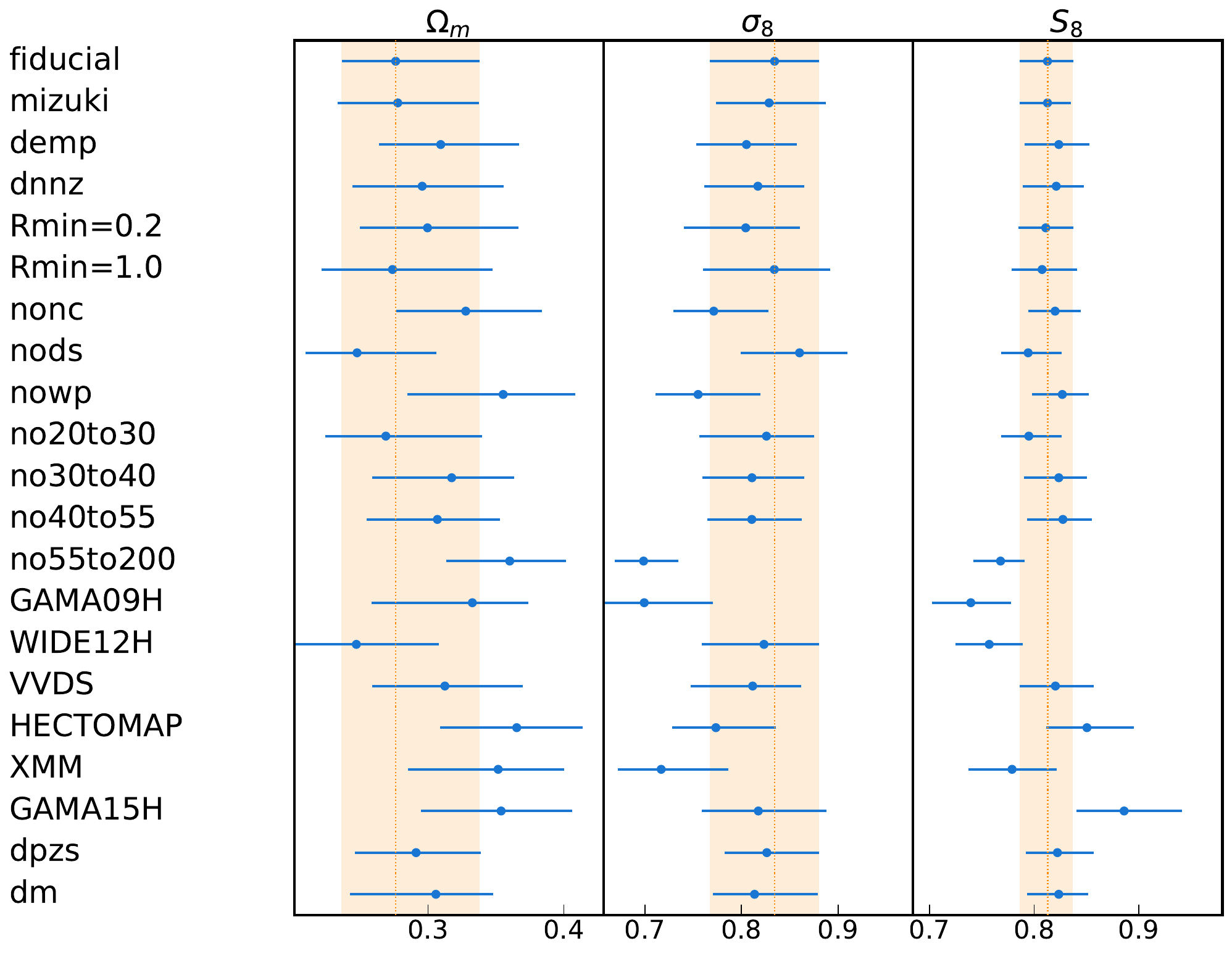}
    \caption{A summary of cosmological parameters constrained from each of the different analysis setups listed in Table~\ref{tab:analysis_setups}. 
    }
    \label{fig:consistency}
\end{figure*}

We first present the cosmological parameter estimation for various setups listed in Table~\ref{tab:analysis_setups} as self-consistency tests.
These are the tests we performed before unblinding to make sure our analysis gives consistent results for different setups.
The results of each setup are shown in Fig.~\ref{fig:consistency}.
We find that the $S_8$ parameter does not have a significant shift in these different setups except in a few cases, while some tests show more scattered shifts in the values of $\Omega_{\rm m}$ and $\sigma_8$. 
However, such shifts are also seen in the tests using the mock catalogs, and therefore the shifts in these parameters are likely due to projection effects of the non-Gaussian posterior distribution in the full dimensional parameter space.

For the case of using different photo-$z$ catalogs (denoted as ``demp'', ``mizuki'', and ``\dnnz''), we find that the constrained values of $S_8$ using these different catalogs agree well with the value constrained in the fiducial analysis. 
Similarly, we find good agreements on the constraint of $S_8$ values for the case of using different minimum scale cuts on the cluster lensing signals. We try the minimum scale cuts of $0.2h^{-1}{\rm Mpc}$ (labeled as ``Rmin=0.2'') and $1.0h^{-1}{\rm Mpc}$ (as ``Rmin=1.0''), and we find that the constraints on cosmological parameters $\Omega_{\rm m}$ and $\sigma_8$ also agree well for the case of ``Rmin=1.0''. 
However, using the minimum scale cut of $0.2h^{-1}{\rm Mpc}$ prefers slightly larger $\Omega_{\rm m}$ and lower $\sigma_8$ by roughly 0.5~$\sigma$.

For the case of dropping any one sector of our data vector (abundance, lensing, and clustering) denoted as "nonc", "nods", and "nowp" respectively, the shifts in $S_8$ parameter are within $0.5~\sigma$.
Interestingly, the cases of "nonc" and "nowp" shift the parameter values of $\Omega_{\rm m}$, $\sigma_8$, and $S_8$ in the same direction, while the case of "nods" shifts in the opposite direction, as a lower $\Omega_{\rm m}$ and higher $\sigma_8$ than the fiducial result.
The result from \citetalias{Park_2022} which tested the same setups (see Fig.~7 in \citetalias{Park_2022}) showed the opposite result that the "nods" case (i.e., dropping the lensing data vector) prefers a larger $\Omega_{\rm m}$ and lower $\sigma_8$ than their fiducial result although these parameters are less constrained compared to the other cases ("nonc" and "nowp").
This is because \citetalias{Park_2022} uses richness-dependent parameters to model the anisotropic boost of clustering and lensing signals and the lensing signals were crucial to solve the degeneracy between the cluster bias and the anisotropic boost parameters.

For the case of dropping one of the richness bins ("no20to30", "no30to40", "no40to55", and "no55to200"), the results are consistent except in the case of "no55to200" where the $S_8$ value is shifted almost $1\sigma$ lower than the fiducial analysis.
To test whether this is due to noise in the data vector, we repeated the same analysis using the mock catalogs described in Sec.~\ref{sec:mock_test}.
However, we did not find any significant shifts in the case of the noiseless mock data vector.
To check whether this shift is statistically significant, we compute the $p$-value of $S_8$ considering all four cases with the assumption that our fiducial result is the true $S_8$ value.
Note that this is a crude estimation because we treat four different setups as independent measurements. The measured $p$-value is 0.461 with $\chi^2=3.61$, indicating that shift in the case of "no55to200" is not statistically significant and consistent in the data vector prediction.
We additionally note that we do not see any significant changes in the values of the cosmological parameters by dropping the lowest richness bin, unlike the result from \cite{desy1cl}.
We also consider the cases of using lensing signals measured only from one of the six HSC-Y3 regions. Areas of these regions vary from 33~${\rm deg}^2$ to 121~${\rm deg}^2$ and some of them only contain a few clusters at the largest richness bin. These results show larger scatters in the value of $S_8$ compared to other setups. However, the measured $p$-value for these cases is 0.148 with $\chi^2=9.49$ and the scatters are not statistically significant. 

Finally, we consider the cases using a flat prior for the residual multiplicative bias (labeled as ``dm'') and randomly sampling the residual photo-$z$ bias (labeled as ``dpzs'') from the chain provided by \cite{hscy3_shear_real}.
The detail of the chain used is described in Appendix~\ref{app:photo-z}.
For both cases, the constraints on $S_8$ agree well with the fiducial result and the $\Omega_{\rm m}$ and $\sigma_8$ are within 0.5~$\sigma$ shift from the fiducial.
Therefore, we conclude that our choices for the residual multiplicative bias and randomly sampling the residual photo-$z$ bias are unlikely to bias our cosmology result. 


\subsection{$\Lambda$CDM constraints}
\label{sec:res:result}
\begin{figure*}
	\includegraphics[width=\textwidth]{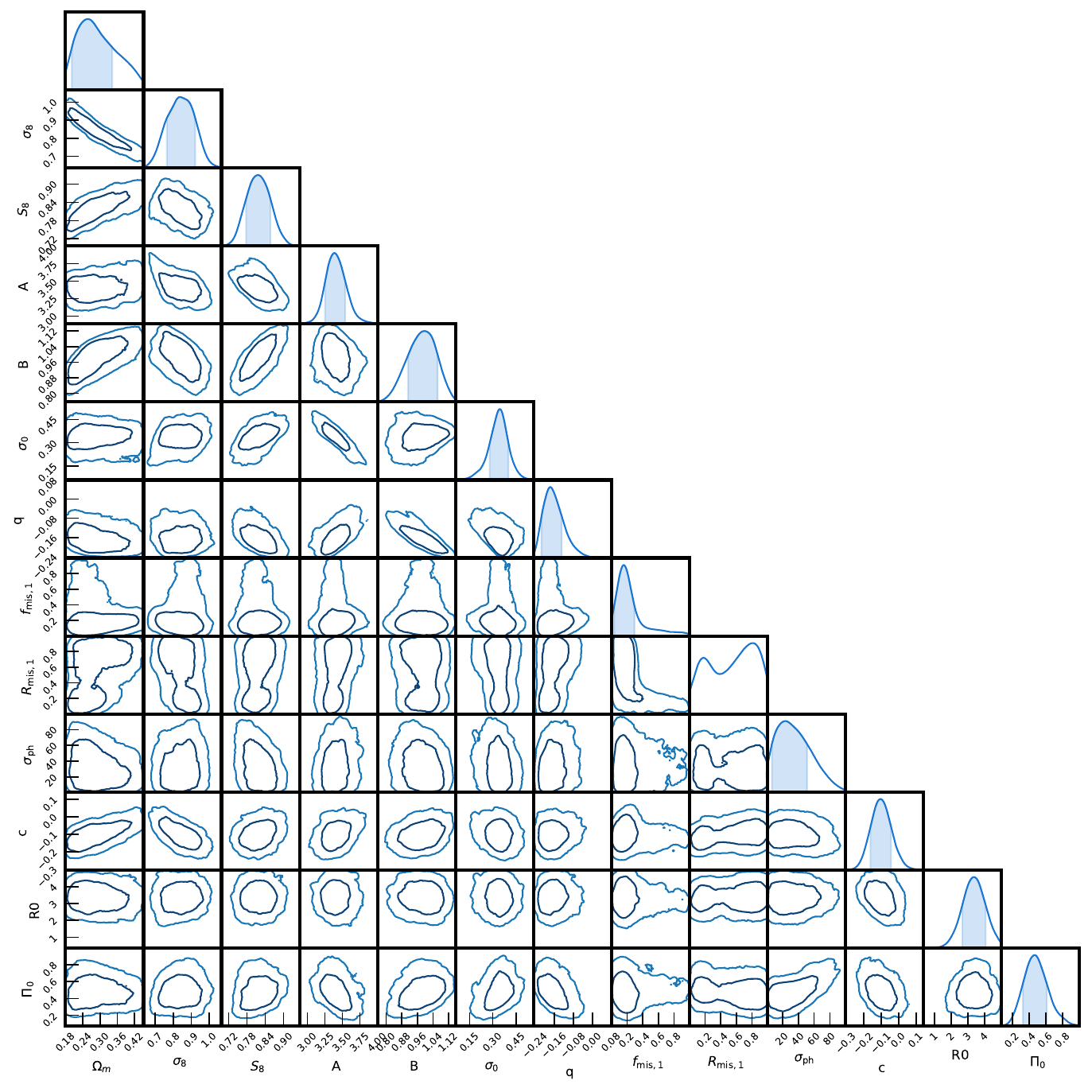}
    \caption{The 68\% and 95\% confidence regions from our fiducial analysis combining the SDSS redMaPPer cluster abundances, clustering, and lensing (measured from the HSC-Y3 catalog). We show three derived cosmological parameters of interest, four MOR parameters (A, B, $\sigma_0$, and q), two mis-centering parameters for the first richness bin ($f_{\rm mis,1}$ and $R_{\rm mis,1}$), one cluster photo-z scatter ($\sigma_{\rm ph}$), and three anisotropic boost parameters (x, R0, and $\Pi$).
    }
    \label{fig:results_all}
\end{figure*}

\begin{figure*}
	\includegraphics[width=0.66\columnwidth]{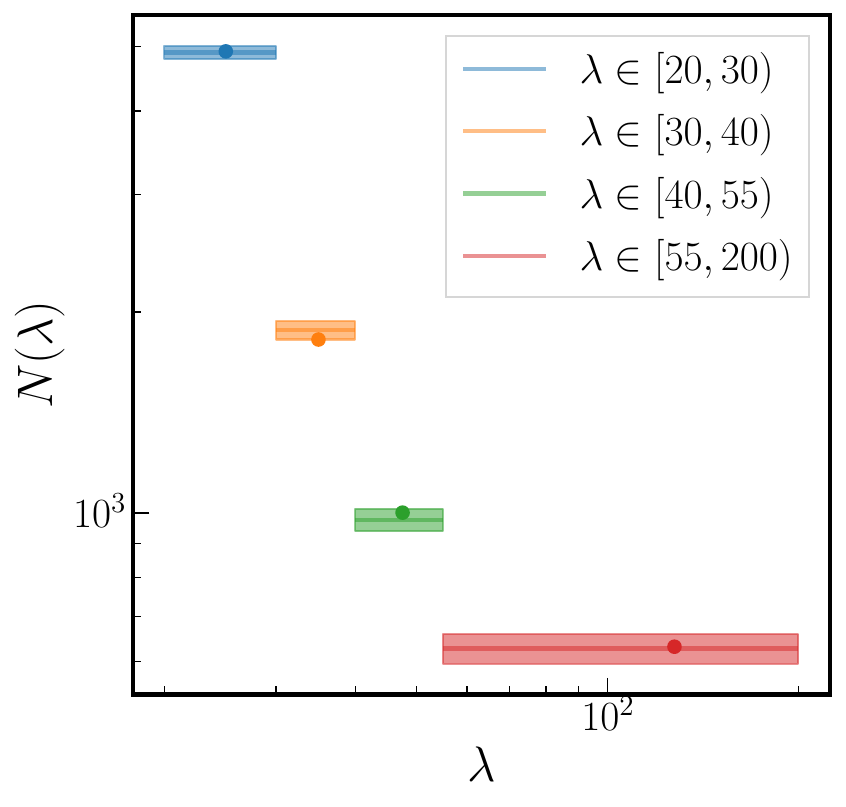}
	\includegraphics[width=0.66\columnwidth]{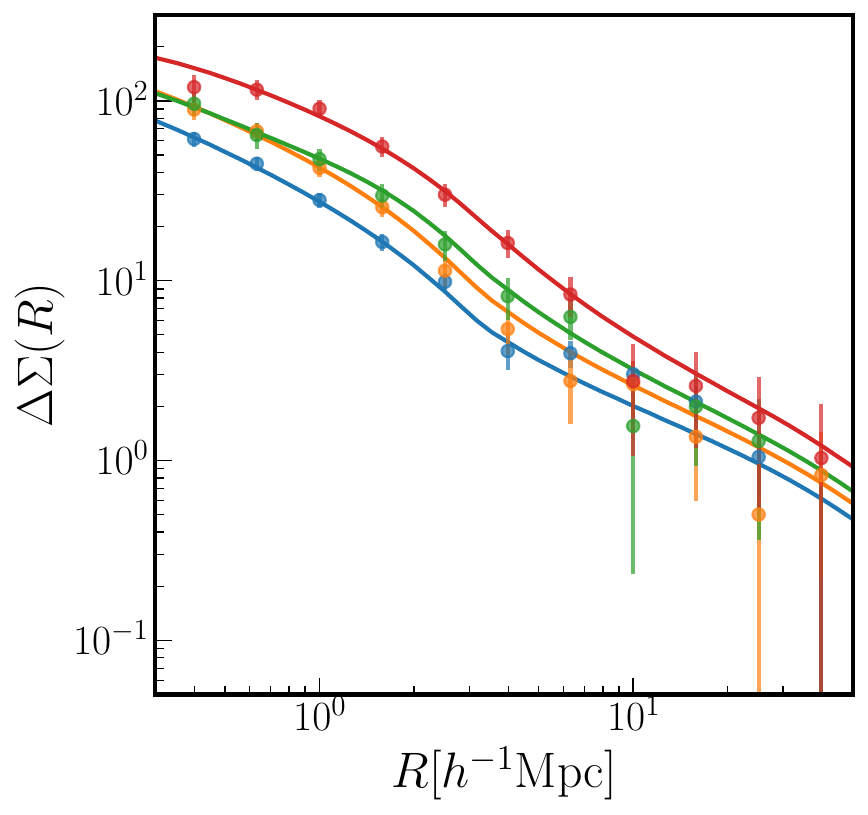}
	\includegraphics[width=0.66\columnwidth]{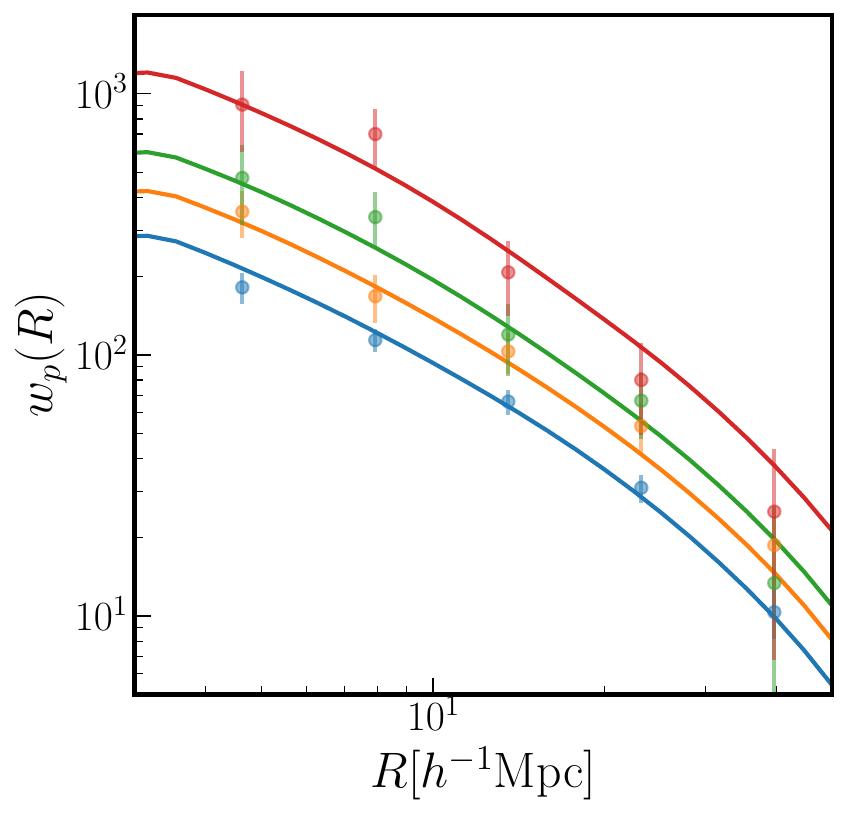}
    \caption{\textit{Left:} Cluster abundance measurements from the SDSS redMaPPer catalog. Horizontal bars represent our number counts, and the shaded regions represent the 1-$\sigma$ uncertainties. The colors correspond to our richness bins. The points are the best-fit predictions from the fiducial analysis. \textit{Center:} Points show cluster lensing measurements from the HSC-Y3 shape catalog and 1~$\sigma$ uncertainties for the four richness bins, and the solid lines are the best-fit predictions. \textit{Right:} Same as the center panel but for cluster clustering measurements.}
    \label{fig:meas}
\end{figure*}

\begin{figure}
	\includegraphics[width=\columnwidth]{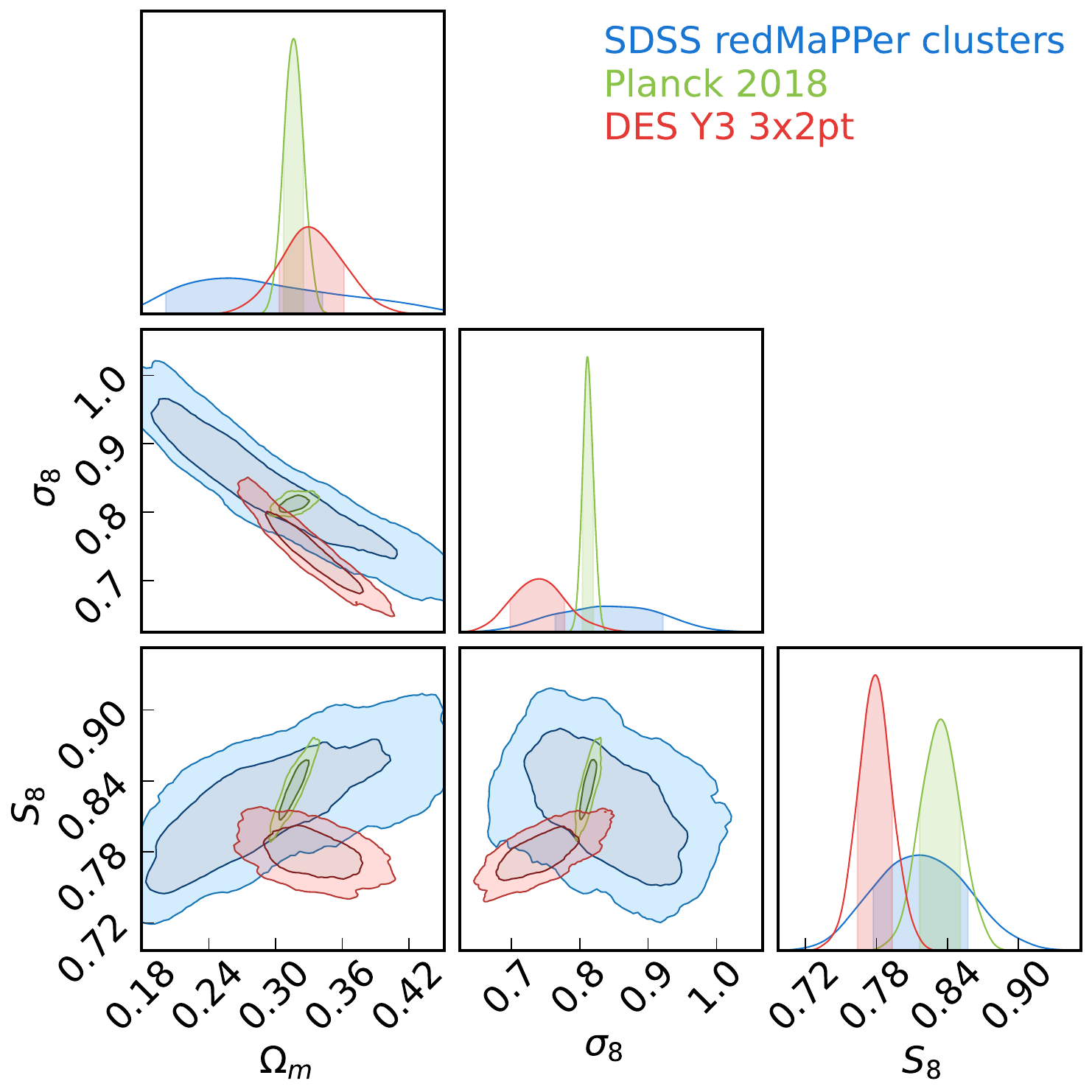}
    \caption{Comparison of 68\% and 95\% confidence regions for cosmological parameters of interest from our fiducial 
    results (blue) with \citet{Planck18} in light green and \citet{des_y3} in red.
    }
    \label{fig:results_fid}
\end{figure}

We show the results from our fiducial analysis using the SDSS redMaPPer cluster and the HSC-Y3 catalogs in Figs.~\ref{fig:results_all} and \ref{fig:results_fid}.
We report the central value and 68\% credible interval for the cosmological parameters of interest:
\begin{ceqn}
\begin{align*}
    \Omega_{\rm m}=  0.258^{+0.085}_{-0.057},\nonumber \\
    \sigma_8= 0.838^{+0.083}_{-0.074},\nonumber \\
    S_8 = 0.816^{+0.041}_{-0.039}.
\end{align*}
\end{ceqn}
Extending the definition of $S_8$ to $S'_8\equiv \sigma_8(\Omega_{\rm m}/0.3)^\alpha$ with $\alpha$ being a free parameter, we find that the best-constrained parameter is $\alpha\simeq 0.357$: with this value, we find $S_8'\simeq 0.823\pm0.027$.
This result is from the \texttt{multinest} without using the ``constant efficiency'' mode, and we confirmed that the results converge when we use two different seeds.
As discussed in Sec.~\ref{sec:mod:param}, we use the \texttt{multinest} with the ``constant efficiency'' mode for the mock tests and the internal consistency tests as the crucial point of these analyses is to make sure that all the results are consistent.
However, underestimation of the errors is a known issue when the \texttt{multinest} with ``constant efficiency'' mode is used \citep{lemos_2023}.
To confirm how much the errors on cosmological parameters of interests are underestimated by using the \texttt{multinest} with the ``constant efficiency'' mode, we repeat the fiducial analysis ten times using different seeds.
We find that the constraints on $S_8$ have the standard deviation of 25\%. 
The constraint on $S_8$ from a chain closest to the mean $S_8$ is $S_8=0.813^{+0.024}_{-0.027}$, which underestimates the error by 40\%.

Fig.~\ref{fig:results_all} shows the 4 MOR parameters (A, B, $\sigma_0$, and q), 2 mis-centering parameters ($f_{\rm mis,1}$ and $R_{\rm mis,1}$) for the lowest richness bin, the photo-$z$ scatter ($\sigma_{\rm ph}$), and 3 anisotropic boost parameters (c,$R_0$, and $\Pi$) in addition to the derived cosmological parameters of interest.
The rest of the mis-centering parameters are presented in Appendix~\ref{app:mis}.
We find that our systematic parameters are mostly self-calibrated.
The degree of cluster photo-$z$ scatter is determined to be $\sim 30 h^{-1}{\rm Mpc}$ is consistent with the performance of the SDSS redMaPPer photo-$z$ estimation.
We also find mis-centering fraction $\sim 15\%$ is consistent with X-ray studies on the same cluster sample \citep[e.g.][]{miyatakecl,zhang18}. However, the mis-centering scale of $\sim 0.7h^{-1}{\rm Mpc}$ is somewhat overestimated compared to these studies, which estimated values of $\sim 0.2h^{-1}{\rm Mpc}$. 
Note that the mis-centering parameters are not correlated with cosmological parameters and therefore overestimation of these parameter values does not bias our cosmology result.
The anisotropic boost amplitude ranging between 40-50\% is somewhat overestimated compared to the independent constraint from \cite{sunayama22}, which is roughly 30-40\%. 

We model the MOR using Eqs.~\ref{eq:mor1} and \ref{eq:mor2} and the best-fit values for these parameters are: $A=3.38^{+0.16}_{-0.12}$, $B=1.0^{+0.062}_{-0.065}$, $\sigma_0=0.344^{+0.055}_{-0.062}$, and $q=-0.176^{+0.048}_{-0.034}$.
Our model for the MOR is slightly different from the one adopted in \cite{costanzisdss,desy1cl}, where these studies model the MOR such as
\begin{equation}
\left<M_{\rm 200m}|\lambda \right>=10^{M_0} \left(\frac{\lambda}{40} \right)^{\alpha}.
\end{equation}
Converting the best-fit values of our MOR parameters to the above parameter model, we obtain roughly $M_0 \sim 14.5$ and $\alpha \sim 1.1$, which are consistent with the values found in \cite{tokrause2} and other cluster studies \citep{baxter_2016,farahi_2016,costanzi_spt_2021}. Note that many of these studies constrain the MOR parameters of the redMaPPer clusters assuming the fixed cosmology.
Since the MOR parameters are inherently correlated with cosmological parameters, the fact that our constrained MOR parameters are in good agreement with other cluster studies using the redMaPPer clusters is encouraging.
In particular, we employed almost the same analysis and used the same cluster catalog as \citetalias{Park_2022}, whose posterior also favored lower $\Omega_{\rm m}$.
The difference between \citetalias{Park_2022} and our result is due to the measured lensing signals.
We compare the measured lensing signals using the SDSS and HSC-Y3 data in Appendix~\ref{app:lensing} and further discuss its implication.
In short, the amplitude of the lensing signal measured from the SDSS data was suppressed \hmrv{at} 
$R>3h^{-1}{\rm Mpc}$ compared to the one measured from the HSC-Y3 data and we suspect that this difference is the main driver of the difference in the resulting cosmological constraints between \citetalias{Park_2022} and this work.
We compare our measurements with the best-fit predictions from our fiducial analysis in Fig.~\ref{fig:meas} and find good agreement across
all sectors between measurements and predictions.

Fig.~\ref{fig:results_fid} compares our result with these from other cosmology surveys.
For the CMB constraints, we consider "Planck 2018" result \citep{Planck18} from the analysis where the primary CMB temperature and E-mode polarization anisotropy information ("TT, EE, TE+lowE") are used. 
Note that the neutrino mass is fixed at 0.06eV in this analysis, consistent with our analysis.
For the posterior distribution of the "DES Y3 3x2pt" result, we used the public chain, which is the result obtained from the 3x2pt cosmological analysis \citep{des_y3}.
Considering the comparison with other cosmology results, Fig.~\ref{fig:results_fid} does not show any significant tensions with the results from both \emph{Planck} 2018 and DES Y3 3x2pt.
In particular, our $S_8$ value agrees with the $S_8$ value from \emph{Planck} 2018 within $1\sigma$. 

\begin{figure}
	\includegraphics[width=\columnwidth]{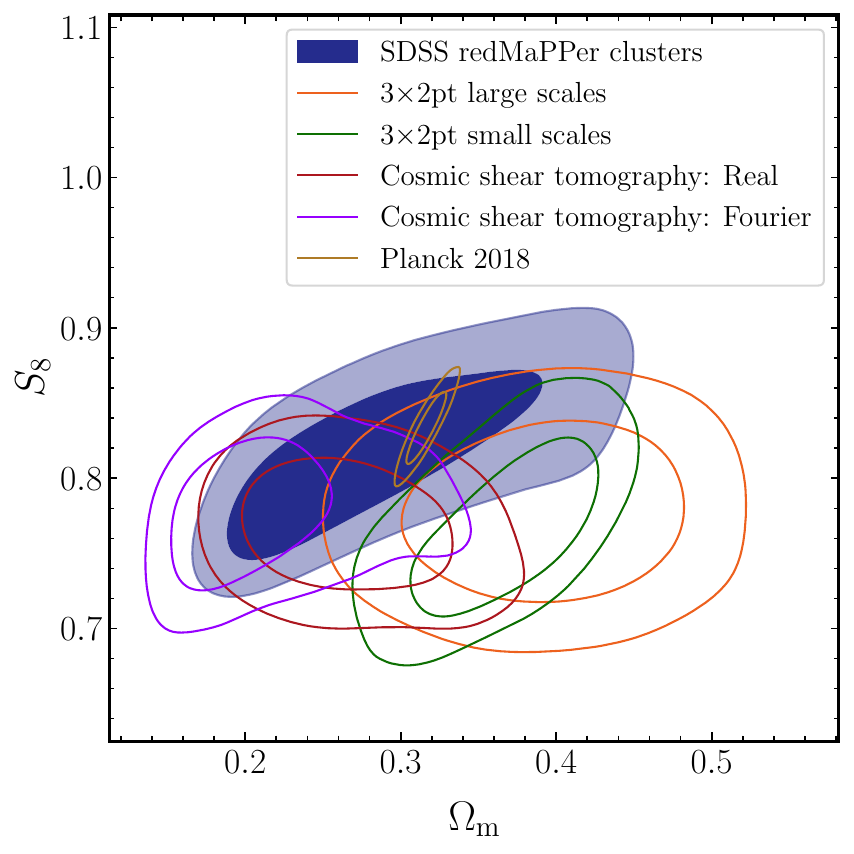}
    \caption{Comparison of 68\% and 95\% confidence regions for cosmological parameters of interest from our fiducial result (blue) with other HSC-Y3 cosmological analyses.
    }
    \label{fig:results_hsc}
\end{figure}

Fig.~\ref{fig:results_hsc} compares the cosmological constraints from different cosmology analyses using the HSC-Y3 data. 
``3x2pt small/large-scale'' \citep{hscy3_3x2_minimal,hscy3_3x2_small} use the same data, but ``3x2pt small scale'' uses the Dark Emulator and HOD to model galaxy clustering and lensing signals while ``3x2pt large scale'' uses the perturbation theory based theoretical template for their modeling.
There are also two cosmic shear tomography analyses. One measures the cosmic shear signals in real-space \citep{hscy3_shear_real}, and the other measures the signals in Fourier-space \citep{hscy3_shear_fourier}.
Our analysis is consistent with other HSC-Y3 cosmology analyses at levels of roughly $1\sigma$.
Even though all these results are consistent on the constraint of the $S_8$ value, the ``3x2pt'' analyses prefer a larger value of $\Omega_{\rm m}$, while cosmic shear results prefer a smaller value of $\Omega_{\rm m}$.
Our posterior on $\Omega_{\rm m}$ is consistent with both the ``3x2pt'' and the cosmic shear results. 
Note that the constraining power of HSC-Y3 analyses on cosmological parameters is almost the same as the HSC Y1 result \citep{hikage_etal2019}, because these analyses adopted the uninformative prior to the residual photo-$z$ error parameters.
Since self-calibration on the residual photo-$z$ error requires tomographic analysis using the same source galaxies \citep{oguritakada11}, our analysis cannot use such uninformative priors. 
So, we adopt the residual photo-$z$ error parameter samples from the HSC-Y3 cosmic shear.

\begin{figure}
	\includegraphics[width=0.95\columnwidth]{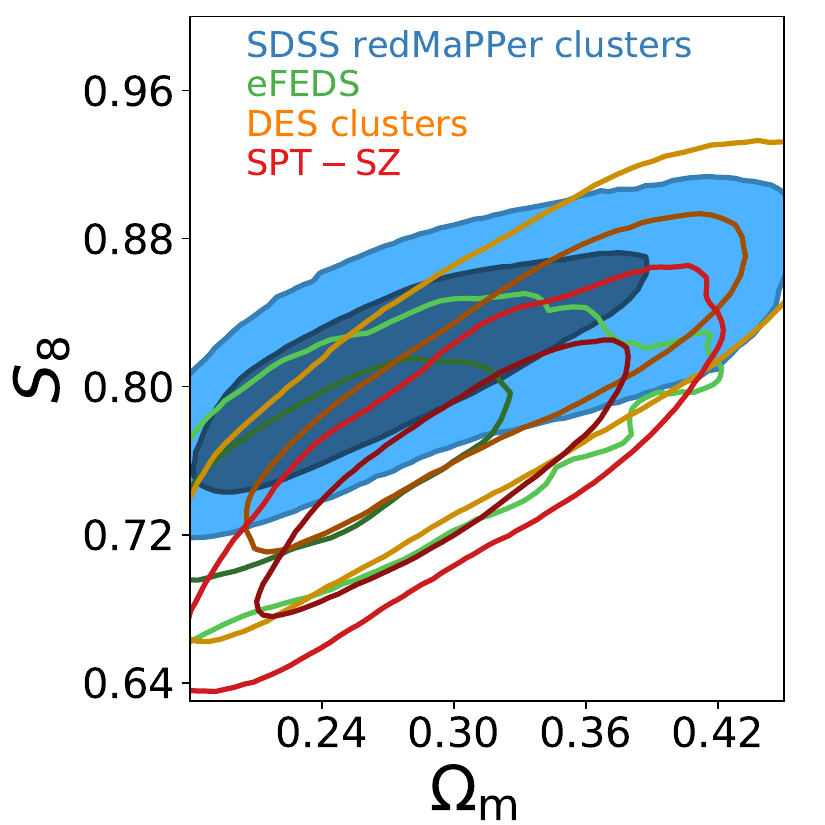}
    \caption{Comparison of 68\% and 95\% confidence regions for cosmological parameters of interest from our fiducial 
    results (blue) with other cluster cosmology analyses using different data: eFEDS X-ray cluster \citep{Inon2023} in green, DES clusters using multi-wavelength data \citep{costanzi_spt_2021} in yellow, and \citep{SPT2019} in red.
    }
    \label{fig:results_cluster}
\end{figure}

Fig.~\ref{fig:results_cluster} compares the cosmological constraints on $\Omega_{\rm m}$ and $S_8$ from other cluster cosmology analyses using clusters from different surveys: SZ clusters from the SPT survey \citep{SPT2019}, X-ray clusters from \emph{eROSITA} Final Equatorial
Depth Survey (eFEDS) \citep{Inon2023}, and optical clusters calibrated by SZ multi-wavelength data from DES and SPT surveys \citep{costanzi_spt_2021}.
Our constraints are in agreement with other cosmology results using differently identified cluster samples.
The consistency with the cosmology results from these cluster catalogs are at levels of roughly $1\sigma$.

\begin{figure}
	\includegraphics[width=0.9\columnwidth]{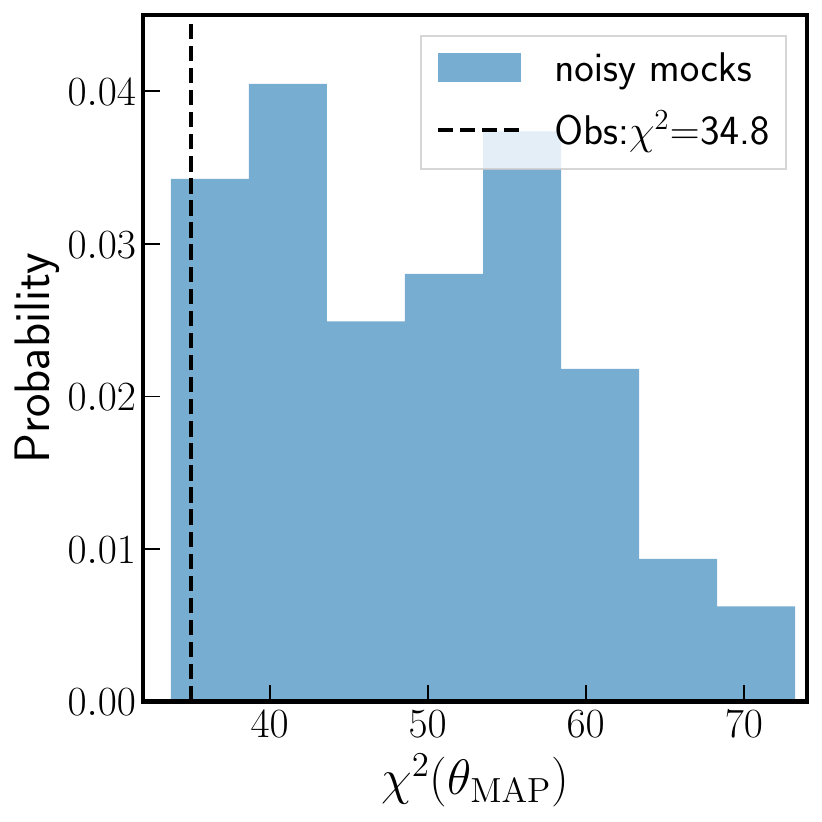}
    \caption{An evaluation of the goodness-of-fit of the best-fit (MAP) model in Figs.~\ref{fig:results_all} and \ref{fig:results_fid}. The histogram shows the probability density distribution of the $\chi^2$ values at the MAP model, obtained by applying the same baseline analysis to 65 noisy mock datasets. The vertical black dashed line denotes the $\chi^2$ value ($\chi^2$=34.8) at MAP for the actual cluster analysis using the SDSS redMaPPer and the HSC-Y3 catalogs.
    }
    \label{fig:goodness}
\end{figure}

Finally, we evaluate the goodness-of-fit of the best-fit model to the measured signal in Fig.~\ref{fig:goodness}.
To do this evaluation, we first generate 65 noisy mock realizations using the covariance matrix used in our fiducial analysis.
The histogram in Fig.~\ref{fig:goodness} shows the distribution of the $\chi^2$ value of the \emph{maximum a posteriori} (MAP) model prediction for each of the 65 realizations.
From the histogram, the effective degrees of freedom is $\nu_{\rm eff}=$ 48.4 while our naive estimation of degrees of freedom is $\nu = 68-21 = 47$.
The actual $\chi^2$ at MAP for our fiducial analysis is $\chi^2=$34.8, corresponding to a $p$-value of 0.919, which indicates that our theoretical model gives an acceptable fit to the data within the error bars.

\section{Summary}
\label{sec:summary}

In this paper, we have presented a cluster cosmology analysis using the SDSS redMaPPer cluster and the HSC-Y3 shape catalog.
We use an analysis that fully forward models the abundance, clustering, and lensing signals of galaxy clusters as developed in \citetalias{Park_2022}.
This previous study used cluster lensing signals measured from the SDSS data, and the result preferred low $\Omega_{\rm m}$ and high $\sigma_8$.
However, in this study, we measure the weak lensing signals of the clusters using the HSC-Y3 shape catalog, which has significantly deeper photometry and better image quality than the SDSS source catalog, enabling source galaxy selections to be more secure and robust. 
With the lensing measurements from the HSC-Y3, the result of our cosmology analysis is consistent with the cosmology constraints from other CMB/LSS methods and surveys.

Our work and findings can be summarized as follows.
\begin{itemize}
\item Our cluster analysis constraints the $S_8$ value, $S_8=0.816^{+0.041}_{-0.039}$. We note that our constraint on $S_8$ agrees with the result from \emph{Planck} 2018 within $1\sigma$ and do not find any significant tensions with CMB or LSS measurements.

\item The difference between our analysis and \citetalias{Park_2022} comes from the weak lensing measurements of the clusters. Using the HSC-Y3 shape catalog enables us to securely select source galaxies. This makes our lensing measurement almost correction-free as shown in Fig.~\ref{fig:boost_cross}.

\item Due to a small area covered by the HSC-Y3 data, the lensing component of the covariance matrix is not shape-noise-dominated on large scales. Due to the loss of constraining power from the lensing signals on large scales (i.e., unable to resolve the degeneracy between the cluster bias and the anisotropic boost parameters), we decided to use the same anisotropic boost model for the clustering and lensing observables in all richness bins.

\item In addition to the validation tests done in \citetalias{Park_2022}, we additionally consider baryonic effects on the lensing signals. For cluster-sized halos, baryonic effects suppress the amplitude of the lensing signals by $\sim 10\%$ at scales $R \leq 1 h^{-1}{\rm Mpc}$. This suppression has negligible effects on the cosmology constraints.

\item Before unblinding, we carried out internal consistency tests with various analysis setups and found that the $S_8$ parameter does not shift significantly in the different setups.
\end{itemize}

In summary, our analysis performed well on the mocks and accurately constrained cosmological parameters against systematics including projection effects, mis-centering, photo-$z$ scatter, and baryonic effects. 
Our cosmology results on real data are consistent with other CMB and LSS-based cosmology results, especially the CMB cosmology result from \emph{Planck} 2018.
In other words, we do not find any evidence of tension in the $S_8$ values between our analysis and the \emph{Planck} CMB analysis.
However, we need to improve the precision of our measurements for this to be a conclusive statement.

This is the first study that cosmological analysis using only optical cluster observables provides consistent results with other CMB/LSS cosmology analyses. In doing so, it was essential to model the anisotropic boosts due to projection effects and robustly measure the lensing signals with source galaxies well apart from lens clusters.
The latter was possible due to superb image quality and depth of the HSC-Y3 data.
However, our cluster cosmology analysis does not strongly constrain $\Omega_{\rm m}$ because our cluster analysis is done at a single richness bin. Evolution of cluster abundance is sensitive to $\Omega_{\rm m}$ \citep{Bahcall1995}. 
An optically-selected cluster catalog from the HSC data \citep{oguri2018} (called CAMIRA cluster catalog) spans the redshift range of $0.1 \leq z \leq 1.1$, which enables us to track down the evolution of cluster abundance and is expected to give a tighter constrain on $\Omega_{\rm m}$ than our current work. 
Furthermore, we can self-calibrate the residual photo-$z$ bias by using cluster samples at multiple redshift bins. Conducting cluster cosmology analysis on the HSC CAMIRA cluster catalog using our analysis method will be our future work.

\section*{Acknowledgements}
We thank Youngsoo Park for helping us to provide the data from his previous analysis and for the useful discussions on our mock tests. We thank Sebastian Bocquet and Matteo Costanzi for providing their cosmological chains to make Figure~\ref{fig:results_cluster}.

This work was supported in part by World Premier International Research Center Initiative (WPI Initiative), MEXT, Japan, and JSPS KAKENHI Grant Numbers JP19K14767,  20H01932, JP20H05855, JP20H05861,  JP21J10314, JP21K03625, 21H05456,  JP22K21349, JP23H00108, and by Tokai Pathways to Global Excellence (T-GEx), part of MEXT Strategic Professional Development Program for Young Researchers.
A part of numerical computations was carried out on Cray XC50 at the Center for Computational Astrophysics in NAOJ.
RD acknowledges support from the NSF Graduate Research Fellowship Program under Grant No.\ DGE-2039656. 
XL is supported by the Department of Energy grant DE-SC0010118.
INC is supported by the National Science and Technology Council in Taiwan (Grant NSTC 111-2112-M-006-037-MY3)

The Hyper Suprime-Cam (HSC) collaboration includes the astronomical communities of Japan and Taiwan, and Princeton University. The HSC instrumentation and software were developed by the National Astronomical Observatory of Japan (NAOJ), the Kavli Institute for the Physics and Mathematics of the Universe (Kavli IPMU), the University of Tokyo, the High Energy Accelerator Research Organization (KEK), the Academia Sinica Institute for Astronomy and Astrophysics in Taiwan (ASIAA), and Princeton University. Funding was contributed by the FIRST program from Japanese Cabinet Office, the Ministry of Education, Culture, Sports, Science and Technology (MEXT), the Japan Society for the Promotion of Science (JSPS), Japan Science and Technology Agency (JST), the Toray Science Foundation, NAOJ, Kavli IPMU, KEK, ASIAA, and Princeton University. This paper makes use of software developed for the Large Synoptic Survey Telescope. We thank the LSST Project for making their code available as free software at \url{http://dm.lsst.org}

The Pan-STARRS1 Surveys (PS1) have been made possible through contributions of the Institute for Astronomy, the University of Hawaii, the Pan-STARRS Project Office, the Max-Planck Society and its participating institutes, the Max Planck Institute for Astronomy, Heidelberg and the Max Planck Institute for Extraterrestrial Physics, Garching, The Johns Hopkins University, Durham University, the University of Edinburgh, Queen's University Belfast, the Harvard-Smithsonian Center for Astrophysics, the Las Cumbres Observatory Global Telescope Network Incorporated, the National Central University of Taiwan, the Space Telescope Science Institute, the National Aeronautics and Space Administration under Grant No. NNX08AR22G issued through the Planetary Science Division of the NASA Science Mission Directorate, the National Science Foundation under Grant No. AST-1238877, the University of Maryland, and Eotvos Lorand University (ELTE) and the Los Alamos National Laboratory.

Based in part on data collected at the Subaru Telescope and retrieved from the HSC data archive system, which is operated by Subaru Telescope and Astronomy Data Center at National Astronomical Observatory of Japan.

Funding for SDSS-III has been provided by the Alfred P. Sloan Foundation, the Participating Institutions, the National Science Foundation, and the U.S. Department of Energy Office of Science. The SDSS-III website is \url{http://www.sdss3.org/.}

SDSS-III is managed by the Astrophysical Research Consortium for the Participating Institutions of the SDSS-III Collaboration including the University of Arizona, the Brazilian Participation Group, Brookhaven National Laboratory, Carnegie Mellon University, University of Florida, the French Participation Group, the German Participation Group, Harvard University, the Instituto de Astrofisica de Canarias, the Michigan State/Notre Dame/JINA Participation Group, Johns Hopkins University, Lawrence Berkeley National Laboratory, Max Planck Institute for Astrophysics, Max Planck Institute for Extraterrestrial Physics, New Mexico State University, New York University, Ohio State University, Pennsylvania State University, University of Portsmouth, Princeton University, the Spanish Participation Group, University of Tokyo, University of Utah, Vanderbilt University, University of Virginia, University of Washington, and Yale University.

\section*{Data Availability}

The SDSS DR8 redMaPPer cluster and cluster member catalogs are publicly available at \url{http://risa.stanford.edu/redmapper}. The HSC-Y3 shape catalog used in this work is not currently publicly available; please contact the HSC Weak Lensing Working Group for inquiries on obtaining access.



\bibliographystyle{mnras}
\bibliography{refs} 

\begin{thebibliography}{}
\makeatletter
\relax
\def\mn@urlcharsother{\let\do\@makeother \do\$\do\&\do\#\do\^\do\_\do\%\do\~}
\def\mn@doi{\begingroup\mn@urlcharsother \@ifnextchar [ {\mn@doi@}
  {\mn@doi@[]}}
\def\mn@doi@[#1]#2{\def\@tempa{#1}\ifx\@tempa\@empty \href
  {http://dx.doi.org/#2} {doi:#2}\else \href {http://dx.doi.org/#2} {#1}\fi
  \endgroup}
\def\mn@eprint#1#2{\mn@eprint@#1:#2::\@nil}
\def\mn@eprint@arXiv#1{\href {http://arxiv.org/abs/#1} {{\tt arXiv:#1}}}
\def\mn@eprint@dblp#1{\href {http://dblp.uni-trier.de/rec/bibtex/#1.xml}
  {dblp:#1}}
\def\mn@eprint@#1:#2:#3:#4\@nil{\def\@tempa {#1}\def\@tempb {#2}\def\@tempc
  {#3}\ifx \@tempc \@empty \let \@tempc \@tempb \let \@tempb \@tempa \fi \ifx
  \@tempb \@empty \def\@tempb {arXiv}\fi \@ifundefined
  {mn@eprint@\@tempb}{\@tempb:\@tempc}{\expandafter \expandafter \csname
  mn@eprint@\@tempb\endcsname \expandafter{\@tempc}}}

\bibitem[\protect\citeauthoryear{{Abbott} et~al.,}{{Abbott}
  et~al.}{2020}]{DES2020}
{Abbott} T.~M.~C.,  et~al., 2020, \mn@doi [\prd] {10.1103/PhysRevD.102.023509},
  \href {https://ui.adsabs.harvard.edu/abs/2020PhRvD.102b3509A} {102, 023509}

\bibitem[\protect\citeauthoryear{{Abbott} et~al.,}{{Abbott}
  et~al.}{2022}]{des_y3}
{Abbott} T.~M.~C.,  et~al., 2022, \mn@doi [\prd] {10.1103/PhysRevD.105.023520},
  \href {https://ui.adsabs.harvard.edu/abs/2022PhRvD.105b3520A} {105, 023520}

\bibitem[\protect\citeauthoryear{{Aihara} et~al.,}{{Aihara}
  et~al.}{2011}]{Aihara_etal2011}
{Aihara} H.,  et~al., 2011, \mn@doi [\apjs] {10.1088/0067-0049/193/2/29}, \href
  {https://ui.adsabs.harvard.edu/abs/2011ApJS..193...29A} {193, 29}

\bibitem[\protect\citeauthoryear{{Aihara} et~al.,}{{Aihara}
  et~al.}{2018}]{HSCOverview:17}
{Aihara} H.,  et~al., 2018, \mn@doi [\pasj] {10.1093/pasj/psx066}, \href
  {https://ui.adsabs.harvard.edu/abs/2018PASJ...70S...4A} {70, S4}

\bibitem[\protect\citeauthoryear{{Amendola} et~al.,}{{Amendola}
  et~al.}{2018}]{euclid2018}
{Amendola} L.,  et~al., 2018, \mn@doi [Living Reviews in Relativity]
  {10.1007/s41114-017-0010-3}, \href
  {https://ui.adsabs.harvard.edu/abs/2018LRR....21....2A} {21, 2}

\bibitem[\protect\citeauthoryear{{Bahcall}}{{Bahcall}}{1995}]{Bahcall1995}
{Bahcall} N.~A.,  1995, in {M{\"u}cket} J.~P.,  {Gottloeber} S.,   {M{\"u}ller}
  V.,  eds, Large Scale Structure in the Universe. p.~M (\mn@eprint {arXiv}
  {astro-ph/9503111}), \mn@doi{10.48550/arXiv.astro-ph/9503111}

\bibitem[\protect\citeauthoryear{{Bardeen}, {Bond}, {Kaiser}  \&
  {Szalay}}{{Bardeen} et~al.}{1986}]{bbks}
{Bardeen} J.~M.,  {Bond} J.~R.,  {Kaiser} N.,   {Szalay} A.~S.,  1986, \mn@doi
  [\apj] {10.1086/164143}, \href
  {https://ui.adsabs.harvard.edu/abs/1986ApJ...304...15B} {304, 15}

\bibitem[\protect\citeauthoryear{{Baxter}, {Rozo}, {Jain}, {Rykoff}  \&
  {Wechsler}}{{Baxter} et~al.}{2016}]{baxter_2016}
{Baxter} E.~J.,  {Rozo} E.,  {Jain} B.,  {Rykoff} E.,   {Wechsler} R.~H.,
  2016, \mn@doi [\mnras] {10.1093/mnras/stw1939}, \href
  {https://ui.adsabs.harvard.edu/abs/2016MNRAS.463..205B} {463, 205}

\bibitem[\protect\citeauthoryear{Behroozi, Wechsler  \& Wu}{Behroozi
  et~al.}{2012}]{rockstar}
Behroozi P.~S.,  Wechsler R.~H.,   Wu H.-Y.,  2012, \mn@doi [The Astrophysical
  Journal] {10.1088/0004-637x/762/2/109}, 762, 109

\bibitem[\protect\citeauthoryear{{Bocquet} et~al.,}{{Bocquet}
  et~al.}{2019}]{SPT2019}
{Bocquet} S.,  et~al., 2019, \mn@doi [\apj] {10.3847/1538-4357/ab1f10}, \href
  {https://ui.adsabs.harvard.edu/abs/2019ApJ...878...55B} {878, 55}

\bibitem[\protect\citeauthoryear{Bosch et~al.,}{Bosch
  et~al.}{2017}]{Bosch_2017}
Bosch J.,  et~al., 2017, \mn@doi [Publications of the Astronomical Society of
  Japan] {10.1093/pasj/psx080}, 70

\bibitem[\protect\citeauthoryear{{Chiu} et~al.,}{{Chiu}
  et~al.}{2022}]{Inon2022}
{Chiu} I.~N.,  et~al., 2022, \mn@doi [\aap] {10.1051/0004-6361/202141755},
  \href {https://ui.adsabs.harvard.edu/abs/2022A&A...661A..11C} {661, A11}

\bibitem[\protect\citeauthoryear{{Chiu}, {Klein}, {Mohr}  \& {Bocquet}}{{Chiu}
  et~al.}{2023}]{Inon2023}
{Chiu} I.~N.,  {Klein} M.,  {Mohr} J.,   {Bocquet} S.,  2023, \mn@doi [\mnras]
  {10.1093/mnras/stad957}, \href
  {https://ui.adsabs.harvard.edu/abs/2023MNRAS.522.1601C} {522, 1601}

\bibitem[\protect\citeauthoryear{{Costanzi} et~al.,}{{Costanzi}
  et~al.}{2019}]{costanzisdss}
{Costanzi} M.,  et~al., 2019, \mn@doi [\mnras] {10.1093/mnras/stz1949}, \href
  {https://ui.adsabs.harvard.edu/abs/2019MNRAS.488.4779C} {488, 4779}

\bibitem[\protect\citeauthoryear{{Costanzi} et~al.,}{{Costanzi}
  et~al.}{2021}]{costanzi_spt_2021}
{Costanzi} M.,  et~al., 2021, \mn@doi [\prd] {10.1103/PhysRevD.103.043522},
  \href {https://ui.adsabs.harvard.edu/abs/2021PhRvD.103d3522C} {103, 043522}

\bibitem[\protect\citeauthoryear{{DES Collaboration} et~al.,}{{DES
  Collaboration} et~al.}{2020}]{desy1cl}
{DES Collaboration} et~al., 2020, \mn@doi [\prd] {10.1103/PhysRevD.102.023509},
  \href {https://ui.adsabs.harvard.edu/abs/2020PhRvD.102b3509A} {102, 023509}

\bibitem[\protect\citeauthoryear{{Dalal} et~al.,}{{Dalal}
  et~al.}{2023}]{hscy3_shear_fourier}
{Dalal} R.,  et~al., 2023, \mn@doi [arXiv e-prints]
  {10.48550/arXiv.2304.00701}, \href
  {https://ui.adsabs.harvard.edu/abs/2023arXiv230400701D} {p. arXiv:2304.00701}

\bibitem[\protect\citeauthoryear{{Dore} et~al.,}{{Dore}
  et~al.}{2019}]{WFIRST2019}
{Dore} O.,  et~al., 2019, \baas, \href
  {https://ui.adsabs.harvard.edu/abs/2019BAAS...51c.341D} {51, 341}

\bibitem[\protect\citeauthoryear{{Farahi}, {Evrard}, {Rozo}, {Rykoff}  \&
  {Wechsler}}{{Farahi} et~al.}{2016}]{farahi_2016}
{Farahi} A.,  {Evrard} A.~E.,  {Rozo} E.,  {Rykoff} E.~S.,   {Wechsler} R.~H.,
  2016, \mn@doi [\mnras] {10.1093/mnras/stw1143}, \href
  {https://ui.adsabs.harvard.edu/abs/2016MNRAS.460.3900F} {460, 3900}

\bibitem[\protect\citeauthoryear{{Feroz} \& {Hobson}}{{Feroz} \&
  {Hobson}}{2008}]{multinest1}
{Feroz} F.,  {Hobson} M.~P.,  2008, \mn@doi [\mnras]
  {10.1111/j.1365-2966.2007.12353.x}, \href
  {https://ui.adsabs.harvard.edu/abs/2008MNRAS.384..449F} {384, 449}

\bibitem[\protect\citeauthoryear{{Feroz}, {Hobson}  \& {Bridges}}{{Feroz}
  et~al.}{2009}]{multinest2}
{Feroz} F.,  {Hobson} M.~P.,   {Bridges} M.,  2009, \mn@doi [\mnras]
  {10.1111/j.1365-2966.2009.14548.x}, \href
  {https://ui.adsabs.harvard.edu/abs/2009MNRAS.398.1601F} {398, 1601}

\bibitem[\protect\citeauthoryear{{Feroz}, {Hobson}, {Cameron}  \&
  {Pettitt}}{{Feroz} et~al.}{2019}]{multinest3}
{Feroz} F.,  {Hobson} M.~P.,  {Cameron} E.,   {Pettitt} A.~N.,  2019, \mn@doi
  [The Open Journal of Astrophysics] {10.21105/astro.1306.2144}, \href
  {https://ui.adsabs.harvard.edu/abs/2019OJAp....2E..10F} {2, 10}

\bibitem[\protect\citeauthoryear{{Giri} \& {Schneider}}{{Giri} \&
  {Schneider}}{2021}]{baryonification3}
{Giri} S.~K.,  {Schneider} A.,  2021, \mn@doi [\jcap]
  {10.1088/1475-7516/2021/12/046}, \href
  {https://ui.adsabs.harvard.edu/abs/2021JCAP...12..046G} {2021, 046}

\bibitem[\protect\citeauthoryear{{Haiman}, {Mohr}  \& {Holder}}{{Haiman}
  et~al.}{2001}]{2001ApJ...553..545H}
{Haiman} Z.,  {Mohr} J.~J.,   {Holder} G.~P.,  2001, \mn@doi [\apj]
  {10.1086/320939}, \href
  {https://ui.adsabs.harvard.edu/abs/2001ApJ...553..545H} {553, 545}

\bibitem[\protect\citeauthoryear{{Hartlap}, {Simon}  \& {Schneider}}{{Hartlap}
  et~al.}{2007}]{hartlap}
{Hartlap} J.,  {Simon} P.,   {Schneider} P.,  2007, \mn@doi [\aap]
  {10.1051/0004-6361:20066170}, \href
  {https://ui.adsabs.harvard.edu/abs/2007A&A...464..399H} {464, 399}

\bibitem[\protect\citeauthoryear{{Hikage}, {Mandelbaum}, {Takada}  \&
  {Spergel}}{{Hikage} et~al.}{2013}]{2013MNRAS.435.2345H}
{Hikage} C.,  {Mandelbaum} R.,  {Takada} M.,   {Spergel} D.~N.,  2013, \mn@doi
  [\mnras] {10.1093/mnras/stt1446}, \href
  {https://ui.adsabs.harvard.edu/abs/2013MNRAS.435.2345H} {435, 2345}

\bibitem[\protect\citeauthoryear{{Hikage} et~al.,}{{Hikage}
  et~al.}{2019}]{hikage_etal2019}
{Hikage} C.,  et~al., 2019, \mn@doi [\pasj] {10.1093/pasj/psz010}, \href
  {https://ui.adsabs.harvard.edu/abs/2019PASJ...71...43H} {71, 43}

\bibitem[\protect\citeauthoryear{{Hinton}}{{Hinton}}{2016}]{chainconsumer}
{Hinton} S.~R.,  2016, \mn@doi [The Journal of Open Source Software]
  {10.21105/joss.00045}, \href
  {http://adsabs.harvard.edu/abs/2016JOSS....1...45H} {1, 00045}

\bibitem[\protect\citeauthoryear{{Hirata} \& {Seljak}}{{Hirata} \&
  {Seljak}}{2003}]{hirataseljak}
{Hirata} C.,  {Seljak} U.,  2003, \mn@doi [\mnras]
  {10.1046/j.1365-8711.2003.06683.x}, \href
  {https://ui.adsabs.harvard.edu/abs/2003MNRAS.343..459H} {343, 459}

\bibitem[\protect\citeauthoryear{{Hsieh} \& {Yee}}{{Hsieh} \&
  {Yee}}{2014}]{DEMP}
{Hsieh} B.~C.,  {Yee} H.~K.~C.,  2014, \mn@doi [\apj]
  {10.1088/0004-637X/792/2/102}, \href
  {https://ui.adsabs.harvard.edu/abs/2014ApJ...792..102H} {792, 102}

\bibitem[\protect\citeauthoryear{{Johnston} et~al.,}{{Johnston}
  et~al.}{2007}]{2007arXiv0709.1159J}
{Johnston} D.~E.,  et~al., 2007, arXiv e-prints, \href
  {https://ui.adsabs.harvard.edu/abs/2007arXiv0709.1159J} {p. arXiv:0709.1159}

\bibitem[\protect\citeauthoryear{{Kaiser}}{{Kaiser}}{1992}]{limberkaiser}
{Kaiser} N.,  1992, \mn@doi [\apj] {10.1086/171151}, \href
  {https://ui.adsabs.harvard.edu/abs/1992ApJ...388..272K} {388, 272}

\bibitem[\protect\citeauthoryear{{Kravtsov} \& {Borgani}}{{Kravtsov} \&
  {Borgani}}{2012}]{kravtsovborgani}
{Kravtsov} A.~V.,  {Borgani} S.,  2012, \mn@doi [\araa]
  {10.1146/annurev-astro-081811-125502}, \href
  {https://ui.adsabs.harvard.edu/abs/2012ARA&A..50..353K} {50, 353}

\bibitem[\protect\citeauthoryear{{Kuijken} et~al.,}{{Kuijken}
  et~al.}{2015}]{KiDs2015}
{Kuijken} K.,  et~al., 2015, \mn@doi [\mnras] {10.1093/mnras/stv2140}, \href
  {https://ui.adsabs.harvard.edu/abs/2015MNRAS.454.3500K} {454, 3500}

\bibitem[\protect\citeauthoryear{{LSST Science Collaboration} et~al.,}{{LSST
  Science Collaboration} et~al.}{2009}]{LSST2009}
{LSST Science Collaboration} et~al., 2009, arXiv e-prints, \href
  {https://ui.adsabs.harvard.edu/abs/2009arXiv0912.0201L} {p. arXiv:0912.0201}

\bibitem[\protect\citeauthoryear{{Landy} \& {Szalay}}{{Landy} \&
  {Szalay}}{1993}]{landyszalay}
{Landy} S.~D.,  {Szalay} A.~S.,  1993, \mn@doi [\apj] {10.1086/172900}, \href
  {https://ui.adsabs.harvard.edu/abs/1993ApJ...412...64L} {412, 64}

\bibitem[\protect\citeauthoryear{{Lemos} et~al.,}{{Lemos}
  et~al.}{2023}]{lemos_2023}
{Lemos} P.,  et~al., 2023, \mn@doi [\mnras] {10.1093/mnras/stac2786}, \href
  {https://ui.adsabs.harvard.edu/abs/2023MNRAS.521.1184L} {521, 1184}

\bibitem[\protect\citeauthoryear{{Lesci} et~al.,}{{Lesci}
  et~al.}{2022}]{KiDS2022}
{Lesci} G.~F.,  et~al., 2022, \mn@doi [\aap] {10.1051/0004-6361/202040194},
  \href {https://ui.adsabs.harvard.edu/abs/2022A&A...659A..88L} {659, A88}

\bibitem[\protect\citeauthoryear{Li et~al.,}{Li et~al.}{2022}]{Li_2022}
Li X.,  et~al., 2022, \mn@doi [Publications of the Astronomical Society of
  Japan] {10.1093/pasj/psac006}, 74, 421

\bibitem[\protect\citeauthoryear{{Li} et~al.,}{{Li}
  et~al.}{2023}]{hscy3_shear_real}
{Li} X.,  et~al., 2023, \mn@doi [arXiv e-prints] {10.48550/arXiv.2304.00702},
  \href {https://ui.adsabs.harvard.edu/abs/2023arXiv230400702L} {p.
  arXiv:2304.00702}

\bibitem[\protect\citeauthoryear{{Lima} \& {Hu}}{{Lima} \& {Hu}}{2005}]{limahu}
{Lima} M.,  {Hu} W.,  2005, \mn@doi [\prd] {10.1103/PhysRevD.72.043006}, \href
  {https://ui.adsabs.harvard.edu/abs/2005PhRvD..72d3006L} {72, 043006}

\bibitem[\protect\citeauthoryear{{Limber}}{{Limber}}{1954}]{limber}
{Limber} D.~N.,  1954, \mn@doi [\apj] {10.1086/145870}, \href
  {https://ui.adsabs.harvard.edu/abs/1954ApJ...119..655L} {119, 655}

\bibitem[\protect\citeauthoryear{Mandelbaum et~al.,}{Mandelbaum
  et~al.}{2018}]{Mandelbaum_2018}
Mandelbaum R.,  et~al., 2018, \mn@doi [Monthly Notices of the Royal
  Astronomical Society] {10.1093/mnras/sty2420}, 481, 3170

\bibitem[\protect\citeauthoryear{{Mantz} et~al.,}{{Mantz} et~al.}{2015}]{wtg4}
{Mantz} A.~B.,  et~al., 2015, \mn@doi [\mnras] {10.1093/mnras/stu2096}, \href
  {https://ui.adsabs.harvard.edu/abs/2015MNRAS.446.2205M} {446, 2205}

\bibitem[\protect\citeauthoryear{{McClintock} et~al.,}{{McClintock}
  et~al.}{2019}]{mcclintock18}
{McClintock} T.,  et~al., 2019, \mn@doi [\mnras] {10.1093/mnras/sty2711}, \href
  {https://ui.adsabs.harvard.edu/abs/2019MNRAS.482.1352M} {482, 1352}

\bibitem[\protect\citeauthoryear{{Miller} et~al.,}{{Miller}
  et~al.}{2013}]{miller13}
{Miller} L.,  et~al., 2013, \mn@doi [\mnras] {10.1093/mnras/sts454}, \href
  {https://ui.adsabs.harvard.edu/abs/2013MNRAS.429.2858M} {429, 2858}

\bibitem[\protect\citeauthoryear{{Miyatake}, {More}, {Takada}, {Spergel},
  {Mandelbaum}, {Rykoff}  \& {Rozo}}{{Miyatake} et~al.}{2016}]{miyatakecl}
{Miyatake} H.,  {More} S.,  {Takada} M.,  {Spergel} D.~N.,  {Mandelbaum} R.,
  {Rykoff} E.~S.,   {Rozo} E.,  2016, \mn@doi [\prl]
  {10.1103/PhysRevLett.116.041301}, \href
  {https://ui.adsabs.harvard.edu/abs/2016PhRvL.116d1301M} {116, 041301}

\bibitem[\protect\citeauthoryear{{Miyatake} et~al.,}{{Miyatake}
  et~al.}{2021}]{2021arXiv211102419M}
{Miyatake} H.,  et~al., 2021, arXiv e-prints, \href
  {https://ui.adsabs.harvard.edu/abs/2021arXiv211102419M} {p. arXiv:2111.02419}

\bibitem[\protect\citeauthoryear{{Miyatake} et~al.,}{{Miyatake}
  et~al.}{2023}]{hscy3_3x2_small}
{Miyatake} H.,  et~al., 2023, \mn@doi [arXiv e-prints]
  {10.48550/arXiv.2304.00704}, \href
  {https://ui.adsabs.harvard.edu/abs/2023arXiv230400704M} {p. arXiv:2304.00704}

\bibitem[\protect\citeauthoryear{{More}}{{More}}{2013}]{2013ApJ...777L..26M}
{More} S.,  2013, \mn@doi [\apjl] {10.1088/2041-8205/777/2/L26}, \href
  {https://ui.adsabs.harvard.edu/abs/2013ApJ...777L..26M} {777, L26}

\bibitem[\protect\citeauthoryear{{More} et~al.,}{{More}
  et~al.}{2023}]{More_etal2023}
{More} S.,  et~al., 2023, \mn@doi [arXiv e-prints] {10.48550/arXiv.2304.00703},
  \href {https://ui.adsabs.harvard.edu/abs/2023arXiv230400703M} {p.
  arXiv:2304.00703}

\bibitem[\protect\citeauthoryear{{Murata}, {Nishimichi}, {Takada}, {Miyatake},
  {Shirasaki}, {More}, {Takahashi}  \& {Osato}}{{Murata}
  et~al.}{2018}]{muratasdss}
{Murata} R.,  {Nishimichi} T.,  {Takada} M.,  {Miyatake} H.,  {Shirasaki} M.,
  {More} S.,  {Takahashi} R.,   {Osato} K.,  2018, \mn@doi [\apj]
  {10.3847/1538-4357/aaaab8}, \href
  {https://ui.adsabs.harvard.edu/\#abs/2018ApJ...854..120M} {854, 120}

\bibitem[\protect\citeauthoryear{{Murata} et~al.,}{{Murata}
  et~al.}{2019}]{2019PASJ...71..107M}
{Murata} R.,  et~al., 2019, \mn@doi [\pasj] {10.1093/pasj/psz092}, \href
  {https://ui.adsabs.harvard.edu/abs/2019PASJ...71..107M} {71, 107}

\bibitem[\protect\citeauthoryear{{Nakajima}, {Mandelbaum}, {Seljak}, {Cohn},
  {Reyes}  \& {Cool}}{{Nakajima} et~al.}{2012}]{nakajima12}
{Nakajima} R.,  {Mandelbaum} R.,  {Seljak} U.,  {Cohn} J.~D.,  {Reyes} R.,
  {Cool} R.,  2012, \mn@doi [\mnras] {10.1111/j.1365-2966.2011.20249.x}, \href
  {https://ui.adsabs.harvard.edu/abs/2012MNRAS.420.3240N} {420, 3240}

\bibitem[\protect\citeauthoryear{{Navarro}, {Frenk}  \& {White}}{{Navarro}
  et~al.}{1997}]{nfw}
{Navarro} J.~F.,  {Frenk} C.~S.,   {White} S. D.~M.,  1997, \mn@doi [\apj]
  {10.1086/304888}, \href
  {https://ui.adsabs.harvard.edu/abs/1997ApJ...490..493N} {490, 493}

\bibitem[\protect\citeauthoryear{{Nishimichi} et~al.,}{{Nishimichi}
  et~al.}{2019}]{darkemu}
{Nishimichi} T.,  et~al., 2019, \mn@doi [\apj] {10.3847/1538-4357/ab3719},
  \href {https://ui.adsabs.harvard.edu/abs/2019ApJ...884...29N} {884, 29}

\bibitem[\protect\citeauthoryear{Nishizawa, Hsieh, Tanaka  \& Takata}{Nishizawa
  et~al.}{2020}]{nishizawa2020photometric}
Nishizawa A.~J.,  Hsieh B.-C.,  Tanaka M.,   Takata T.,  2020, Photometric
  Redshifts for the Hyper Suprime-Cam Subaru Strategic Program Data Release 2
  (\mn@eprint {arXiv} {2003.01511})

\bibitem[\protect\citeauthoryear{{Oguri}}{{Oguri}}{2014}]{camira}
{Oguri} M.,  2014, \mn@doi [\mnras] {10.1093/mnras/stu1446}, \href
  {https://ui.adsabs.harvard.edu/abs/2014MNRAS.444..147O} {444, 147}

\bibitem[\protect\citeauthoryear{{Oguri} \& {Takada}}{{Oguri} \&
  {Takada}}{2011}]{oguritakada11}
{Oguri} M.,  {Takada} M.,  2011, \mn@doi [\prd] {10.1103/PhysRevD.83.023008},
  \href {https://ui.adsabs.harvard.edu/abs/2011PhRvD..83b3008O} {83, 023008}

\bibitem[\protect\citeauthoryear{{Oguri} et~al.,}{{Oguri}
  et~al.}{2018}]{oguri2018}
{Oguri} M.,  et~al., 2018, \mn@doi [\pasj] {10.1093/pasj/psx042}, \href
  {https://ui.adsabs.harvard.edu/abs/2018PASJ...70S..20O} {70, S20}

\bibitem[\protect\citeauthoryear{Park, Sunayama, Takada, Kobayashi, Miyatake,
  More, Nishimichi  \& Sugiyama}{Park et~al.}{2022}]{Park_2022}
Park Y.,  Sunayama T.,  Takada M.,  Kobayashi Y.,  Miyatake H.,  More S.,
  Nishimichi T.,   Sugiyama S.,  2022, \mn@doi [Monthly Notices of the Royal
  Astronomical Society] {10.1093/mnras/stac3410}, 518, 5171

\bibitem[\protect\citeauthoryear{{Planck Collaboration} et~al.,}{{Planck
  Collaboration} et~al.}{2016}]{PlanckSZ:16}
{Planck Collaboration} et~al., 2016, \mn@doi [\aap]
  {10.1051/0004-6361/201525833}, \href
  {http://adsabs.harvard.edu/abs/2016A%26A...594A..24P} {594, A24}

\bibitem[\protect\citeauthoryear{{Planck Collaboration} et~al.,}{{Planck
  Collaboration} et~al.}{2020}]{Planck18}
{Planck Collaboration} et~al., 2020, \mn@doi [\aap]
  {10.1051/0004-6361/201833910}, \href
  {https://ui.adsabs.harvard.edu/abs/2020A&A...641A...6P} {641, A6}

\bibitem[\protect\citeauthoryear{Rau et~al.,}{Rau et~al.}{2022}]{rau2022weak}
Rau M.~M.,  et~al., 2022, Weak Lensing Tomographic Redshift Distribution
  Inference for the Hyper Suprime-Cam Subaru Strategic Program three-year shape
  catalogue (\mn@eprint {arXiv} {2211.16516})

\bibitem[\protect\citeauthoryear{{Rozo} \& {Rykoff}}{{Rozo} \&
  {Rykoff}}{2014}]{rm2}
{Rozo} E.,  {Rykoff} E.~S.,  2014, \mn@doi [\apj] {10.1088/0004-637X/783/2/80},
  \href {https://ui.adsabs.harvard.edu/abs/2014ApJ...783...80R} {783, 80}

\bibitem[\protect\citeauthoryear{{Rozo} et~al.,}{{Rozo} et~al.}{2010}]{rozo10}
{Rozo} E.,  et~al., 2010, \mn@doi [\apj] {10.1088/0004-637X/708/1/645}, \href
  {https://ui.adsabs.harvard.edu/abs/2010ApJ...708..645R} {708, 645}

\bibitem[\protect\citeauthoryear{{Rozo}, {Rykoff}, {Bartlett}  \&
  {Melin}}{{Rozo} et~al.}{2015a}]{rm3}
{Rozo} E.,  {Rykoff} E.~S.,  {Bartlett} J.~G.,   {Melin} J.-B.,  2015a, \mn@doi
  [\mnras] {10.1093/mnras/stv605}, \href
  {https://ui.adsabs.harvard.edu/abs/2015MNRAS.450..592R} {450, 592}

\bibitem[\protect\citeauthoryear{{Rozo}, {Rykoff}, {Becker}, {Reddick}  \&
  {Wechsler}}{{Rozo} et~al.}{2015b}]{rm4}
{Rozo} E.,  {Rykoff} E.~S.,  {Becker} M.,  {Reddick} R.~M.,   {Wechsler} R.~H.,
   2015b, \mn@doi [\mnras] {10.1093/mnras/stv1560}, \href
  {https://ui.adsabs.harvard.edu/abs/2015MNRAS.453...38R} {453, 38}

\bibitem[\protect\citeauthoryear{{Rykoff} et~al.,}{{Rykoff} et~al.}{2014}]{rm1}
{Rykoff} E.~S.,  et~al., 2014, \mn@doi [\apj] {10.1088/0004-637X/785/2/104},
  \href {https://ui.adsabs.harvard.edu/abs/2014ApJ...785..104R} {785, 104}

\bibitem[\protect\citeauthoryear{{Schneider} \& {Teyssier}}{{Schneider} \&
  {Teyssier}}{2015}]{baryonification1}
{Schneider} A.,  {Teyssier} R.,  2015, \mn@doi [\jcap]
  {10.1088/1475-7516/2015/12/049}, \href
  {https://ui.adsabs.harvard.edu/abs/2015JCAP...12..049S} {2015, 049}

\bibitem[\protect\citeauthoryear{{Schneider}, {Teyssier}, {Stadel}, {Chisari},
  {Le Brun}, {Amara}  \& {Refregier}}{{Schneider}
  et~al.}{2019}]{baryonification2}
{Schneider} A.,  {Teyssier} R.,  {Stadel} J.,  {Chisari} N.~E.,  {Le Brun} A.
  M.~C.,  {Amara} A.,   {Refregier} A.,  2019, \mn@doi [\jcap]
  {10.1088/1475-7516/2019/03/020}, \href
  {https://ui.adsabs.harvard.edu/abs/2019JCAP...03..020S} {2019, 020}

\bibitem[\protect\citeauthoryear{Shirasaki \& Yoshida}{Shirasaki \&
  Yoshida}{2014}]{Shirasaki_2014}
Shirasaki M.,  Yoshida N.,  2014, \mn@doi [The Astrophysical Journal]
  {10.1088/0004-637x/786/1/43}, 786, 43

\bibitem[\protect\citeauthoryear{Shirasaki, Takada, Miyatake, Takahashi,
  Hamana, Nishimichi  \& Murata}{Shirasaki et~al.}{2017}]{Shirasaki_2017}
Shirasaki M.,  Takada M.,  Miyatake H.,  Takahashi R.,  Hamana T.,  Nishimichi
  T.,   Murata R.,  2017, \mn@doi [Monthly Notices of the Royal Astronomical
  Society] {10.1093/mnras/stx1477}, 470, 3476

\bibitem[\protect\citeauthoryear{Shirasaki, Hamana, Takada, Takahashi  \&
  Miyatake}{Shirasaki et~al.}{2019}]{Shirasaki_2019}
Shirasaki M.,  Hamana T.,  Takada M.,  Takahashi R.,   Miyatake H.,  2019,
  \mn@doi [Monthly Notices of the Royal Astronomical Society]
  {10.1093/mnras/stz791}, 486, 52

\bibitem[\protect\citeauthoryear{{Shirasaki}, {Miyatake}, {Hilton}
  et~al.}{{Shirasaki} et~al.}{prep}]{Shirasaki_etal_prep}
{Shirasaki} M.,  {Miyatake} H.,  {Hilton} M.,   et~al., in prep., Masses of
  Sunyaev-Zel’dovich Galaxy Clusters Detected by The Atacama Cosmology
  Telescope: Stacked Lensing Measurements with Subaru HSC Year 3 data

\bibitem[\protect\citeauthoryear{{Simet}, {McClintock}, {Mandelbaum}, {Rozo},
  {Rykoff}, {Sheldon}  \& {Wechsler}}{{Simet} et~al.}{2017}]{simet17}
{Simet} M.,  {McClintock} T.,  {Mandelbaum} R.,  {Rozo} E.,  {Rykoff} E.,
  {Sheldon} E.,   {Wechsler} R.~H.,  2017, \mn@doi [\mnras]
  {10.1093/mnras/stw3250}, \href
  {https://ui.adsabs.harvard.edu/abs/2017MNRAS.466.3103S} {466, 3103}

\bibitem[\protect\citeauthoryear{{Sugiyama} et~al.,}{{Sugiyama}
  et~al.}{2023}]{hscy3_3x2_minimal}
{Sugiyama} S.,  et~al., 2023, \mn@doi [arXiv e-prints]
  {10.48550/arXiv.2304.00705}, \href
  {https://ui.adsabs.harvard.edu/abs/2023arXiv230400705S} {p. arXiv:2304.00705}

\bibitem[\protect\citeauthoryear{{Sunayama}}{{Sunayama}}{2023}]{sunayama22}
{Sunayama} T.,  2023, \mn@doi [\mnras] {10.1093/mnras/stad786}, \href
  {https://ui.adsabs.harvard.edu/abs/2023MNRAS.521.5064S} {521, 5064}

\bibitem[\protect\citeauthoryear{{Sunayama} \& {More}}{{Sunayama} \&
  {More}}{2019}]{sunayamamore}
{Sunayama} T.,  {More} S.,  2019, \mn@doi [\mnras] {10.1093/mnras/stz2832},
  \href {https://ui.adsabs.harvard.edu/abs/2019MNRAS.490.4945S} {490, 4945}

\bibitem[\protect\citeauthoryear{{Sunayama} et~al.,}{{Sunayama}
  et~al.}{2020}]{projeff}
{Sunayama} T.,  et~al., 2020, \mn@doi [\mnras] {10.1093/mnras/staa1646}, \href
  {https://ui.adsabs.harvard.edu/abs/2020MNRAS.496.4468S} {496, 4468}

\bibitem[\protect\citeauthoryear{{Takada} \& {Bridle}}{{Takada} \&
  {Bridle}}{2007}]{takadabridle}
{Takada} M.,  {Bridle} S.,  2007, \mn@doi [New Journal of Physics]
  {10.1088/1367-2630/9/12/446}, \href
  {https://ui.adsabs.harvard.edu/abs/2007NJPh....9..446T} {9, 446}

\bibitem[\protect\citeauthoryear{Takahashi, Hamana, Shirasaki, Namikawa,
  Nishimichi, Osato  \& Shiroyama}{Takahashi et~al.}{2017}]{Takahashi_2017}
Takahashi R.,  Hamana T.,  Shirasaki M.,  Namikawa T.,  Nishimichi T.,  Osato
  K.,   Shiroyama K.,  2017, \mn@doi [The Astrophysical Journal]
  {10.3847/1538-4357/aa943d}, 850, 24

\bibitem[\protect\citeauthoryear{Tanaka et~al.,}{Tanaka
  et~al.}{2017}]{Tanaka_2017}
Tanaka M.,  et~al., 2017, \mn@doi [Publications of the Astronomical Society of
  Japan] {10.1093/pasj/psx077}, 70

\bibitem[\protect\citeauthoryear{{The Dark Energy Survey Collaboration}}{{The
  Dark Energy Survey Collaboration}}{2005}]{DES2005}
{The Dark Energy Survey Collaboration} 2005, arXiv e-prints, \href
  {https://ui.adsabs.harvard.edu/abs/2005astro.ph.10346T} {pp
  astro--ph/0510346}

\bibitem[\protect\citeauthoryear{{To} et~al.,}{{To} et~al.}{2021}]{tokrause2}
{To} C.,  et~al., 2021, \mn@doi [\prl] {10.1103/PhysRevLett.126.141301}, \href
  {https://ui.adsabs.harvard.edu/abs/2021PhRvL.126n1301T} {126, 141301}

\bibitem[\protect\citeauthoryear{{Weinberg}, {Mortonson}, {Eisenstein},
  {Hirata}, {Riess}  \& {Rozo}}{{Weinberg} et~al.}{2013}]{weinberg13}
{Weinberg} D.~H.,  {Mortonson} M.~J.,  {Eisenstein} D.~J.,  {Hirata} C.,
  {Riess} A.~G.,   {Rozo} E.,  2013, \mn@doi [\physrep]
  {10.1016/j.physrep.2013.05.001}, \href
  {https://ui.adsabs.harvard.edu/abs/2013PhR...530...87W} {530, 87}

\bibitem[\protect\citeauthoryear{{Zhang} et~al.,}{{Zhang}
  et~al.}{2019}]{zhang18}
{Zhang} Y.,  et~al., 2019, \mn@doi [\mnras] {10.1093/mnras/stz1361}, \href
  {https://ui.adsabs.harvard.edu/abs/2019MNRAS.487.2578Z} {487, 2578}

\bibitem[\protect\citeauthoryear{{Zuntz} et~al.,}{{Zuntz}
  et~al.}{2015}]{cosmosis}
{Zuntz} J.,  et~al., 2015, \mn@doi [Astronomy and Computing]
  {10.1016/j.ascom.2015.05.005}, \href
  {https://ui.adsabs.harvard.edu/abs/2015A&C....12...45Z} {12, 45}

\bibitem[\protect\citeauthoryear{van~den Bosch, More, Cacciato, Mo  \&
  Yang}{van~den Bosch et~al.}{2013}]{vdBosch2013}
van~den Bosch F.~C.,  More S.,  Cacciato M.,  Mo H.,   Yang X.,  2013, \mn@doi
  [Monthly Notices of the Royal Astronomical Society] {10.1093/mnras/sts006},
  430, 725–746

\bibitem[\protect\citeauthoryear{{von der Linden} et~al.,}{{von der Linden}
  et~al.}{2014}]{wtg1}
{von der Linden} A.,  et~al., 2014, \mn@doi [\mnras] {10.1093/mnras/stt1945},
  \href {https://ui.adsabs.harvard.edu/abs/2014MNRAS.439....2V} {439, 2}

\makeatother
\end{thebibliography}




\appendix

\section{Mis-centering Parameters}
\label{app:mis}
\begin{figure*}
	\includegraphics[width=\textwidth]{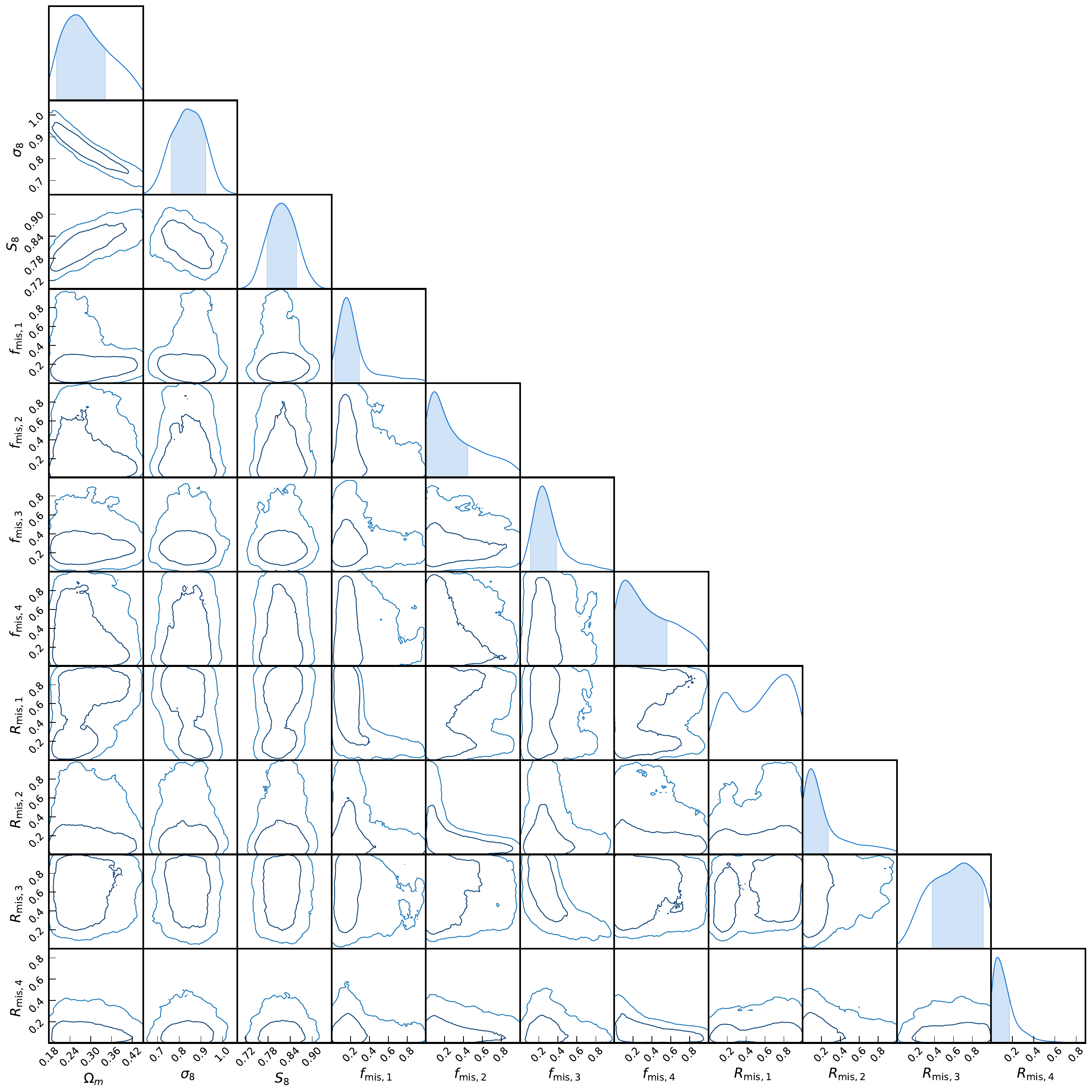}
    \caption{The 68\% and 95\% confidence regions for the full set of mis-centering parameters from our fiducial analysis along with the three derived cosmological parameters of interest.
    }
    \label{fig:miscentering}
\end{figure*}
We show the full constraints from our fiducial analysis.
Fig.~\ref{fig:miscentering} shows the constrained mis-centering parameters along with the cosmological parameters of interest.
While all the miscetering parameters were well constrained in \citetalias{Park_2022} and agree with the values from the X-ray studies on the same cluster sample \citep{zhang18} even with flat priors, our constraints on mis-centering parameters are not consistent. For the cases of the clusters with $\lambda \in [20,30)$ and $[40,55)$, the mis-centering fractions are well constrained and agree well with the X-ray studies. 
However, the mis-centering scales are overestimated.
For other cluster samples with $\lambda \in [30,40)$ and $[55,200)$, the mis-centering fraction parameters are not well-constrained, but the mis-centering locations are consistent with the previous studies.
Note that we use the minimum scale cut of $R_{\rm min}=0.5h^{-1}{\rm Mpc}$ for the cluster lensing signals to avoid the biased result due to baryonic effects, while the previous study by \citetalias{Park_2022} used $R_{\rm min}=0.2h^{-1}{\rm Mpc}$. Therefore, we cannot do an apple-to-apple comparison.

\section{Estimating residual photo-$z$ uncertainty}
\label{app:photo-z}
To measure the residual photo-$z$ uncertainty of our source galaxies, we first match our sample to the source galaxies used in \cite{hscy3_shear_real} \tianqing{and \cite{hscy3_shear_fourier}}, which are classified into four tomographic redshift bins with a stacked probability distribution $P_i(z)$.
The residual photo-$z$ uncertainty of each tomographic bin $i$ is constrained and given as $\Delta z_{\rm ph}^{i}$.

\begin{figure}
	\includegraphics[width=0.9\columnwidth]{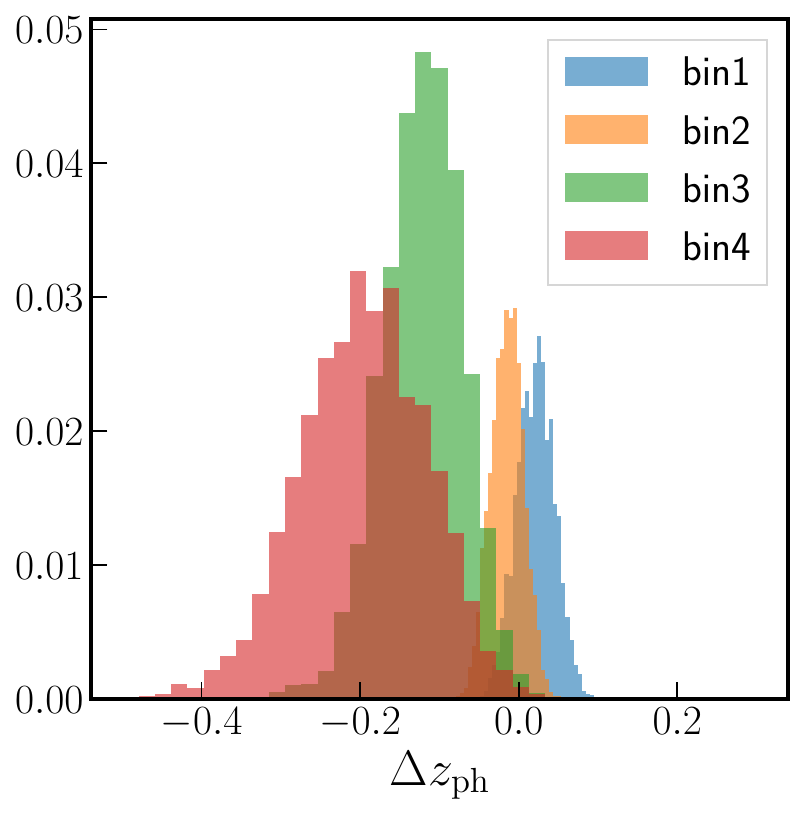}
    \caption{Stacked probability density distribution of $\Delta z_{\rm ph}$ of the source galaxies classified into four tomographic redshift bins used in \citep{hscy3_shear_real, hscy3_shear_fourier}. The stacked probability distributions of two high-z tomographic bins (bin3 and bin4) show larger residual photo-$z$ bias values.}
    \label{fig:photoz_v1}
\end{figure}

Fig.~\ref{fig:photoz_v1} shows stacked probability density distributions of our source galaxies classified into four tomographic redshift bins. As is reported by \cite{hscy3_shear_real}, the stacked probability density distributions of two high-z tomographic bins show larger photo-$z$ biases, while the two low-z tomographic bins have smaller photo-$z$ biases because they are calibrated by \cite{rau2022weak}.

\begin{figure}
	\includegraphics[width=0.9\columnwidth]{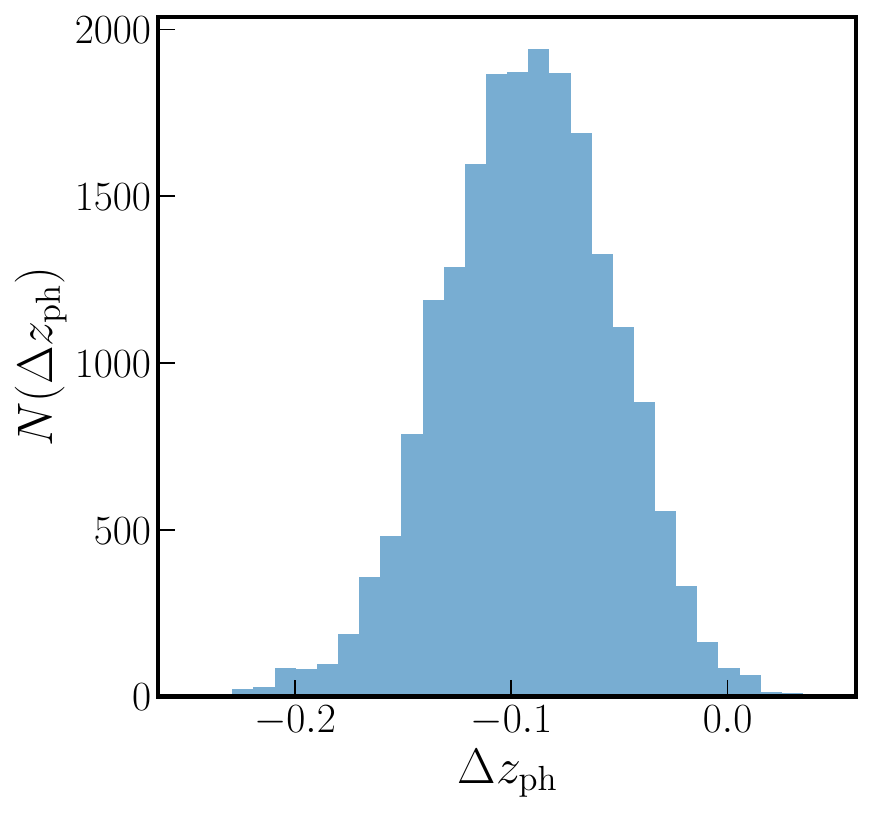}
    \caption{A histogram of $\Delta z_{\rm ph}$ of our source galaxies.
    }
    \label{fig:photoz_v2}
\end{figure}

We compute the photo-$z$ probability distribution of our source galaxies as
\begin{equation}
    p^{\rm corr}(z) = \sum_{i=1,2,3,4}f_i p_i(z),
\end{equation}
where $p_i(z)$ is given by \cite{hscy3_shear_fourier}.
Some source galaxies are not used in \cite{hscy3_shear_real} due to different selection criteria used. 
However, the fraction of such source galaxies is a few percent and we can safely omit these galaxies to compute the residual photo-$z$ bias.
Fig.~\ref{fig:photoz_v2} shows the histogram of photo-$z$ bias of our sample. 
The mean photo-$z$ bias $\Delta z_{\rm ph}$ is -0.093 and we use this value for our fiducial analysis.

\begin{figure}
	\includegraphics[width=0.9\columnwidth]{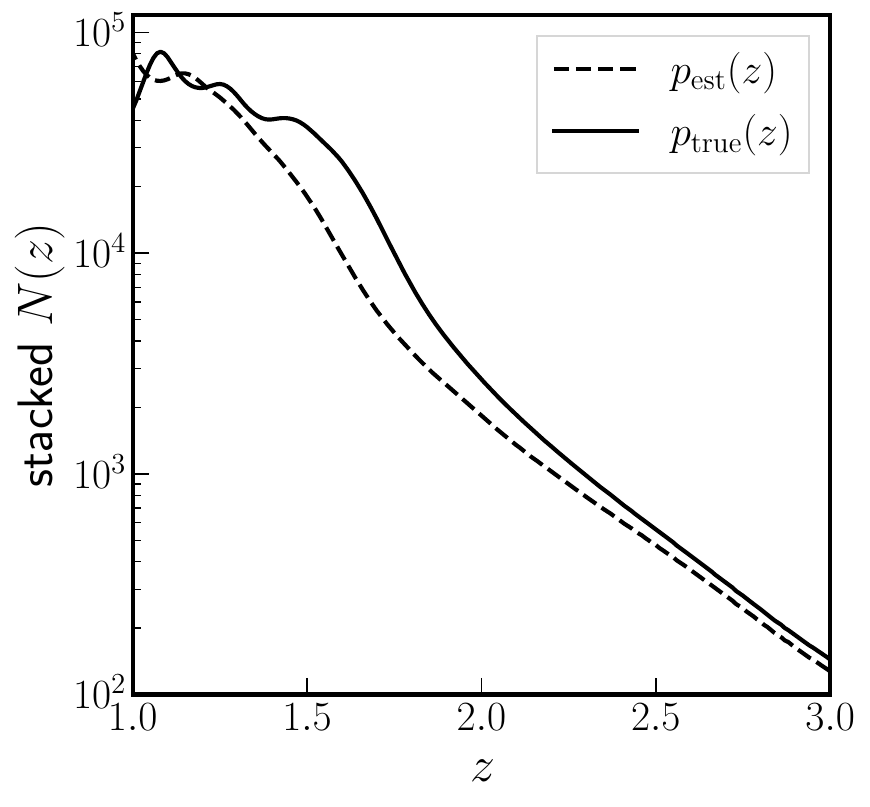}
    \caption{A stacked N(z) of our source galaxies estimated from photo-$z$ (dashed) and corrected using $\Delta z_{\rm ph}$ (solid).
    }
    \label{fig:photoz_v3}
\end{figure}

We can correct the photo-$z$ distribution using $\Delta z_{\rm ph}^{i}$ from the estimated distribution such as
\begin{equation}
p^{\rm true}(z|{\Delta z_{\rm ph}^i})=\sum_i f_i p^{\rm est}_i(z+\Delta z_{\rm ph}^i),
\end{equation}
where $f_i$ is the fraction of the source galaxies in the redshift bin $z_i$.
The corrected photo-$z$ distribution is shown in Fig.~\ref{fig:photoz_v3}. 
\hmcm{$\Delta z_{\rm ph}$ is defined such that $p^{\rm true}(z|{\Delta z_{\rm ph}})=p^{\rm est}_i(z+\Delta z_{\rm ph})$ (see eq. 25 in https://arxiv.org/pdf/2304.00704.pdf), so the negative $\Delta z_{\rm ph}$ means the photo-$z$ distribution is shifted towards high redshift. It seems Fig. B3 is opposite}

\tianqing{This test validates the assessment of the photometric redshift estimation systematics in the HSC-Y3 cosmic shear analysis \citep{hscy3_shear_real,hscy3_shear_fourier}. Using the galaxy-galaxy lensing, we obtain consistent mean redshift bias $\Delta z_{\rm ph}$ for tomographic bins 3 and 4, compared to the values from the cosmic shear analysis. This result motivates further investigation on the source sample redshift estimation and redshift distribution inference, especially for the high redshift bins of HSC.}

\section{Comparison of Cluster Lensing measured from SDSS and HSC-Y3}
\label{app:lensing}
\begin{figure}
	\includegraphics[width=0.9\columnwidth]{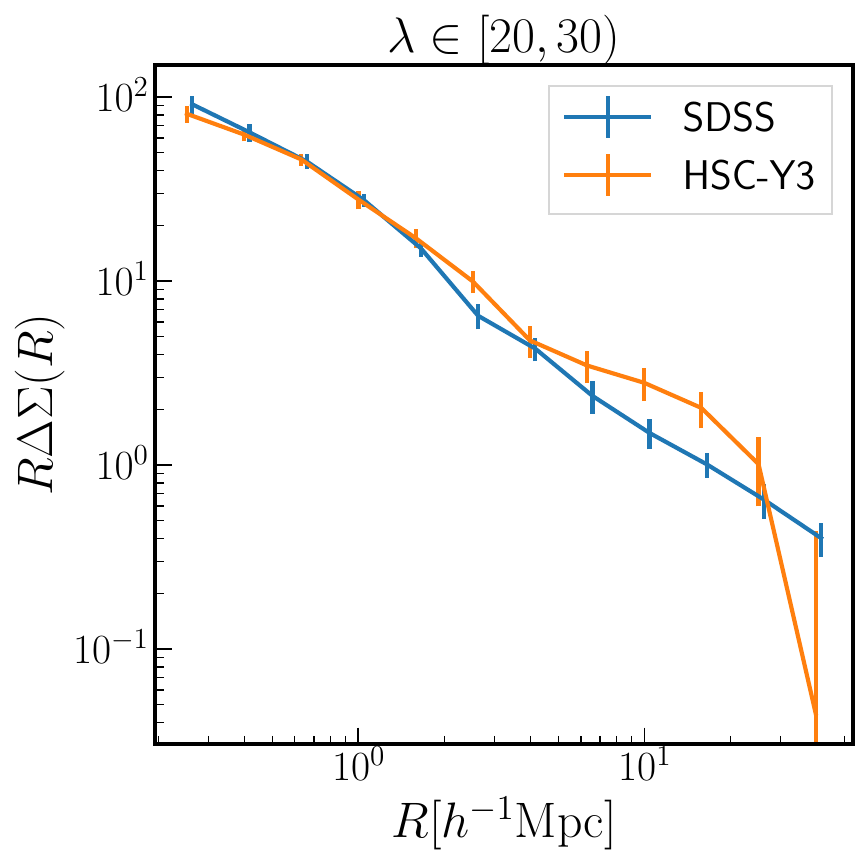}
    \caption{The cluster lensing signals of the SDSS redMaPPer clusters ($\lambda \in [20,30)$) using the SDSS (blue) and HSC-Y3 (orange) source galaxies. We applied the corrections to both measured lensing signals. On small scales, the lensing signals are consistent. However the lensing amplitude measured from the SDSS source galaxies is suppressed on large scales compared to the one measured from the HSC-Y3 source galaxies.
    }
    \label{fig:lens_sdss}
\end{figure}

\begin{figure}
	\includegraphics[width=0.9\columnwidth]{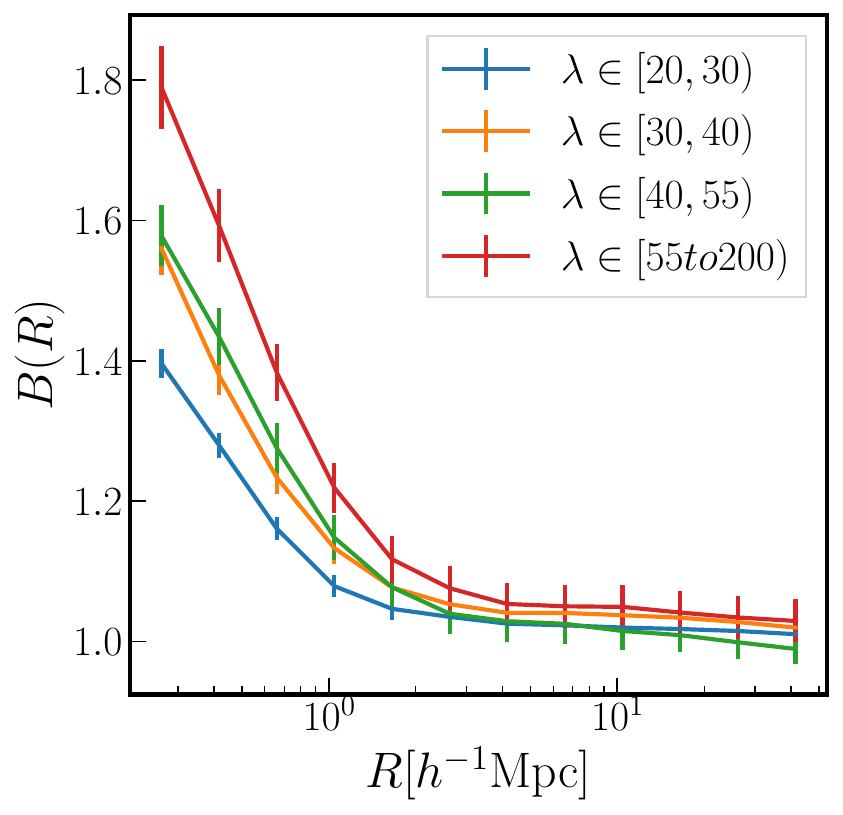}
    \caption{The ``boost factor'' $B(R)$, which measures and excess number of the SDSS redMaPPer clusters (lens) and the SDSS source galaxies pairs using random points and the source pairs. Note that all the source galaxies behind the lens clusters are selected. Compared to Fig.~\ref{fig:boost_cross}, the amplitude of the boost factor becomes larger than one as $R$ goes to a smaller scale. This implies the physical correlation between the lens-source pairs on small scales due to member dilutions.
    }
    \label{fig:boost_sdss}
\end{figure}

We compare the measured cluster lensing signal from the SDSS data used in \citetalias{Park_2022} and from the HSC\hmrv{-}Y3 data used in this study, as shown in Fig.~\ref{fig:lens_sdss}.
Note that the lensing signals in Fig.~\ref{fig:lens_sdss} are both after the boost and random signal corrections. For the lensing signal measured from the SDSS data, we additionally calibrate the measurements for source photo-$z$ biases following the method proposed by \cite{nakajima12}. The details of this correction is in \citetalias{Park_2022}.
As is clear from the figure, the lensing amplitudes of one-halo term are consistent, while the two-halo term exhibit the difference in amplitude.
The two-halo terms measured from the SDSS data is suppressed compared to the one measured from the HSC-Y3 data.
The major difference between these signals is the selection of the source galaxies.
Since the mean redshift of the SDSS source galaxies is \hmrv{$\langle z_s \rangle \sim 0.4$}
, we select all the source galaxies behind the lens clusters. 
On the other hand, the mean redshift of the HSC-Y3 source galaxies is \hmrv{$\langle z_s \rangle\sim 1.2$}
and therefore we could select the source galaxies used in our measurements with $z_s>0.6$. 
This difference in selection gives different boost factors as shown in Fig.~\ref{fig:boost_sdss}.
Fig.~\ref{fig:boost_sdss} shows the boost signals measured from the SDSS source galaxies.
Compared to Fig.~\ref{fig:boost_cross}, the boost factors measured from the SDSS data deviate from 1 on small scales due to the \hmrv{contamination to the source galaxies from physically-associated galaxies to the lens clusters.}

At last, the difference in the measured lensing signals explain the difference in the resulting cosmological constraints between \citetalias{Park_2022} and our work.
\citetalias{Park_2022} discussed the possibility of having a problem in the lensing data vector.
Fig.~7 of \citetalias{Park_2022} demonstrated that removing the lensing data vector in their analysis can shift $\Omega_{\rm m}$ to a larger value and therefore a higher $S_8$ value.
Additionally, Fig.~10 of \citetalias{Park_2022} compared the measurements with the best-fit predictions from the analysis fixed to the Planck 2018 cosmology. 
The predicted lensing signals from the best-fit parameters deviate from the measured lensing signals on large scales and have higher amplitudes.
All these findings support our finding that the measured lensing signal from the SDSS data is a cause of lower $\Omega_{\rm m}$ and higher $\sigma_8$ in \citetalias{Park_2022}.

\section{Off-diagonal components of covariance matrix}
\label{app:off-diag}
In \citetalias{Park_2022}, we decided to use a block diagonal covariance matrix assuming each observable is independent. This is because we computed the covariance matrix using a jackknife resampling and a finite number of jackknife regions made the choice of radial binning for clustering and lensing difficult without this assumption.
Here, we explore how off-diagonal components of the covariance matrix between abundance and clustering affect the cosmological parameter constraints.
Since the HSC-Y3 region is small compared to the SDSS region, we still assume that errors of lensing and abundance/clustering are not correlated.

\begin{figure}
	\includegraphics[width=\columnwidth]{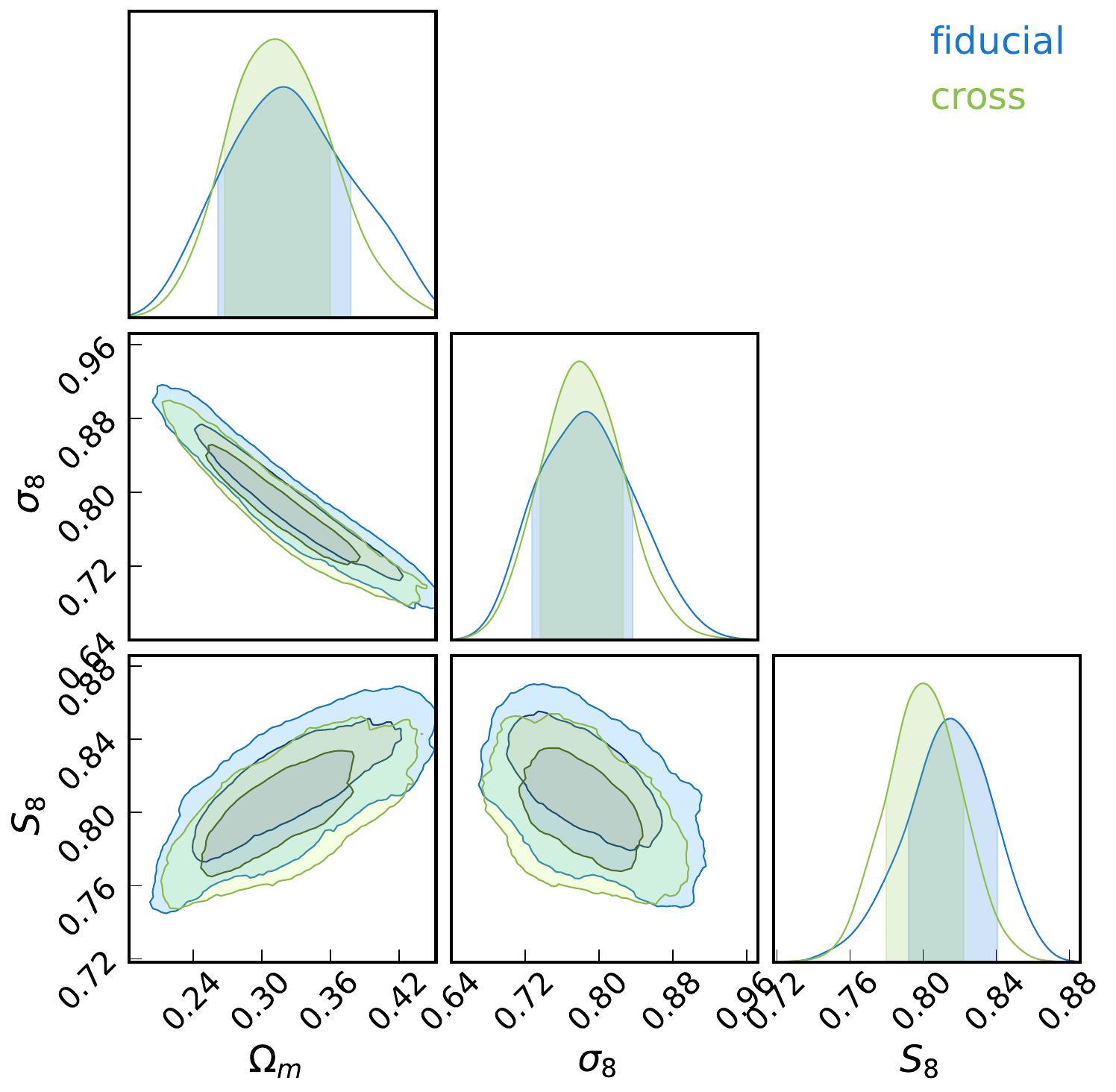}
    \caption{The 68\% and 95\% confidence regions for cosmological parameters of interest using the covariance matrix used in our fiducial analysis (denoted as "fiducial") as well as the covariance matrix including the off-diagonal components between abundance and clustering (denoted as "cross"). 
    }
    \label{fig:off-diag}
\end{figure}

Fig.~\ref{fig:off-diag} shows the 68\% and 95\% confidence regions for cosmological parameters of interest using the covariance matrix with and without the off-diagonal components of the covariance between abundance and clustering, labeled as ``cross'' and ``fiducial'' respectively.
Including the off-diagonal component of the covariance slightly improves the constraint on $\Omega_{\rm m}$ and shifting the value of $S_8$, but we do not find any significant effects on cosmological parameter constraints and therefore decide to follow the analysis setup used in \citetalias{Park_2022}.



\clearpage

\bsp	
\label{lastpage}
\end{document}